\documentclass[11pt, a4paper]{article}
\PassOptionsToPackage{unicode}{hyperref}
\usepackage[yyyymmdd,hhmmss]{datetime}
\usepackage{amssymb}
\usepackage{jheppub}
\usepackage{upgreek}
\usepackage{slashed}

\title{Disk, interval, point: on constructions of quantum field
  theories with holomorphic action functionals}

\author[1,2]{Nafiz Ishtiaque}
\author[2]{and Junya Yagi}

\date{Compiled on \today \; at \currenttime}

\affiliation[1]{Institute for Advanced Study,
  Princeton, NJ,
  USA}

\affiliation[2]{Perimeter Institute for Theoretical Physics, Waterloo,
  ON,
  Canada}

\abstract{Bosonic quantum field theories with holomorphic action
  functionals are realized by two types of constructions involving
  supersymmetric quantum field theories, compactified on an interval
  in one type and compactified on a disk and deformed in the other.
  We establish the equivalence between the two types of constructions
  by reducing the disk to the interval and the interval to a point.
  As examples, we discuss constructions of zero-dimensional gauged
  sigma model, gauged quantum mechanics, gauged symplectic bosons in
  two dimensions, and Chern--Simons theory and its higher-dimensional
  variants.}

\keywords{}

\arxivnumber{}

\let\U\relax
\let\C\relax

\newcommand{\gf}{\mathfrak{g}}

\newcommand{\hf}{\mathfrak{h}}

\newcommand{\del}{\partial}
\newcommand{\delb}{{\bar\partial}}

\newcommand{\vol}{\mathrm{vol}}
\def\ie{\begin{equation}\begin{aligned}}
\def\fe{\end{aligned}\end{equation}}

\newcommand{\Map}{\mathop{\mathrm{Map}}\nolimits}

\newcommand{\Ad}{\mathop{\mathrm{Ad}}\nolimits}

\renewcommand{\Im}{\mathop{\mathrm{Im}}\nolimits}
\renewcommand{\Re}{\mathop{\mathrm{Re}}\nolimits}

\newcommand{\Tr}{\mathop{\mathrm{Tr}}\nolimits}

\newcommand{\crit}{\mathop{\mathrm{Crit}}}

\newcommand{\SU}{\mathrm{SU}}

\newcommand{\Spin}{\mathrm{Spin}}

\newcommand{\U}{\mathrm{U}}

\newcommand{\iso}{\cong}

\newcommand{\Z}{\mathbb{Z}}

\newcommand{\R}{\mathbb{R}}
\newcommand{\C}{\mathbb{C}}

\renewcommand{\P}{\mathbb{P}}

\newcommand{\D}{\mathbb{D}}

\renewcommand{\S}{\mathbb{S}}

\let\nc\newcommand
\let\renc\renewcommand
\nc{\wbar}{\overline}
\let\td\tilde
\let\wtd\widetilde
\let\wht\widehat
\let\mcl\mathcal

\nc{\ab}{{\bar{a}}} \nc{\at}{\tilde{a}} \nc{\ah}{\hat{a}}
\nc{\bb}{{\bar{b}}} \nc{\bt}{\tilde{b}} \nc{\bh}{\hat{b}}
\nc{\cb}{{\bar{c}}} \nc{\ct}{\tilde{c}} 
\nc{\db}{{\bar{d}}} \nc{\dt}{\tilde{d}} \renc{\dh}{\hat{d}}
\nc{\eb}{{\bar{e}}} \nc{\et}{\tilde{e}} \nc{\eh}{\hat{e}}
\nc{\fb}{{\bar{f}}} \nc{\ft}{\tilde{f}} \nc{\fh}{\hat{f}}
\nc{\gb}{{\bar{g}}} \nc{\gt}{\tilde{g}} \nc{\gh}{\hat{g}}
\nc{\hb}{{\bar{h}}} \nc{\hh}{\hat{h}} 
\nc{\ib}{{\bar{\imath}}} \nc{\ih}{\hat{\imath}} 
\nc{\jb}{{\bar{\jmath}}} \nc{\jt}{\tilde{\jmath}} \nc{\jh}{\hat{\jmath}}
\nc{\kb}{{\bar{k}}} \nc{\kt}{\tilde{k}} \nc{\kh}{\hat{k}}
\nc{\lb}{{\bar{l}}} \nc{\lt}{\tilde{l}} \nc{\lh}{\hat{l}}
\nc{\mb}{{\bar{m}}} \nc{\mt}{\tilde{m}} \nc{\mh}{\hat{m}}
\nc{\nb}{{\bar{n}}} \nc{\nt}{\tilde{n}} \nc{\nh}{\hat{n}}
\nc{\ob}{{\bar{o}}} \nc{\ot}{\tilde{o}} \nc{\oh}{\hat{o}}
\nc{\pb}{{\bar{p}}} \nc{\pt}{\tilde{p}} \nc{\ph}{\hat{p}}
\nc{\qb}{{\bar{q}}} \nc{\qt}{\tilde{q}} \nc{\qh}{\hat{q}}
\nc{\rb}{{\bar{r}}} \nc{\rt}{\tilde{r}} \nc{\rh}{{\hat{r}}}
\renc{\sb}{{\bar{s}}} \nc{\st}{\tilde{s}} \nc{\sh}{{\hat{s}}}
\nc{\tb}{{\bar{t}}} \renc{\th}{{\hat{t}}} 
\nc{\ub}{{\bar{u}}} \nc{\ut}{\tilde{u}} \nc{\uh}{\hat{u}}
\nc{\vb}{{\bar{v}}} \nc{\vt}{\tilde{v}} \nc{\vh}{\hat{v}}
\nc{\wb}{{\bar{w}}} \nc{\wt}{\tilde{w}} \nc{\wh}{\hat{w}}
\nc{\xb}{{\bar{x}}} \nc{\xt}{\tilde{x}} \nc{\xh}{\hat{x}}
\nc{\yb}{{\bar{y}}} \nc{\yt}{\tilde{y}} \nc{\yh}{\hat{y}}
\nc{\zb}{{\bar{z}}} \nc{\zt}{\tilde{z}} \nc{\zh}{\hat{z}}

\nc{\Ab}{{\wbar{A}}} \nc{\At}{{\wtd{A}}} \nc{\Ah}{{\wht{A}}}
\nc{\Bb}{{\wbar{B}}} \nc{\Bt}{{\wtd{B}}} \nc{\Bh}{{\wht{B}}}
\nc{\Cb}{{\wbar{C}}} \nc{\Ct}{{\wtd{C}}} \nc{\Ch}{{\wht{C}}}
\nc{\Db}{{\wbar{D}}} \nc{\Dt}{{\wtd{D}}} \nc{\Dh}{{\wht{D}}}
\nc{\Eb}{{\wbar{E}}} \nc{\Et}{{\wtd{E}}} \nc{\Eh}{{\wht{E}}}
\nc{\Fb}{{\wbar{F}}} \nc{\Ft}{{\wtd{F}}} \nc{\Fh}{{\wht{F}}}
\nc{\Gb}{{\wbar{G}}} \nc{\Gt}{{\wtd{G}}} \nc{\Gh}{{\wht{G}}}
\nc{\Hb}{{\wbar{H}}} \nc{\Ht}{{\wtd{H}}} \nc{\Hh}{{\wht{H}}}
\nc{\Ib}{{\bar{I}}} \nc{\It}{{\wtd{I}}} \nc{\Ih}{{\wht{I}}}
\nc{\Jb}{{\wbar{J}}} \nc{\Jt}{{\wtd{J}}} \nc{\Jh}{{\wht{J}}}
\nc{\Kb}{{\wbar{K}}} \nc{\Kt}{{\wtd{K}}} \nc{\Kh}{{\wht{K}}}
\nc{\Lb}{{\wbar{L}}} \nc{\Lt}{{\wtd{L}}} \nc{\Lh}{{\wht{L}}}
\nc{\Mb}{{\wbar{M}}} \nc{\Mt}{{\wtd{M}}} \nc{\Mh}{{\wht{M}}}
\nc{\Nb}{{\wbar{N}}} \nc{\Nt}{{\wtd{N}}} \nc{\Nh}{{\wht{N}}}
\nc{\Ob}{{\wbar{O}}} \nc{\Ot}{{\wtd{O}}} \nc{\Oh}{{\wht{O}}}
\nc{\Pb}{{\wbar{P}}} \nc{\Pt}{{\wtd{P}}} \nc{\Ph}{{\wht{P}}}
\nc{\Qb}{{\wbar{Q}}} \nc{\Qt}{{\wtd{Q}}} \nc{\Qh}{{\wht{Q}}}
\nc{\Rb}{{\wbar{R}}} \nc{\Rt}{{\wtd{R}}} \nc{\Rh}{{\wht{R}}}
\nc{\Sb}{{\wbar{S}}} \nc{\St}{{\wtd{S}}} \nc{\Sh}{{\wht{S}}}
\nc{\Tb}{{\wbar{T}}} \nc{\Tt}{{\wtd{T}}} \nc{\Th}{{\wht{T}}}
\nc{\Ub}{{\wbar{U}}} \nc{\Ut}{{\wtd{U}}} \nc{\Uh}{{\wht{U}}}
\nc{\Vb}{{\wbar{V}}} \nc{\Vt}{{\wtd{V}}} \nc{\Vh}{{\wht{V}}}
\nc{\Wb}{{\wbar{W}}} \nc{\Wt}{{\wtd{W}}} \nc{\Wh}{{\wht{W}}}
\nc{\Xb}{{\wbar{X}}} \nc{\Xt}{{\wtd{X}}} \nc{\Xh}{{\wht{X}}}
\nc{\Yb}{{\wbar{Y}}} \nc{\Yt}{{\wtd{Y}}} \nc{\Yh}{{\wht{Y}}}
\nc{\Zb}{{\wbar{Z}}} \nc{\Zt}{{\wtd{Z}}} \nc{\Zh}{{\wht{Z}}}

\nc{\CA}{{\mcl{A}}} \nc{\CAb}{{\wbar{\CA}}} \nc{\CAt}{{\wtd{\CA}}} \nc{\CAh}{{\wht{\CA}}}
\nc{\CB}{{\mcl{B}}} \nc{\CBb}{{\wbar{\CB}}} \nc{\CBt}{{\wtd{\CB}}} \nc{\CBh}{{\wht{\CB}}}
\nc{\CC}{{\mcl{C}}} \nc{\CCb}{{\wbar{\CC}}} \nc{\CCt}{{\wtd{\CC}}} \nc{\CCh}{{\wht{\CC}}}
\nc{\cD}{{\mcl{D}}} \nc{\cDb}{{\wbar{\cD}}} \nc{\cDt}{{\wtd{\cC}}} \nc{\cDh}{{\wht{\cD}}}
\nc{\CE}{{\mcl{E}}} \nc{\CEb}{{\wbar{\CE}}} \nc{\CEt}{{\wtd{\CE}}} \nc{\CEh}{{\wht{\CE}}}
\nc{\CF}{{\mcl{F}}} \nc{\CFb}{{\wbar{\CF}}} \nc{\CFt}{{\wtd{\CF}}} \nc{\CFh}{{\wht{\CF}}}
\nc{\CG}{{\mcl{G}}} \nc{\CGb}{{\wbar{\CG}}} \nc{\CGt}{{\wtd{\CG}}} \nc{\CGh}{{\wht{\CG}}}
\nc{\CH}{{\mcl{H}}} \nc{\CHb}{{\wbar{\CH}}} \nc{\CHt}{{\wtd{\CH}}} \nc{\CHh}{{\wht{\CH}}}
\nc{\CI}{{\mcl{I}}} \nc{\CIb}{{\wbar{\CI}}} \nc{\CIt}{{\wtd{\CI}}} \nc{\CIh}{{\wht{\CI}}}
\nc{\CJ}{{\mcl{J}}} \nc{\CJb}{{\wbar{\CJ}}} \nc{\CJt}{{\wtd{\CJ}}} \nc{\CJh}{{\wht{\CJ}}}
\nc{\CK}{{\mcl{K}}} \nc{\CKb}{{\wbar{\CK}}} \nc{\CKt}{{\wtd{\CK}}} \nc{\CKh}{{\wht{\CK}}}
\nc{\CL}{{\mcl{L}}} \nc{\CLb}{{\wbar{\CL}}} \nc{\CLt}{{\wtd{\CL}}} \nc{\CLh}{{\wht{\CL}}}
\nc{\CM}{{\mcl{M}}} \nc{\CMb}{{\wbar{\CM}}} \nc{\CMt}{{\wtd{\CM}}} \nc{\CMh}{{\wht{\CM}}}
\nc{\CN}{{\mcl{N}}} \nc{\CNb}{{\wbar{\CN}}} \nc{\CNt}{{\wtd{\CN}}} \nc{\CNh}{{\wht{\CN}}}
\nc{\CO}{{\mcl{O}}} \nc{\COb}{{\wbar{\CO}}} \nc{\COt}{{\wtd{\CO}}} \nc{\COh}{{\wht{\CO}}}
\nc{\CP}{{\mcl{P}}} \nc{\CPb}{{\wbar{\CP}}} \nc{\CPt}{{\wtd{\CP}}} \nc{\CPh}{{\wht{\CP}}}
\nc{\CQ}{{\mcl{Q}}} \nc{\CQb}{{\wbar{\CQ}}} \nc{\CQt}{{\wtd{\CQ}}} \nc{\CQh}{{\wht{\CQ}}}
\nc{\CR}{{\mcl{R}}} \nc{\CRb}{{\wbar{\CR}}} \nc{\CRt}{{\wtd{\CR}}} \nc{\CRh}{{\wht{\CR}}}
\nc{\CS}{{\mcl{S}}} \nc{\CSb}{{\wbar{\CS}}} \nc{\CSt}{{\wtd{\CS}}} \nc{\CSh}{{\wht{\CS}}}
\nc{\CT}{{\mcl{T}}} \nc{\CTb}{{\wbar{\CT}}} \nc{\CTt}{{\wtd{\CT}}} \nc{\CTh}{{\wht{\CT}}}
\nc{\CU}{{\mcl{U}}} \nc{\CUb}{{\wbar{\CU}}} \nc{\CUt}{{\wtd{\CU}}} \nc{\CUh}{{\wht{\CU}}}
\nc{\CV}{{\mcl{V}}} \nc{\CVb}{{\wbar{\CV}}} \nc{\CVt}{{\wtd{\CV}}} \nc{\CVh}{{\wht{\CV}}}
\nc{\CW}{{\mcl{W}}} \nc{\CWb}{{\wbar{\CW}}} \nc{\CWt}{{\wtd{\CW}}} \nc{\CWh}{{\wht{\CW}}}
\nc{\CX}{{\mcl{X}}} \nc{\CXb}{{\wbar{\CX}}} \nc{\CXt}{{\wtd{\CX}}} \nc{\CXh}{{\wht{\CX}}}
\nc{\CY}{{\mcl{Y}}} \nc{\CYb}{{\wbar{\CY}}} \nc{\CYt}{{\wtd{\CY}}} \nc{\CYh}{{\wht{\CY}}}
\nc{\CZ}{{\mcl{Z}}} \nc{\CZb}{{\wbar{\CZ}}} \nc{\CZt}{{\wtd{\CZ}}} \nc{\CZh}{{\wht{\CZ}}}

\let\eps\epsilon
\let\ups\upsilon
\let\veps\varepsilon
\let\vtht\vartheta

\let\vsgm\varsigma
\let\vphi\varphi
\let\vrho\varrho

\nc{\alphab}{{\bar{\alpha}}} \nc{\alphat}{{\td{\alpha}}} \nc{\alphah}{{\hat{\alpha}}}
\nc{\betab}{{\bar{\beta}}}   \nc{\betat}{{\td{\beta}}}   \nc{\betah}{{\hat{\beta}}} 
\nc{\gammab}{{\bar{\gamma}}} \nc{\gammat}{{\td{\gamma}}} \nc{\gammah}{{\hat{\gamma}}} 
\nc{\deltab}{{\bar{\delta}}} \nc{\deltat}{{\td{\delta}}} \nc{\deltah}{{\hat{\delta}}} 
\nc{\epsilonb}{{\bar{\eps}}} \nc{\epsilont}{{\td{\eps}}} \nc{\epsilonh}{{\hat{\eps}}} 
\nc{\vepsb}{{\bar{\veps}}}   \nc{\vepst}{{\td{\veps}}}   \nc{\vepsh}{{\hat{\veps}}} 
\nc{\zetab}{{\bar{\zeta}}}   \nc{\zetat}{{\td{\zeta}}}   \nc{\zetah}{{\hat{\zeta}}} 
\nc{\etab}{{\bar{\eta}}}     \nc{\etat}{{\td{\eta}}}     \nc{\etah}{{\hat{\eta}}} 
\nc{\thetab}{{\bar{\theta}}} \nc{\thetat}{{\td{\theta}}} \nc{\thetah}{{\hat{\theta}}} 
\nc{\vthetab}{{\bar{\vtht}}} \nc{\vthetat}{{\td{\vtht}}} \nc{\vthetah}{{\hat{\vtht}}} 
\nc{\lambdab}{{\bar{\lambda}}} \nc{\lambdat}{{\td{\lambda}}} \nc{\lambdah}{{\hat{\lambda}}} 
\nc{\iotab}{{\bar{\iota}}}   \nc{\iotat}{{\td{\iota}}}   \nc{\iotah}{{\hat{\iota}}} 
\nc{\kappab}{{\bar{\kappa}}} \nc{\kappat}{{\td{\kappa}}} \nc{\kappah}{{\hat{\kapvpa}}} 
\nc{\lmdb}{{\bar{\lmd}}}     \nc{\lmdt}{{\td{\lmd}}}     \nc{\lmdh}{{\hat{\lmd}}} 
\nc{\mub}{{\bar{\mu}}}       \nc{\mut}{{\td{\mu}}}       \nc{\muh}{{\hat{\mu}}} 
\nc{\nub}{{\bar{\nu}}}       \nc{\nut}{{\td{\nu}}}       \nc{\nuh}{{\hat{\nu}}} 
\nc{\xib}{{\bar{\xi}}}       \nc{\xit}{{\td{\xi}}}       \nc{\xih}{{\hat{\xi}}} 
\nc{\pib}{{\bar{\pi}}}       \nc{\pit}{{\td{\pi}}}       \nc{\pih}{{\hat{\pi}}} 
\nc{\vpib}{{\bar{\vpi}}}     \nc{\vpit}{{\td{\vpi}}}     \nc{\vpih}{{\hat{\vpi}}} 
\nc{\rhob}{{\bar{\rho}}}     \nc{\rhot}{{\td{\rho}}}     \nc{\rhoh}{{\hat{\rho}}} 
\nc{\vrhob}{{\bar{\vrho}}}   \nc{\vrhot}{{\td{\vrho}}}   \nc{\vrhoh}{{\hat{\vrho}}} 
\nc{\sigmab}{{\bar{\sigma}}} \nc{\sigmat}{{\td{\sigma}}} \nc{\sigmah}{{\hat{\sigma}}} 
\nc{\vsigmab}{{\bar{\vsgm}}} \nc{\vsigmat}{{\td{\vsgm}}} \nc{\vsigmah}{{\hat{\vsgm}}} 
\nc{\taub}{{\bar{\tau}}}     \nc{\taut}{{\td{\tau}}}     \nc{\tauh}{{\hat{\tau}}} 
\nc{\upsb}{{\bar{\ups}}} \nc{\upst}{{\td{\ups}}} \nc{\upsh}{{\hat{\ups}}} 
\nc{\phib}{{\bar{\phi}}}     \nc{\phit}{{\td{\phi}}}     \nc{\phih}{{\hat{\phi}}} 
\nc{\varphib}{{\bar{\vphi}}}   \nc{\varphit}{{\td{\vphi}}}   \nc{\varphih}{{\hat{\vphi}}} 
\nc{\chib}{{\bar{\chi}}}     \nc{\chit}{{\td{\chi}}}     \nc{\chih}{{\hat{\chi}}} 
\nc{\psib}{{\bar{\psi}}}     \nc{\psit}{{\td{\psi}}}     \nc{\psih}{{\hat{\psi}}} 
\nc{\omegab}{{\bar{\omega}}} \nc{\omegat}{{\td{\omega}}} \nc{\omegah}{{\hat{\omega}}} 

\nc{\Gammab}{{\wbar{\Gamma}}}     \nc{\Gammat}{{\wtd{\Gamma}}}     \nc{\Gammah}{{\wht{\Gamma}}}
\nc{\Deltab}{{\wbar{\Delta}}}     \nc{\Deltat}{{\wtd{\Delta}}}     \nc{\Deltah}{{\wht{\Delta}}}
\nc{\Thetab}{{\wbar{\Theta}}}     \nc{\Thetat}{{\wtd{\Theta}}}     \nc{\Thetah}{{\wht{\Theta}}}
\nc{\Lambdab}{{\wbar{\Lambda}}}   \nc{\Lambdat}{{\wtd{\Lambda}}}   \nc{\Lambdah}{{\wht{\Lambda}}}
\nc{\Xib}{{\wbar{\Xi}}}           \nc{\Xit}{{\wtd{\Xi}}}           \nc{\Xih}{{\wht{\Xi}}}
\nc{\Pib}{{\wbar{\Pi}}}           \nc{\Pit}{{\wtd{\Pi}}}           \nc{\Pih}{{\wht{\Pi}}}
\nc{\Sigmab}{{\wbar{\Sigma}}}     \nc{\Sigmat}{{\wtd{\Sigma}}}     \nc{\Sigmah}{{\wht{\Sigma}}}
\nc{\Upsilonb}{{\wbar{\Upsilon}}} \nc{\Upsilont}{{\wtd{\Upsilon}}} \nc{\Upsilonh}{{\wht{\Upsilon}}}
\nc{\Phib}{{\wbar{\Phi}}} \nc{\Phit}{{\wtd{\Phi}}} \nc{\Phih}{{\wht{\Phi}}}
\nc{\Psib}{{\wbar{\Psi}}}         \nc{\Psit}{{\wtd{\Psi}}}         \nc{\Psih}{{\wht{\Psi}}}
\nc{\Omegab}{{\wbar{\Omega}}}     \nc{\Omegat}{{\wtd{\Omega}}}     \nc{\Omegah}{{\wht{\Omega}}}

\newcommand{\rmd}{\mathrm{d}}
\newcommand{\epsb}{\epsilonb}
\newcommand{\ChS}{\mathrm{CS}}

\newcommand{\auxF}{\mathsf{F}}
\newcommand{\auxFb}{\overline{\auxF}}
\newcommand{\auxG}{\mathsf{G}}
\newcommand{\auxGb}{\overline{\auxG}}
\newcommand{\auxD}{\mathsf{D}}

\newcommand{\auxH}{\mathsf{H}}

\newcommand{\iu}{\mathrm{i}}

\let\starx\star
\let\star\relax
\newcommand{\star}{\mathop{\starx}\nolimits}

\newcommand{\nablat}{{\widetilde{\nabla}}}
\newcommand{\nablah}{{\widehat{\nabla}}}
\renewcommand{\th}{t}

\newcommand{\mm}{\upmu} 
\newcommand{\sslash}{\mathord{/\mkern-5mu/}} 
\newcommand{\ssslash}{\mathord{/\mkern-5mu/\mkern-5mu/}} 

\usepackage[mathscr]{euscript}

\newcommand{\SL}{\mathscr{L}}
\newcommand{\SM}{\mathscr{M}}
\newcommand{\SX}{\mathscr{X}}

\newcommand{\Morse}{h}
\newcommand{\wsmetric}{\gamma}

\newcommand{\BLag}{L}
\newcommand{\BX}{X} 
\newcommand{\Bomega}{\omega}
\newcommand{\Bnabla}{\nabla}
\newcommand{\BW}{W}
\newcommand{\BWb}{\Wb}

\newcommand{\fg}{a} 

\newcommand{\LA}{f}
\newcommand{\LE}{E}
\newcommand{\LI}{I}
\newcommand{\LJ}{J}

\newcommand{\I}{\mathbb{I}}
\newcommand{\M}{\mathbb{M}}

\newcommand{\QQ}{\smash{\slashed{Q}}}
\newcommand{\QQb}{\smash{\overline{\slashed{Q}}}}
\renewcommand{\Qb}{\smash{\overline{Q}}}

\newcommand{\cstr}{j} 

\begin{document}
\maketitle

\section{Introduction}

The purpose of this paper is to build a bridge between two types of
constructions of quantum field theories, typically bosonic ones, from
supersymmetric quantum field theories in higher spacetime dimensions.
These constructions have appeared in such diverse contexts as
quantization of symplectic manifolds~\cite{Gukov:2008ve,
  Yagi:2014toa}, vertex algebras in three-~\cite{Gaiotto:2017euk,
  Costello:2018fnz} and four-dimensional~\cite{Beem:2013sza,
  Oh:2019bgz, Jeong:2019pzg} supersymmetric field theories, analytic
continuation of knot invariants~\cite{Witten:2010cx, Witten:2010zr}
and their categorification~\cite{Witten:2011zz}, the 3d-3d
correspondence~\cite{Dimofte:2010tz, Terashima:2011qi,
  Terashima:2011xe, Cecotti:2011iy, Dimofte:2011jd, Dimofte:2011ju,
  Dimofte:2011py, Luo:2014sva}, and emergence of integrable systems
from supersymmetric gauge theories~\cite{Nekrasov:2009uh,
  Nekrasov:2009ui, Chen:2011sj, Nekrasov:2012xe, Nekrasov:2013xda,
  Yamazaki:2012cp, Yagi:2015lha, Costello:2013zra, Costello:2013sla,
  Costello:2017dso, Costello:2018gyb, Costello:2018txb}.

We will refer the two types of constructions as the A-type and the
B-type.  There are some similarities between them, but also crucial
differences.

In an A-type construction, one starts with a $(d+1)$-dimensional
supersymmetric field theory, formulated on a product $\I \times \M_d$,
where $\I$ is an interval $[0,\ell]$ and $\M_d$ is a $d$-manifold.
The theory is topologically twisted, either fully or partially, so
that it is topological on $\I$.  Let $s$ be the coordinate on $\I$.

At one end of $\I$, say at $s = 0$, one imposes Neumann-like boundary
conditions on the bosonic matter fields.  At the other end, at
$s = \ell$, boundary conditions constrain the bosonic matter fields to
be, roughly speaking, valued in a Lagrangian submanifold $M$ of their
K\"ahler%
\footnote{More generally, $Y$ may be complex and almost symplectic but
  non-K\"ahler.  We will consider such target spaces in
  sections~\ref{sec:B-SQM-0dSM} and~\ref{sec:RW-A-QM}.}
target space $Y$.

Since the theory is topological on $\I$, one can shorten $\I$ until it
collapses to a point.  One thus obtains a $d$-dimensional theory on
$\M_d$.  It turns out that the action functional of this theory on
$\M_d$ is holomorphic in complex-valued bosonic fields, but the path
integral makes sense for appropriate choices of $M$ as it is performed
over a middle-dimensional cycle determined by $M$.

In a prototypical example~\cite{Gukov:2008ve} of an A-type
construction, $d = 1$ and the two-dimensional theory on
$\I \times \M_1$ is the A-model~\cite{Witten:1988xj}, which may be
obtained from two-dimensional $\CN = (2,2)$ supersymmetric sigma model
by the A-twist.  The target space $Y$ is a complex symplectic
manifold.  Located at $s = 0$ is a canonical coisotropic
brane~\cite{Kapustin:2001ij}, supported on all of $Y$.  At $s = \ell$,
one chooses a brane whose support $M$ is Lagrangian with respect to
two of the real symplectic structures of $Y$, but symplectic with
respect to another.  The one-dimensional theory on $\M_1$ is quantum
mechanics whose phase space is $M$, regarded as a symplectic manifold.

More elaborate examples arise for $d \geq 2$.

For $d = 2$, one may take the three-dimensional theory to be an
$\CN = 4$ supersymmetric field theory in the
A-twist~\cite{Blau:1996bx} (which is the ``mirror'' of the
Rozansky--Witten twist).  The two-dimensional theory produced by
reduction on $\I$ is a chiral conformal field theory (CFT) on a
Riemann surface $\M_2$~\cite{Gaiotto:2017euk, Costello:2018fnz}.  If
the theory one picks is a gauge theory constructed from vector
multiplets and hypermultiplets in a complex symplectic representation
$Y$ of the gauge group, the CFT is the system of gauged symplectic
bosons valued in $Y$.

An example with $d = 3$ is provided by the
GL-twist~\cite{Yamron:1988qc, Marcus:1995mq, Kapustin:2006pk} of
$\CN = 4$ super Yang--Mills theory in four dimensions.  In this case,
the three-dimensional theory is Chern--Simons theory on a
three-manifold $\M_3$, whose gauge group is the complexification
$H_\C$ of the gauge group $H$ of the Yang--Mills
theory~\cite{Gukov:2008ve, Witten:2010cx, Witten:2010zr}.

A B-type construction starts with a $(d+2)$-dimensional supersymmetric
field theory on $\D \times \M_d$, where $\D$ is a disk.  The
construction involves a special kind of deformation of the theory,
called the $\Omega$-deformation~\cite{Nekrasov:2002qd,
  Nekrasov:2003rj}.  The theory is topologically twisted so that it is
topological on $\D$, but the $\Omega$-deformation reduces the
topological invariance to the invariance under deformations that leave
the rotation symmetry of $\D$ unbroken.  On the boundary $\del\D$ of
the disk, one imposes a boundary condition such that the bosonic
matter fields take values in a Lagrangian submanifold $M$ of the
K\"ahler target space $Y$.

As in A-type constructions, one can shrink $\D$ to a point to obtain a
$d$-dimensional theory on $\M_d$.  Again, the resulting theory turns
out to have a holomorphic action functional, and the path integral is
performed over a middle-dimensional cycle specified by $M$.

In an example~\cite{Yagi:2014toa} with $d = 1$, the three-dimensional
theory on $\D \times \M_1$ is Rozansky--Witten
theory~\cite{Rozansky:1996bq}, which arises from $\CN = 4$
supersymmetric sigma model by the B-twist (or the Rozansky--Witten
twist).  The target space $Y$ is a complex symplectic manifold, and
the boundary condition on $\del\D$ requires the bosonic field to lie
in a submanifold $M$ that is either Lagrangian or symplectic,
depending on which real symplectic structure one refers to.  The
one-dimensional theory on $\M_1$ that arises after the
$\Omega$-deformation and reduction on $\D$ is quantum mechanics,
quantizing the symplectic manifold $M$.

For an example with $d = 2$, one may consider Kapustin's
holomorphic--topological twist~\cite{Kapustin:2006hi} of an $\CN = 2$
supersymmetric field theory in four dimensions.  The
$\Omega$-deformation and reduction on $\D$ result in a chiral CFT,
which, for a gauge theory with hypermultiplets in a complex symplectic
representation $Y$, is the system of gauged symplectic bosons valued
in $Y$~\cite{Oh:2019bgz,Jeong:2019pzg}.

Finally, for $d = 3$, the $\Omega$-deformation and reduction of a
topological twist of five-dimensional $\CN = 2$ super Yang--Mills
theory with gauge group $H$ yields Chern--Simons theory with gauge
group $H_\C$~\cite{Luo:2014sva}.

Above we have described examples of A-type and B-type constructions
for $d = 1$, $2$, $3$.  The parallel between the two types is
conspicuous: for every $d$, there is a pair of A-type and B-type
constructions for one and the same $d$-dimensional bosonic theory.

This observation suggests that for each $d$, the theory used in the
A-type construction should be somehow related to the one used in the
corresponding B-type construction.  The expectation is further
strengthened if one notices the fact that the two theories may be
obtained by topological twist from two physical theories with the same
amount of supersymmetry.

In fact, there is a natural way to produce a $(d+1)$-dimensional
theory on $\I \times \M_d$ from a $(d+2)$-dimensional theory on
$\D \times \M_d$: one deforms the disk $\D$ into the shape of a cigar
and performs circle reduction of the latter theory, considering the
cigar as a circle fibration over the interval $\I$.  Such a
deformation of $\D$ is allowed in the theory for the B-type
construction since it preserves rotation symmetry.  A similar
reduction has been studied before in a different but related
setting~\cite{Nekrasov:2010ka}.

Presumably, the B-type theory turns into its A-type counterpart
through the $\Omega$-deformation and the cigar reduction of $\D$ to
$\I$.  In this paper we demonstrate that this is indeed true.

Our strategy is to first establish the correspondence between A-type
and B-type constructions in the most basic case, namely the case in
which $d = 0$.  Then, we apply the results to higher-dimensional
examples.

For $d = 0$, the two-dimensional theory relevant for B-type
construction is the gauged B-model~\cite{Baptista:2007ap} on $\D$,
which may originate from $\CN = (2,2)$ supersymmetric gauged sigma
model via the B-twist~\cite{Vafa:1990mu, Witten:1991zz}.  The target
space of the model is a K\"ahler manifold $X$ with a holomorphic
$G_\C$-action, where $G$ is the gauge group.  To this theory we apply
the $\Omega$-deformation~\cite{Yagi:2014toa, Luo:2014sva}.  On
$\del\D$, we place a brane whose support $L$ is $G_\C$-invariant and
defines a Lagrangian submanifold $\SL$ inside the K\"ahler quotient
$X \sslash G$.

We will show that upon cigar reduction, the $\Omega$-deformed gauged
B-model on $\D$ becomes topologically twisted supersymmetric gauged
quantum mechanics whose target space is $X$.  The theory thus obtained
on $\I$ is the A-type theory for $d = 0$.  The brane on $\del\D$
descends to a similar boundary condition at $s = \ell$, while the
center of $\D$, or the ``tip'' of the cigar, becomes the boundary at
$s = 0$ where the Neumann-like boundary conditions for the bosonic
matter fields emerge.

We will also show that in the limit where $\I$ shrinks to a point,
this supersymmetric gauged quantum mechanics reduces to a gauged sigma
model on a point $\M_0$ with target $L$, provided that $L$ is chosen
appropriately.  This is the bosonic theory for $d = 0$.  Remarkably,
the gauge group complexifies to $G_\C$ in the process of this
reduction.

To establish the link between the A-type and B-type constructions for
$d \geq 1$, we merely apply the results we have obtained for $d = 0$
to various infinite-dimensional target spaces.  The point is that the
B-type theory on $\D \times \M_d$ may be regarded as the gauged
B-model on $\D$, whereas the A-type theory on $\I \times \M_d$ may be
regarded as supersymmetric gauged quantum mechanics on $\I$, and the
two have the same target space $X$.  The bosonic theory on $\M_d$ may
be viewed as a zero-dimensional gauged sigma model with target $L$.

For example, for $d = 1$, we take $X = \Map(\M_1, Y)$, the space of
maps from $\M_1$ to a complex symplectic manifold $Y$.  For the brane
on $\del\D$, we choose an appropriate submanifold $M \subset Y$ and
set $L = \Map(\M_1, M)$.  The A-model on $\I \times \M_1$ with target
$Y$ is topologically twisted supersymmetric quantum mechanics on $\I$
with target $X$, while Rozansky--Witten theory on $\D \times \M_1$
with target $Y$ is the B-model on $\D$ with target $X$.  Hence, these
two theories are related by cigar reduction, and realize the same
zero-dimensional sigma model with target $L$.%
\footnote{We have learned from Dylan Butson that this statement and
  related results may also be understood from the point of view of
  equivariant factorization algebras.}
The last theory describes maps from $\M_1$ to $M$, so it is quantum
mechanics with phase space $M$.

Similarly, for $d = 2$, we take $X = \Map(\M_2, Y)$, the space of maps
from $\M_2$ to the vector space $Y$ for a complex symplectic
representation of a gauge group.  For $d = 3$, we take $X$ to be the
space of $H_\C$ gauge fields on $\M_3$.  The relation between the
A-type and B-type constructions for $d = 1$, $2$, $3$ are thus
established.

If one knows either an A-type or B-type construction of a certain
bosonic theory, one may exploit the relation just explained to arrive
at the corresponding construction of the other type.  In this way we
will deduce an A-type construction of a four-dimensional variant of
Chern--Simons theory~\cite{Costello:2013zra, Costello:2013sla,
  Costello:2017dso} based on five-dimensional $\CN = 2$ super
Yang--Mills theory, starting from the B-type construction using
six-dimensional $\CN = (1,1)$ super Yang--Mills
theory~\cite{Costello:2018txb}.  This A-type construction was
essentially proposed in~\cite{Ashwinkumar:2018tmm,
  Ashwinkumar:2019mtj}.

In the examples with $d = 3$, $4$ described above, $d$-dimensional
Chern--Simons theory is realized by $(d+1)$- and $(d+2)$-dimensional
maximally supersymmetric Yang--Mills theories.  This pattern continues
to hold for $d = 5$, $6$, and we will explain the A-type and B-type
constructions in these cases.  Six-dimensional Chern--Simons theory is
more commonly known as holomorphic Chern--Simons
theory~\cite{Witten:1992fb}.  Five-dimensional Chern--Simons theory
was introduced in~\cite{Costello:2016nkh}.

Table~\ref{table:A-B-constructions} summarizes the A-type and B-type
constructions treated in this paper.

\begin{table}
  \centering
  \begin{tabular}{cccc}
    \hline\hline
    $d$ & Bosonic theory  & A-type theory & B-type theory
    \\
    \hline
    0 & Sigma model & Topological SQM  & B-model \\
    1 & Quantum mechanics & A-model & Rozansky--Witten \\
    2 & CFT & A-twisted $\CN = 4$ & Kapustin \\
    3 & Chern--Simons & GL-twisted $\CN = 4$ SYM & Topological $\CN = 2$ SYM \\
    4 & 4d Chern--Simons & Twisted $\CN = 2$ SYM & Twisted $\CN = (1,1)$ SYM \\
    5 & 5d Chern--Simons & Twisted $\CN = (1,1)$ SYM & Twisted SYM \\
    6 & 6d Chern--Simons & Twisted SYM & Twisted SYM
    \\
    \hline\hline
  \end{tabular}
  \caption{Examples of A-type and B-type constructions.  SQM and SYM
    are abbreviations for ``supersymmetric quantum mechanics'' and
    ``super Yang--Mills theory,'' respectively.}
  \label{table:A-B-constructions}
\end{table}

In concluding this introduction, a few of remarks are in order.

First, for the constructions of higher-dimensional Chern--Simons
theories to be completely satisfactory, super Yang--Mills theories in
dimension greater than four should probably be provided with
ultraviolet completion.  One way to do so is to embed them into string
theory using branes.  This approach proves to be fruitful, as it
allows one to exploit the rich structure of dualities in string
theory~\cite{Witten:2011zz, Costello:2018txb}.

Second, it seems that a large class of bosonic gauge theories with
holomorphic action functionals and complex gauge groups admit A-type
and B-type constructions, at least formally, since they can always be
reformulated as zero-dimensional gauged sigma models.  What
distinguishes the examples we consider is that the corresponding
A-type and B-type theories are physically natural and interesting.  In
contrast, the A-type and B-type theories will not be so nice if one
takes a generic bosonic gauge theory.  Their actions will lack Lorentz
invariance and contain higher-derivative terms.

Last, although our treatment of infinite-dimensional target spaces may
seem naive, it is justified.  These spaces are essentially the field
spaces of the bosonic theories on $\M_d$.  In general, the definition
of a quantum field theory comes with a regularization of ultraviolet
divergences.  Whatever the choice of a regularization we make for a
bosonic theory, we use the same regularization for the corresponding
A-type and B-type theories.  For instance, one may latticize $\M_d$;
the twisted supercharges are compatible with a lattice regularization
since they do not generate translations on $\M_d$.  Then, the field
space becomes the product of copies of a finite-dimensional space.

This paper is organized as follows.  We begin in
section~\ref{sec:B-SQM-0dSM} by studying the A-type and B-type
constructions for $d = 0$ without gauge symmetry.  Then, in
section~\ref{sec:RW-A-QM}, we apply these constructions to demonstrate
the equivalences between the $\Omega$-deformed Rozansky--Witten
theory, the A-model and quantum mechanics.  We incorporate gauge
symmetry into the picture in section~\ref{sec:gauge}.  Finally, in
section~\ref{sec:gauge-applications}, we discuss constructions of
gauged quantum mechanics, gauged symplectic bosons, and Chern--Simons
theory and its higher-dimensional variants by supersymmetric gauge
theories.  Appendix~\ref{sec:8dSYM} explains the formulation of
eight-dimensional super Yang--Mills theory as an $\CN = (2,2)$
supersymmetric gauge theory in two dimensions.

\section{B-model, supersymmetric quantum mechanics and
  zero-dimensional sigma model}
\label{sec:B-SQM-0dSM}

In this section we discuss the fundamental A-type and B-type
constructions, which realize sigma model on a point $\M_0$ within
supersymmetric quantum mechanics on an interval $\I$ and the
$\Omega$-deformed B-model on a cigar $\D$, respectively.  To alleviate
technicalities, we will not consider gauge symmetry yet.

After we formulate the $\Omega$-deformed B-model with a complex target
space, we will show that its circle reduction gives supersymmetric
quantum mechanics.  Then, we will discuss boundary conditions in the
respective theories, and reduction to a zero-dimensional theory.
Finally, we will explain how to construct good boundary conditions
using a gradient flow.

\subsection{\texorpdfstring{$\Omega$-deformation of the
    B-model}{Ω-deformation of the B-model}}

The B-model is a topological quantum field theory of cohomological
type, and may be constructed from $\CN = (2,2)$ supersymmetric sigma
model by the B-type topological twist~\cite{Vafa:1990mu,
  Witten:1991zz}.  For the presence of $\CN = (2,2)$ supersymmetry,
the target space of the model must be a K\"ahler manifold, and for the
B-twist to make sense, it must moreover be Calabi--Yau.  If, however,
one does not require the B-model to originate from a physical theory
via topological twist, the target space can be more general.

For us, the target space is a complex manifold $\BX$ with complex
structure $\LI$, which is not necessarily K\"ahler.  We still require
its first Chern class to vanish, $c_1(\BX) = 0$, so that the theory
suffers no anomalies.  If $\BX$ is a K\"ahler manifold, this
requirement means that $\BX$ is Calabi--Yau.  We will use letters
$\mu$, $\nu$, $\dotsc$ for real indices, $i$, $j$, $\dotsc$ for
holomorphic indices, and $\ib$, $\jb$, $\dotsc$ for antiholomorphic
indices.

The spacetime (or worldsheet) of the B-model is a surface $\Sigma$,
endowed with a Riemannian metric $\wsmetric$.  In this paper we will take
$\Sigma$ to have rotation symmetry and $\wsmetric$ to be rotation invariant.

The final input data of the B-model is the superpotential $\BW$, which
is a holomorphic function on $\BX$.

The fields of the B-model are a bosonic field
\begin{equation}
  \label{eq:B-varphi}
  \varphi \in \Map(\Sigma, \BX)
\end{equation}
and fermionic fields
\begin{align}
  \eta &\in \Pi\Omega^0(\Sigma, \varphi^*T^{0,1}\BX) \,,
  \\
  \rho &\in \Pi\Omega^1(\Sigma, \varphi^*T^{1,0}\BX) \,,
  \\
  \mu &\in \Pi\Omega^2(\Sigma, \varphi^*T^{0,1}\BX) \,.    
\end{align}
Here $\Omega^p(\Sigma,\LE)$ is the space of $p$-forms on $\Sigma$ with
values in the vector bundle $\LE$ over $\Sigma$, and $T^{1,0}\BX$ and
$T^{0,1}\BX$ are the holomorphic and antiholomorphic tangent bundles
of $\BX$, respectively; $\Pi$ denotes parity reversal.  In the
off-shell formulation which we will employ, the theory also has
auxiliary bosonic two-form fields
\begin{align}
  \auxG \in \Omega^2(\Sigma, \varphi^*T^{1,0}\BX) \,,
  \\
  \label{eq:B-Gb}
  \auxGb \in \Omega^2(\Sigma, \varphi^*T^{0,1}\BX) \,.
\end{align}
All of these fields come from a chiral multiplet of $\CN = (2,2)$
supersymmetry.

The B-model has supersymmetry generated by a fermionic conserved
charge $Q_0$.  The supercharge $Q_0$ squares to zero,
\begin{equation}
  Q_0^2 = 0 \,,
\end{equation}
and is used to define cohomology in the space of states and in the
space of operators.  The path integral of the theory, with
$Q_0$-closed operators inserted on $\Sigma$ and $Q_0$-closed states
specified on $\del\Sigma$, depends only on the $Q_0$-cohomology
classes of those operators and states.

The essential point is that the metric $\wsmetric$ on $\Sigma$ enters
the theory only through $Q_0$-exact terms in the action.  As a
consequence, under deformations of $\wsmetric$, the integrand of the
path integral varies by $Q_0$-exact terms and its $Q_0$-cohomology
class remains intact.  Thus, by passing to $Q_0$-cohomology, the
theory becomes invariant under deformations of $\wsmetric$.  In this
sense the B-model is a topological theory.

We will not explain here how $Q_0$ acts on the fields and how the
action functional of the B-model is constructed.  Rather, we directly
proceed to describe the $\Omega$-deformation of the B-model.

The $\Omega$-deformation of the B-model~\cite{Yagi:2014toa} is a
deformation that may be applied whenever $\Sigma$ admits an isometry.
Let $V$ be a Killing vector field generating this isometry.  The
$\Omega$-deformed B-model has deformed supersymmetry generated by a
supercharge $Q_V$.  The $\Omega$-deformed supercharge $Q_V$ reduces to
$Q_0$ for $V = 0$, and squares to the generator $\CL_V$ of the
isometry:
\begin{equation}
  Q_V^2 = \CL_V \,.
\end{equation}
On fields, $\CL_V$ acts by the Lie derivative by $V$.

Slightly more generally, we allow $V$ to be a complex linear
combination of Killing vector fields, provided that it commutes with
its complex conjugate $\Vb$:
\begin{equation}
  [V, \Vb] = 0 \,.
\end{equation}
This condition ensures that $\CL_V$ commutes with $\iota_\Vb$.%
\footnote{A quick way to see this is to note that since $V + \Vb$ and
  $\iu(V - \Vb)$ are two commuting real vector fields, one can find
  local coordinates $(x,y)$ such that $V = \del_x + \iu\del_y$ (or
  $V = c\del_x$ for some $c \in \C$ if $V$ and $\Vb$ are linearly
  dependent).  In terms of these coordinates, $\CL_V$ acts on tensors
  by $\del_x + \iu\del_y$, and the commutativity is obvious.}
The fact that $V$ generates a complexified isometry implies that
$\CL_V$ commutes with the Hodge star operator $\star$.

As in the case of the ordinary B-model, in the $\Omega$-deformed
B-model one considers the $Q_V$-cohomologies of states and operators.
The difference is that in the $\Omega$-deformed case, one has to
restrict the action of the supercharge to $V$-invariant states and
operators, for only in the spaces of such states and operators does
one have the relation $Q_V^2 = 0$.  By restriction to the
$Q_V$-invariant sector, the $\Omega$-deformed B-model becomes
quasi-topological: the path integral is invariant under deformations
of $\wsmetric$ as long as $V$ remains as a Killing vector field.

Now we describe the $\Omega$-deformation more explicitly.

The field content of the $\Omega$-deformed B-model is the same as that
of the ordinary B-model.  On the fields, $Q_V$ acts by the variations
\begin{align}
  \label{eq:OB-SUSY-varphi}
  \delta\varphi^i &= \iota_V\rho^i,
  \\
  \label{eq:OB-SUSY-rho}
  \delta\rho^i &= \rmd\varphi^i + \iota_V \auxG^i,
  \\
  \delta \auxG^i &= \rmd\rho^i \,,
  \\
  \delta\varphib^\ib &= \eta^\ib,
  \\
  \delta\eta^\ib &= V(\varphib^\ib),
  \\
  \delta\mu^\ib &= \auxGb^\ib,
  \\
  \label{eq:OB-SUSY-Gb}
  \delta\auxGb^\ib &= \rmd \iota_V\mu^\ib \,,
\end{align}
where we have described the map $\varphi$ locally on $\BX$ with a
tuple of complex functions $(\varphi^i, \varphib^\ib)$, corresponding
to holomorphic and antiholomorphic coordinates.  We see that the
vector field $\delta_{Q_V}$ in the field space representing the action
of $Q_V$ satisfies
\begin{equation}
  \label{eq:OB-delta2}
  \delta_{Q_V}^2 = \rmd\iota_V + \iota_V\rmd \,.
\end{equation}
The right-hand side is the Lie derivative by $V$ on differential
forms.

While the
transformations~\eqref{eq:OB-SUSY-varphi}--\eqref{eq:OB-SUSY-Gb}
satisfy the desired supersymmetry algebra, they are not covariant
under diffeomorphisms of $\BX$ because $\delta_{Q_V}$ and $\rmd$ are
not covariant derivatives.  For the construction of the model, it will
be more convenient to rewrite the above formulas in manifestly
covariant forms.  This can be achieved as follows.

Choose a torsion-free connection $\nabla$ on $X$ that preserves the
complex structure:
\begin{equation}
  \nabla\LI = 0 \,.
\end{equation}
The connection coefficients $\Gamma$ of $\nabla$ are symmetric,
\begin{equation}
  \Gamma^\mu_{\nu\rho} = \Gamma^\mu_{\rho\nu} \,,
\end{equation}
and has no mixed components, that is,
\begin{equation}
  \Gamma^\mu_{\nu\rho} = 0
\end{equation}
unless $\mu$, $\nu$, $\rho$ are all holomorphic or all antiholomorphic
indices.  When $\BX$ is K\"ahler, many of the formulas that follow
will simplify greatly if one chooses $\nabla$ to be the Levi-Civita
connection associated with the K\"ahler metric.

Using $\nabla$, we define the covariant exterior derivative
$\rmd_\nabla$ as the exterior derivative $\rmd$ coupled to the
pullback of $\nabla$ by $\varphi$.  For example,
\begin{equation}
  \rmd_\nabla \rho^i
  = \rmd\rho^i + \rmd\varphi^k \Gamma^i_{kj} \wedge \rho^j \,.
\end{equation}
Likewise, we define the covariant variation by
\begin{equation}
  \delta_\nabla \rho^i
  = \delta\rho^i + \delta\varphi^k \Gamma^i_{kj} \wedge \rho^j
\end{equation}
and so on.  Also, we introduce new auxiliary fields $\auxF^i$,
$\auxFb^\ib$ by
\begin{align}
  \label{eq:auxF}
  \auxF^i
  &= \auxG^i + \frac12 \Gamma_{jk}^i \rho^j \wedge \rho^k \,,
  \\
  \label{eq:auxFb}
  \auxFb^\ib
  &= \auxGb^\ib + \Gamma_{\jb\kb}^\ib \eta^\jb \mu^\kb \,.
\end{align}
Then, the variations~\eqref{eq:OB-SUSY-varphi}--\eqref{eq:OB-SUSY-Gb}
can be written covariantly as
\begin{align}
  \label{eq:OB-SUSY-covariant}
  \delta\varphi^i &= \iota_V\rho^i \,,
  \\
  \delta_\nabla\rho^i
  &= \rmd\varphi^i + \iota_V \auxF^i \,,
  \\
  \delta_\nabla \auxF^i
  &= \rmd_\nabla\rho^i
    - \biggl(\frac13 (R_\nabla)^i{}_{jkl}  \iota_V\rho^l
      + \frac12 (R_\nabla)^i{}_{jk\lb}   \eta^\lb\biggr)
  \rho^j \wedge \rho^k \,,
  \\
  \delta\varphib^\ib &= \eta^\ib \,,
  \\
  \delta_\nabla\eta^\ib &= V(\varphib^\ib) \,,
  \\
  \delta_\nabla\mu^\ib &= \auxFb^\ib \,,
  \\
  \delta_\nabla\auxFb^\ib
  &= \rmd_\nabla \iota_V\mu^\ib
  - \biggl((R_\nabla)^\ib{}_{\jb\kb l} \iota_V\rho^l
  + \frac12 (R_\nabla)^\ib{}_{\kb\jb\lb} \eta^\lb\biggr)
  \eta^\jb \mu^\kb \,.
\end{align}
Here $R_\nabla$ is the curvature tensor for $\nabla$:
\begin{equation}
  (R_\nabla)^\mu{}_{\nu\rho\sigma}
  =
  \del_\rho\Gamma^\mu_{\sigma\nu} - \del_\sigma\Gamma^\mu_{\rho\nu}
  + \Gamma^\mu_{\rho\tau}\Gamma^\tau_{\rho\nu}
  - \Gamma^\mu_{\sigma\tau}\Gamma^\tau_{\rho\nu} \,.
\end{equation}
In particular,
$(R_\nabla)^i{}_{jkl} = \del_k\Gamma^i_{lj} - \del_l\Gamma^i_{kj} +
\Gamma^i_{km}\Gamma^m_{lj} - \Gamma^i_{lm}\Gamma^m_{kj}$ and
$(R_\nabla)^i{}_{jk\lb} = -\del_\lb\Gamma^i_{kj}$.

To write down the action $S_{\mathrm{\Omega B}}$ for the
$\Omega$-deformed B-model, we pick a Riemannian metric $g$ on $\BX$
that is compatible with the complex structure.  The action is a sum of
two pieces:
\begin{equation}
  S_{\mathrm{\Omega B}}
  = S_{\mathrm{\Omega B}, \mathrm{C}} + S_{\mathrm{\Omega B}, W} \,.
\end{equation}
The main part $S_{\mathrm{\Omega B}, \mathrm{C}}$ of the action is
$Q_V$-exact and contains the kinetic terms:
\begin{equation}
  \label{eq:OB-SC}
  S_{\mathrm{\Omega B}, \mathrm{C}}
  = \delta_{Q_V} \int_\Sigma g_{i\jb} \Bigl(\rho^i
    \wedge \star\bigl(\rmd\varphib^\jb + \iota_\Vb\auxFb^\jb\bigr)
    + \auxF^i \wedge \star\mu^\jb\Bigr)
    \,.
\end{equation}
This is $Q_V$-invariant because $\CL_V$ commutes with $\star$ and
$\iota_\Vb$, and the integrand is $V$-invariant.

The second piece $S_{\mathrm{\Omega B},W}$ is constructed from the
superpotential.  Assume that $V$ is nonvanishing on the boundary of
$\Sigma$ (which is a collection of circles if $\Sigma$ is compact),
and let $\theta$ be a coordinate on $\del\Sigma$.  Then,
\begin{equation}
  \label{eq:OB-SW}
  S_{\mathrm{\Omega B},W}
  = \int_\Sigma \Bigl(
    \auxF^i \del_i \BW
    + \frac{1}{2} \rho^i \wedge \rho^j \nabla_i\del_j \BW
    - \delta_{Q_V}\bigl(\mu^\ib \del_\ib\BWb\bigr)
    \Bigr)
    - \int_{\del\Sigma} \BW \frac{\rmd\theta}{V^\theta}
    \,.
\end{equation}
The $Q_V$-invariance of $S_{\mathrm{\Omega B},W}$ is easily checked if
$S_{\mathrm{\Omega B},W}$ is expressed with the original auxiliary
fields $\auxG$, $\auxGb$.

Although the above action depends on the choice of the connection
$\nabla$, the $Q_V$-invariant sector of the theory does not.  Written
in terms of $\auxG$ and $\auxGb$, the supersymmetry
transformations~\eqref{eq:OB-SUSY-varphi}--\eqref{eq:OB-SUSY-Gb} are
independent of $\nabla$, while the action depends on $\nabla$ only
through $Q_V$-exact terms.  By the same token, the $Q_V$-invariant
sector is independent of the choice of the target metric~$g$.

\subsection{Reduction to supersymmetric quantum mechanics}

As a preliminary step to understanding the cigar reduction of the
$\Omega$-deformed B-model, let us establish the relation between the
circle reduction of the $\Omega$-deformed B-model and supersymmetric
quantum mechanics.  In the following analysis we will not take into
account the effects of the boundary of $\Sigma$.

Suppose that the B-model is placed on the product
$\Sigma = \R \times \S^1$ of the real line $\R$ and a circle $\S^1$,
endowed with coordinates $(s,\theta)$ and a rotation invariant metric
\begin{equation}
  \wsmetric(s,\theta)
  = \wsmetric_{ss}(s) \rmd s^2 + \wsmetric_{\theta\theta}(s) \rmd\theta^2 \,,
\end{equation}
and we apply the $\Omega$-deformation with respect to the vector field
\begin{equation}
  V = \eps \del_\theta
\end{equation}
generating rotations, where $\eps$ is a complex constant.  We will use
the indices $\sh$, $\thetah$ to denote components of tensors with
respect to the orthonormal vectors
$\del_\sh = \sqrt{\wsmetric^{ss}} \del_s$,
$\del_\thetah = \sqrt{\wsmetric^{\theta\theta}} \del_\theta$ and
one-forms $\rmd\sh = \sqrt{\wsmetric_{ss}} \rmd s$,
$\rmd\thetah = \sqrt{\wsmetric_{\theta\theta}} \rmd\theta$.  In this
notation, the norm $\|V\|$ of $V$ is equal to $|V^\thetah|$.

In general, the path integral whose integrand is supersymmetric
localizes to the field configurations such that the supersymmetry
variations of fermions vanish: away from this locus in the field
space, the parameter of supersymmetry transformations serves as a
fermionic coordinate, but the supersymmetric integrand is by
definition independent of this coordinate and hence the Grassmannian
integration vanishes.  Since
$\delta_{Q_V}(\iota_V\rho^i) = V(\varphi^i)$ and
$\delta_{Q_V}\eta^\ib = V(\varphib^\ib)$, in the case at hand the path
integral localizes to rotation invariant maps.  This means that the
$\Omega$-deformed B-model on $\R \times \S^1$ can be described as a
one-dimensional theory on $\R$.

To understand this reduction to one dimension in a more down-to-earth
manner, we can add to the action the $Q_V$-exact terms
\begin{multline}
  \label{eq:localizing-terms}
  \delta_{Q_V} \int_{\R \times \S^1}
  u g_{i\jb} \bigl(\Vb(\varphi^i) \wedge \star \eta^\jb
  + \CL_V \auxF^i \wedge \star\CL_\Vb\mu^\jb\bigr)
  \\
  = \int_{\R \times \S^1}
  u g_{i\jb} \bigl(\Vb(\varphi^i) \wedge \star V(\varphib^\jb)
    + \CL_V \auxF^i \wedge \star \CL_\Vb \auxFb^\jb
    \\
    + \Vb(\iota_V\rho^i) \wedge \star\eta^\jb
    + \CL_V \rmd\rho^i \wedge \star\CL_\Vb\mu^\jb\bigr)
    + \dotsb \,,
\end{multline}
where $\dotsb$ indicates higher-order terms.  Expand each field $\Psi$
in Fourier modes along $\S^1$ as
$\Psi = \sum_{n \in \Z} \Psi_n e^{in\theta}$.  If we take the limit
$u \to \infty$, the above modification gives infinitely large mass to
all nonzero modes $\Psi_n$, $n \neq 0$, thereby suppressing their
contributions.  Thus, we are left with only the zero modes $\Psi_0$,
which describe fields in the one-dimensional theory.

We now show that this one-dimensional theory is supersymmetric quantum
mechanics with target space $\BX$, in the presence of a potential that
is determined by $\BW$.

The version of supersymmetric quantum mechanics that we will find is a
topologically twisted one, relevant for Morse
theory~\cite{Witten:1982im}.  The theory consists of bosonic fields
\begin{align}
  \label{eq:SQM-chiral-phi}
  \phi  &\in \Map(\R, \BX) \,,
  \\
  \auxH &\in \Omega^1(\R, \phi ^*T\BX)
\end{align}
and fermionic fields
\begin{align}
  \psi &\in \Pi\Omega^0(\R, \phi ^*T\BX) \,,
  \\
  \label{eq:SQM-chiral-psi}
  \chi &\in \Pi\Omega^1(\R, \phi ^*T\BX) \,,
\end{align}
and has supersymmetry transforming them as
\begin{align}
  \label{eq:SQM-SUSY-phi}
  \delta \phi ^\mu &= \psi^\mu \,,
  \\
  \delta_{\nabla'} \psi^\mu &= 0 \,,
  \\
  \delta_{\nabla'} \chi^\mu &= \rmd \phi ^\mu + \iu\auxH^\mu \,,
  \\
  \label{eq:SQM-SUSY-H}
  \delta_{\nabla'} \auxH^\mu
  &= \iu \, \rmd_{\nabla'}\psi^\mu
     - \frac{\iu}{2} (R_{\nabla'})^\mu{}_{\nu\rho\sigma} \chi^\nu \psi^\rho \psi^\sigma \,.
\end{align}
Here $\nabla'$ is any torsion-free connection on $\BX$.  If one
wishes, one could absorb $\nabla'$ by a redefinition of the auxiliary
field $\auxH$.

The action of the theory depends on a real function $\Morse$ and a
flat abelian gauge field $\fg$ on $\BX$.  It is given by
\begin{equation}
  \label{eq:SQM-S}
  S_{\text{SQM}}
  =
  \frac{1}{\hbar} \delta_Q \int_\R
  \frac12 g_{\mu\nu} \chi^\mu \star(\rmd \phi ^\nu - \iu\auxH^\nu)
  + \frac{1}{\hbar} \int_\R
  \bigl(-\delta_Q(\chi^\mu \del_\mu\Morse)
  + \rmd\Morse + \iu \phi ^*\fg\bigr)
  \,,
\end{equation}
where $\hbar$ is the Planck constant and $\delta_Q$ denotes the
supersymmetry variation.  The flatness of $\fg$ is necessary for the
action to be supersymmetric.  (As mentioned already, we neglect the
effects of the boundary of $\R$.  Appropriate boundary conditions are
assumed so that $\int_\R \rmd\Morse$ is supersymmetric.)

The equations of motion for $\auxH$ is
\begin{equation}
  \auxH_\sh^\mu
  = \iu g^{\mu\nu} \del_\nu\Morse
  + \frac{\iu}{2} \delta \phi ^\rho
    g^{\mu\sigma} \nabla'_\rho g_{\sigma\nu} \chi_\sh^\nu \,.
\end{equation}
After $\auxH$ is integrated out, the bosonic part of the action becomes
\begin{equation}
  \label{eq:SQM-S-boson}
  \frac{1}{\hbar} \int_\R \rmd\sh
  \biggl(\frac12 g_{\mu\nu} \del_\sh \phi ^\mu \del_\sh \phi ^\nu
  + \frac12 g^{\mu\nu} \del_\mu\Morse \del_\nu\Morse
  \biggr)
  + \frac{\iu}{\hbar} \int_\R \phi ^*\fg
  \,.
\end{equation}
We see that $h$ provides a potential energy.  This is an analog of the
superpotential $W$ in $\CN = (2,2)$ supersymmetric sigma model.

The on-shell supersymmetry variation of $\chi$ is
\begin{equation}
  \label{eq:SQM-SUSY-on-shell}
  \delta_{\nablat'} \chi_\sh^\mu
  =
  \del_\sh \phi ^\mu - g^{\mu\nu} \del_\nu\Morse \,.
\end{equation}
The connection $\nablat'$ which appears in this formula is different
from $\nabla'$ and preserves the metric:%
\footnote{The connection in the on-shell supersymmetry variation of
  $\chi$ can be changed to any other connection that preserves the
  metric as follows.  Let us add to the action the $Q$-exact term
  $-\delta_Q\int_\R \rmd\sh \, \frac12\Upsilon_{\mu\nu\rho} \delta
  \phi ^\mu \chi_\sh^\nu \chi_\sh^\rho$, where
  $\Upsilon_{\mu\nu\rho} = -\Upsilon_{\mu\rho\nu}$.  Then, the
  equation of motion for $\auxH$ is modified, and the connection in
  question becomes $\nablah'$ whose coefficients are
  $\Gammah'^\mu_{\nu\rho} = \Gammat'^\mu_{\nu\rho} + g^{\mu\sigma}
  \Upsilon_{\nu\sigma\rho}$.  The new connection $\nablah'$ again
  preserves the metric (though may not be torsion-free any longer),
  and any connection preserving the metric can be obtained in this
  way.  A similar modification can be made for the $\Omega$-deformed
  B-model for $\|V\| = 1$ so that $\nablat'$ is modified to
  $\nablah'$.

  For example, $\nablah'$ can be the Levi-Civita connection
  $\nabla^{\text{LC}}$.  In this case we can also take
  $\nabla' = \nabla^{\text{LC}}$ since $\nabla^{\text{LC}}$ is
  torsion-free.  (However, for a metric $g$ that is compatible with
  the complex structure but not K\"ahler, $\Gamma^{\text{LC}}$ has
  mixed components and we cannot take $\nabla = \nabla^{\text{LC}}$.)
  Another example is
  $\nablah' = \nabla^{\text{LC}} + \nabla^{\text{LC}} \LJ \LJ$, where
  $\LJ$ is an almost complex structure compatible with $g$.  This
  connection preserves both $g$ and $\LJ$, and is used
  in~\cite{Witten:1988xj}.  }
\begin{equation}
  \nablat' g = 0 \,.
\end{equation}
Its coefficients are
\begin{equation}
  \label{eq:Gammat'}
  \Gammat'^\mu_{\nu\rho}
  = \Gamma'^\mu_{\nu\rho} + \frac12 g^{\mu\sigma} \nabla'_\nu g_{\sigma\rho} \,.
\end{equation}
Note that setting $\delta_{\nablat'} \chi = 0$ gives the equation for
the gradient flow generated by $h$.  This is how the relation to Morse
theory arises.

The supercharge $Q$ generating the
transformations~\eqref{eq:SQM-SUSY-phi}--\eqref{eq:SQM-SUSY-H}
satisfies $Q^2 = 0$.  Since the metric on $\R$ appears only inside the
$Q$-exact part of the action, the $Q$-cohomology defines a topological
theory.

Let us go back to the $\Omega$-deformed B-model.  The equations of
motion for the auxiliary fields are
\begin{align}
    \auxF^i_{\sh\thetah}
    &=
    \frac{1}{1 + \|V\|^2}
    \bigl(\Vb^\thetah \del_\sh\varphi^i
    + g^{i\jb} \del_\jb\BWb
    + \Vb^\thetah g^{i\kb} \delta\varphi^\mu \nabla_\mu g_{\kb j} \rho_\sh^j
    \bigr)
    \,,
    \\
    \auxFb^\ib_{\sh\thetah}
    &=
    \frac{1}{1 + \|V\|^2}
    \bigl(V^\thetah \del_\sh\varphib^\ib
    - g^{\ib j} \del_j \BW
    - g^{\ib k} \delta\varphi^\mu \nabla_\mu g_{k\jb} \mu_{\sh\thetah}^\jb
    \bigr)
    \,.
\end{align}
Plugging these into the supersymmetry variations of $\rho^i_\sh$ and
$\mu^\ib$, we find
\begin{align}
  \label{eq:OB-SUSY-on-shell-rho}
    \delta_\nablat\rho^i_\sh
    &= \frac{1}{1+\|V\|^2} \bigl(\del_\sh\varphi^i
       - V^\thetah g^{i\jb} \del_\jb\BWb\bigr)
    \,,
    \\
  \label{eq:OB-SUSY-on-shell-mu}
    \delta_\nablat\bigl(\Vb^\thetah \mu^\ib_{\sh\thetah}\bigr)
    &= \frac{\|V\|^2}{1+\|V\|^2} \biggl(\del_\sh\varphib^\ib
      - \frac{\Vb^\thetah}{\|V\|^2} g^{\ib j} \del_j \BW\biggr)
    \,,
\end{align}
where the connection $\nablat$ is given by
\begin{align}
  \Gammat^i_{\mu j}
  &= \Gamma^i_{\mu j}
    + \frac{\|V\|^2}{1 + \|V\|^2} g^{i\kb} \nabla_\mu g_{\kb j} \,,
  \\
  \Gammat^\ib_{\mu\jb}
  &= \Gamma^\ib_{\mu\jb} + \frac{1}{1 + \|V\|^2} g^{\ib k} \nabla_\mu g_{k\jb} \,.
\end{align}
(Adding the deformation terms~\eqref{eq:localizing-terms} affects the
equations of motion and the on-shell supersymmetry variations only for
the nonzero modes.)

The quantities that appear on the right-hand sides of the above
equations are complex conjugate of each other if
\begin{equation}
  \|V\| = 1 \,,
\end{equation}
that is, if we choose
\begin{equation}
  \wsmetric_{\theta\theta} = \frac{1}{|\eps|^2} \,.
\end{equation}
%
%
From now on we assume that this choice is made.  Thus, we can write
\begin{equation}
  V^\thetah = e^{\iu\alpha}
\end{equation}
for some $\alpha \in \R/2\pi\Z$.  We set
\begin{equation}
  \label{eq:Planck}
  \hbar = \frac{|\eps|}{\pi}
\end{equation}
and define real functions $\Morse$, $\LA$ by
\begin{equation}
  \frac{2\pi}{\eps} \BW = \frac{1}{\hbar} (\Morse + \iu\LA) \,,
\end{equation}
or
\begin{align}
  \label{eq:Morse}
  \Morse &= 2\Re(e^{-\iu\alpha} \BW) \,,
  \\
  \LA &= 2\Im(e^{-\iu\alpha} \BW) \,.
\end{align}

Comparing various formulas, especially \eqref{eq:SQM-SUSY-on-shell},
\eqref{eq:OB-SUSY-on-shell-rho} and \eqref{eq:OB-SUSY-on-shell-mu}, we
find that for the zero modes, the on-shell supersymmetry of the
$\Omega$-deformed B-model is the same as that of supersymmetric
quantum mechanics with potential $h$ and the connection
$\nabla' = \nabla$, under the following identification between the
fields:
\begin{align}
  \label{eq:phi-varphi}
  \phi  &= \varphi_0 \,,
  \\
  \psi^i &= \iota_V\rho_0^i \,,
  \\
  \psi^\ib &= \eta_0^\ib \,,
  \\
  \star\chi^i &= 2e^{-\iu\alpha} \iota_V \star\rho_0^i \,,
  \\
  \label{eq:psi-mu}
  \star\chi^\ib &= 2e^{-\iu\alpha} \star\mu_0^\ib \,.
\end{align}
In components, the last two equations are
$\chi_\sh^i = 2(\rho^i_0)_\sh$ and
$\chi_\sh^\ib = 2\Vb^\thetah (\mu^\ib_0)_{\sh\thetah}$.
Note that we have $\nablat = \nablat'$ for $\|V\| = 1$.

Let us check that the actions also agree between the two theories.
The part of the $\Omega$-deformed B-model action that contains $W$ can
be written as
\begin{equation}
  \delta_{Q_V} \int_{\R \times \S^1} \frac{\rmd\theta}{V^\theta} \rho^i \del_i \BW
  + \int_{\R \times \S^1} \rmd \BW \wedge \frac{\rmd\theta}{V^\theta} \,.
\end{equation}
Thus, up to $Q_V$-exact terms, the zero mode action is given by
\begin{equation}
  \frac{2\pi}{\eps} \int_\R \rmd \BW
  =
  \frac{1}{\hbar} \int_\R (\rmd\Morse + \iu \, \rmd\LA) \,.
\end{equation}
This is the action for supersymmetric quantum mechanics in the
presence of the potential $\Morse$ and the gauge field
\begin{equation}
  \fg = \rmd\LA \,,
\end{equation}
with the Planck constant $\hbar$ given by the
expression~\eqref{eq:Planck}.

Even more directly, one can show that the supersymmetry
transformations and the actions match at the off-shell level if the
auxiliary fields are identified as
\begin{align}
  \label{eq:H-F}
  \auxH^i &= -\iu(\rmd\varphi_0^i + 2 \iota_V \auxF_0^i) \,,
  \\
  \label{eq:H-Fb}
  \auxH^\ib &= +\iu(\rmd\varphib_0^\ib + 2 \iota_\Vb \auxFb_0^\ib) \,.
\end{align}
Thus, we have shown that the circle reduction of the $\Omega$-deformed
B-model is the topological twist of supersymmetric quantum mechanics.

\subsection{\texorpdfstring{Cigar reduction of the $\Omega$-deformed
    B-model}{Cigar reduction of the Ω-deformed B-model}}

Having understood the circle reduction of the $\Omega$-deformed
B-model, let us now consider the cigar reduction.  We take $\Sigma$ to
be a cigar $\D$, consisting of a finite cylinder capped at one end.
As before, we endow $\D$ with a rotation invariant metric, with
$\wsmetric_{\theta\theta}(s) = 1/|\eps|^2$ on the cylinder part.  The
boundary of $\D$ is located at $s = \ell$, and the flat cylinder
region continues till a small value of $s$, where the cylinder is
curved abruptly inward to cover the hole.  The tip of the cigar is at
$s = 0$.

Reduction on the circle fibers of $\D$ produces supersymmetric quantum
mechanics on the interval $\I = [0,\ell]$.  The space has two
boundaries, one coming from the boundary circle of $\D$ and the other
from the region near the tip.  We wish to understand what happens at
these points.

At $s = \ell$, the boundary conditions for supersymmetric quantum
mechanics is simply the reduction of the boundary conditions chosen on
$\del\D$ in the $\Omega$-deformed B-model.  We take the latter
boundary conditions to be $Q_V$-invariant.  Then, the former are
$Q$-invariant.

In general, in the $\Omega$-deformed B-model on $\Sigma$, one imposes
a brane-type boundary condition such that the bosonic field maps the
boundary to a chosen submanifold $\BLag \subset \BX$:
\begin{equation}
  \label{eq:brane}
  \varphi(\del\Sigma) \subset \BLag \,.
\end{equation}
If $\Sigma$ has more than one boundary components, one picks such a
submanifold for each component of $\del\Sigma$.

Furthermore, one may turn on a boundary superpotential $\BW_0$, a
locally constant function on $\BLag$.  This introduces an additional
term to the boundary part of the action, which is now given by
\begin{equation}
  \label{eq:B-action-boundary}
  -\int_{\del\Sigma} (\BW - \BW_0) \frac{\rmd\theta}{V^\theta} \,.
\end{equation}

Finally, for the path integral to not diverge, one must choose $\BLag$
in such a way that the integrand of the boundary action is bounded
from above on $\BLag$:
\begin{equation}
  \label{eq:boundedness}
  \Re\biggl(\frac{\BW - \BW_0}{V^\theta}\biggr)\biggr|_\BLag < +\infty \,.
\end{equation}

The rest of the boundary conditions are as follows.  Since the
boundary conditions should preserve $Q_V$, the supersymmetry variation
of the condition~\eqref{eq:brane} must also hold:
\begin{equation}
  \iota_V\rho^i \del_i + \eta^\ib \del_\ib
  \in T_\varphi\BLag \otimes \C \,.
\end{equation}
Requiring that no boundary term arises when we vary fermions in the
action, we find
\begin{equation}
  \iota_V \star\rho^i \del_i + \star\mu^\ib \del_\ib
  \in N_\varphi\BLag \otimes \C
\end{equation}
on the boundary, where $N_\varphi\BLag$ is the normal space to $\BLag$
at $\varphi$.  The supersymmetry variation of this condition gives
\begin{equation}
  (\del_{\sh} \varphi^\mu - g^{\mu\nu} \del_\nu\Morse) \del_\mu
  \in N_\varphi\BLag \,.
\end{equation}
Here we used the fact that $\nablat$ is a metric connection and hence
the covariant variation $\delta_\nablat$ maps a normal vector to a
normal vector.

The above boundary conditions reduce to the following boundary
conditions in supersymmetric quantum mechanics:
\begin{align}
  \label{eq:SQM-BC-l-phi}
  \phi &\in \BLag \,,
  \\
  \label{eq:SQM-BC-l-chi}
  \psi &\in T_\phi \BLag \otimes \C \,,
  \\
  \label{eq:SQM-BC-l-psi}
  \star\chi &\in N_\phi \BLag \otimes \C \,,
  \\
  \label{eq:SQM-BC-l-delphi}
  \star\rmd\phi  - \phi^*(g^{-1} \rmd\Morse) &\in N_\phi \BLag \,.
\end{align}

The boundary conditions at $s = 0$ have a different flavor as they
comes from the tip of the cigar, which is not a boundary in two
dimensions.  Since $\varphi$ is unconstrained at the tip, $\phi$ can
also take any values in $X$.  At the tip $V$ vanishes, so we have
\begin{equation}
  \label{eq:SQM-BC-0-fermi}
  \psi^i = \chi^i = 0
\end{equation}
at $s = 0$.  From the point of view of the $\Omega$-deformed B-model,
the origin of these boundary conditions is the positive curvature near
the tip of the cigar.  The curvature makes the one-form $\rho$
massive, thereby eliminating it from the effective description.

Taking the supersymmetry variations of these equations, we find
\begin{equation}
  \label{eq:SQM-BC-0-phi}
  \del_\sh \phi^\mu  - g^{\mu\nu} \del_\nu\Morse = 0 \,.
\end{equation}
As we will see, the path integral for supersymmetric quantum mechanics
may be localized to the solutions of this equation.  Hence, the
boundary condition at $s = 0$ does not really constrain $\phi $.

Let us calculate the action.  The tip of $\D$ is a special point, so
we first excise a small disk $\D_0$ around it.  On
$\D \setminus \D_0$, the calculation is the same as in the case of
$\R \times \S^1$, except this time we want to take boundary terms into
account.  Up to $Q_V$-exact terms, the action is
\begin{equation}
  \frac{1}{\eps} \int_{\D \setminus \D_0} \rmd \BW \wedge \rmd\theta
  - \frac{1}{\eps} \int_{\del(\D \setminus \D_0)} \BW \rmd\theta
  + \frac{1}{\eps} \int_{\del\D} \BW_0 \rmd\theta
  = \frac{1}{\eps} \int_{\del\D} \BW_0 \rmd\theta \,.
\end{equation}

To this we add the contribution from $\D_0$.  We can evaluate it by
taking the radius of $\D_0$ to zero.  The bulk integral then vanishes
since the Lagrangian has no singularity at the tip, but the boundary
integral remains and gives
\begin{equation}
  - \frac{2\pi}{\eps} \BW(0) \,.
\end{equation}

Therefore, the action for supersymmetric quantum mechanics on $\I$
is, up to $Q$-exact terms,
\begin{equation}
  \label{eq:SQM-action-D-0}
  \frac{2\pi}{\eps} \bigl(\BW_0 - \BW(0)\bigr)
  =
  \frac{1}{\hbar} \bigl(\Morse_0 + \iu\LA_0 - \Morse(0) - \iu\LA(0)\bigr)
  \,,
\end{equation}
where we have defined the constants $h_0$, $f_0$ by
$2\pi \BW_0/\eps = h_0 + \iu\LA_0$.  This is nothing but
\begin{equation}
  \label{eq:SQM-action-I}
  \frac{1}{\hbar} \int_{\I} (\rmd\Morse + \iu \, \rmd\LA)
  + \frac{1}{\hbar} \bigl(\Morse_0 + \iu\LA_0 - \Morse(\ell)
    - \iu \LA(\ell)\bigr) \,,
\end{equation}
namely the non-$Q$-exact part of the action~\eqref{eq:SQM-S} plus
boundary terms at $s = \ell$ which come from the boundary term
\eqref{eq:B-action-boundary} in the $\Omega$-deformed B-model.  Note
that the boundary conditions at $s = 0$ make this expression
$Q$-invariant since $\Morse + \iu\LA$ is a holomorphic function.

\subsection{Reduction to zero-dimensional sigma model}

In~\cite{Yagi:2014toa}, it was shown that if $\BX$ is K\"ahler and the
support $\BLag$ of the brane on $\del\D$ is a Lagrangian submanifold
of $\BX$, the path integral for the $\Omega$-deformed B-model on $\D$
is equivalent to the path integral for a zero-dimensional bosonic
sigma model.  Here we derive a slightly more general result,
applicable to the case in which $\BX$ is not necessarily K\"ahler,
starting from the description in terms of supersymmetric quantum
mechanics.

Let us rescale the metric of $\I$ by a very small factor.  Then, the
kinetic term for $\phi$ in the action \eqref{eq:SQM-S-boson} becomes
very large, whereas the potential term becomes very small.  The path
integral thus localizes to constant maps.  By the boundary condition
at $s = \ell$, these maps must be valued in $\BLag$.  The boundary
condition on $\phi$ at $s = 0$ reduces to $\del_s \phi = 0$ and is
satisfied by constant maps.

Now, suppose that $\BLag$ is middle-dimensional and the pullback of
the two-form $\omega = g\LI$ by the inclusion map
$i_\BLag\colon \BLag \hookrightarrow \BX$ vanishes:
\begin{equation}
  i_L^*\omega = 0 \,.
\end{equation}
If $g$ is a K\"ahler metric (that is, if $\rmd\omega = 0$), then
$\omega$ defines a symplectic structure on $X$ and this condition
means that $\BLag$ is a Lagrangian submanifold.  In general, $\omega$
is only an almost symplectic structure, but we will still call such a
submanifold $\BLag$ Lagrangian, and refer to the corresponding
boundary conditions at $s = \ell$ as a Lagrangian brane with support
$L$.

When $\BLag$ is Lagrangian, the fermionic fields have no zero modes,
as can be shown as follows.  The boundary
conditions~\eqref{eq:SQM-BC-0-fermi} at $s = 0$ kill the constant
modes of $\psi^i$ and $\chi^i$.  Then, by the boundary
conditions~\eqref{eq:SQM-BC-l-chi} and~\eqref{eq:SQM-BC-l-psi} at
$s = \ell$, the constant modes of $\psi^\ib \del_\ib$ and
$\chi_\sh^\ib \del_\ib$ must belong to $T_\phi\BLag \otimes \C$ and
$N_\phi\BLag \otimes \C$, respectively.  Since $i_\BLag^*\omega = 0$,
we have $g(v, \LI w) = 0$ for any $v$, $w \in T_\phi\BLag$ and
therefore $\LI(T_\phi\BLag) \subset N_\phi\BLag$.  As $\BLag$ is
middle-dimensional, $\LI$ exchanges $T_\phi\BLag$ and $N_\phi\BLag$.
Thus, we have
$\LI(\psi^\ib \del_\ib) = -\iu\psi^\ib \del_\ib \in N_\phi\BLag
\otimes \C$ and
$-\iu\chi_\sh^\ib \del_\ib \in T_\phi\BLag \otimes \C$.  It follows
that both $\psi^\ib \del_\ib$ and $\chi_\sh^\ib \del_\ib$ belong to
$T_\phi\BLag \otimes \C$ and $N_\phi\BLag \otimes \C$ simultaneously,
hence they are actually zero.

Finally, the integration over the fluctuations of the bosonic and
fermionic fields produce one-loop determinants that cancel each other
out.  This is because of supersymmetry and the fact that the
fluctuations of $\phi$, $\psi$ and $I\star\chi$ all obey the same
boundary conditions: they vanish at $s = 0$ and belong to
$T_\phi L \otimes \C$ at $s = \ell$.

With the fluctuations integrated out, the only remaining integration
variables is the zero mode of $\phi$.  We conclude that the path
integral for supersymmetric quantum mechanics on $\I$ reduces to the
bosonic integral
\begin{equation}
  \label{eq:localization}
  \int_\BLag \vol_\BLag
  \exp\biggl(\frac{2\pi}{\eps} (\BW - \BW_0)\biggr)
  =
  \int_\BLag \vol_\BLag
  \exp\biggl(\frac{1}{\hbar} (\Morse + \iu\LA - \Morse_0 - \iu\LA_0)\biggr) \,,
\end{equation}
where $\vol_\BLag$ is a volume form on $\BLag$.  This is the path
integral for a zero-dimensional sigma model with target $\BLag$.

\subsection{Multivalued superpotentials}
\label{sec:multivalued-W}

There is an important generalization of the above story that will be
relevant for many applications.  A crucial observation is the
following: if $\Sigma$ has no boundary, $\BW$ does not have to be
single-valued.  Indeed, all we need to write down the action for the
$\Omega$-deformed B-model is $\rmd \BW$, not $\BW$ itself.

If $\Sigma$ does have a boundary, $\BW$ may still be multivalued but
one must make sense of the boundary term \eqref{eq:B-action-boundary},
in which $\BW$ appears directly.  Hence, in the presence of boundary,
one must define $\BW$ on $\BLag$.  The definition of $W$ needs to be
given only modulo $\iu \eps \Z$ since the action appears as $e^{-S}$
in the path integral.

Given a 1-form $\rmd \BW$ on $\BX$, one may try to define $\BW$ on
$\BLag$ as follows.  In each path-connected component $\BLag_a$ of
$\BLag$, one picks a reference point $p_a$ and declares
$\BW(p_a) = 0$.  Then, at any point $p$ of $\BLag_a$, one defines
$\BW(p)$ by choosing a path $P$ from $p_a$ to $p$ and setting
\begin{equation}
  \BW(p) - \BW(p_a) = \int_P \rmd \BW \,.
\end{equation}

The resulting function is generally multivalued since this definition
depends on the choice of the path.  This is not a problem if
$\rmd \BW$ satisfies
\begin{equation}
  \label{eq:quantization}
  [i_\BLag^*\rmd \BW] \in \iu \eps H^1(\BLag;\Z) \,,
\end{equation}
for then $\BW$ is well-defined modulo $\iu\eps\Z$.  Therefore, this
construction provides a good definition of $\BW$ on $\BLag$, as long
as the above quantization condition is obeyed.  Note that the real
part of the quantization condition says that $\rmd\Morse$ is trivial
in the cohomology, so $\Morse$ is single-valued on $\BLag$.

If $\BW$ does not exist as a well-defined function modulo $\iu\eps\Z$
on all of $\BX$, the expression~\eqref{eq:SQM-action-D-0} of the
action is not valid because the meaning of $W(0)$ is ambiguous.  If
either the real or imaginary part of $\rmd \BW/\eps$ obeys the
quantization condition on the entire $\BX$, then the Stokes theorem
can be applied to that part in the expression~\eqref{eq:SQM-action-I}.

At any rate, the localization formula~\eqref{eq:localization} remains
valid since the path integral localizes to $\BLag$, where $\BW$ is
well-defined modulo $\iu\eps\Z$.

The situation that will arise when we discuss the relation between the
$\Omega$-deformed Rozansky--Witten theory and the A-model is that the
imaginary part of $\rmd \BW/\eps$ obeys the quantization condition on
$\BX$.  In terms of the flat connection
$\fg = 2\pi\hbar\Im(\rmd W/\eps)$, the quantization condition is
\begin{equation}
  [\fg] \in 2\pi\hbar H^1(\BX;\Z) \,.
\end{equation}
Moreover, we will find that for the condition~\eqref{eq:boundedness}
to be satisfied, we must have $\Morse - \Morse_0 = 0$ on $\BLag$.
This condition can be satisfied by an appropriate choice of $h_0$ if
and only if $\Morse$ is locally constant on $\BLag$:
\begin{equation}
  i_\BLag^*(\rmd\Morse) = 0 \,.
\end{equation}

In this situation, we can define the function $\LA$ modulo $2\pi\hbar$
on $\BX$, and the action can be written as
\begin{equation}
  \label{eq:SQM-action-D-0-f}
  \frac{1}{\hbar} \int_{\I} \rmd\Morse
  + \frac{\iu}{\hbar} \bigl(\LA_0 - \LA(0)\bigr) \,,
\end{equation}
up to $Q$-exact terms.  We may think of this expression as the action
for supersymmetric quantum mechanics with potential $\Morse$, with
$\LA(0)$ and $\LA_0$ being zero-form boundary gauge fields turned on
at $s = 0$ and $s = \ell$, respectively.

\subsection{Lagrangian branes from the gradient flow}
\label{sec:Lag-grad}

If $M$ is K\"ahler, there is a way to construct good Lagrangian
submanifolds, which at the same time provides a definition of $\BW$
without help of any quantization condition~\cite{Nekrasov:2018pqq,
  Costello:2018txb, Witten:2010zr}.  These Lagrangian branes are the
boundary conditions produced at $s = 0$ by supersymmetric quantum
mechanics on the half-line $[0,+\infty)$.

Consider supersymmetric quantum mechanics on $[0,+\infty)$, and
rescale the $Q$-exact part of the action~\eqref{eq:SQM-S} by a large
factor $u$.  Then, the bosonic part of the action becomes
\begin{equation}
  \frac{u}{\hbar} \int_{[0,+\infty)} \rmd\sh
  \frac12 g_{\mu\nu}
  (\del_\sh \phi ^\mu - g^{\mu\rho} \del_\rho\Morse)
  (\del_\sh \phi ^\nu - g^{\nu\sigma} \del_\sigma\Morse)
  + \frac{1}{\hbar} \int_{[0,+\infty)} (\rmd\Morse + \iu \phi ^*\fg)
  \,.
\end{equation}
The real part of the action remains positive semidefinite, so this is
a valid deformation.

We see that in the limit $u \to \infty$, the path integral localizes
to the solutions of the gradient flow equation
\begin{equation}
  \del_\sh \phi ^\mu - g^{\mu\nu} \del_\nu\Morse = 0 \,.
\end{equation}
For a solution $\phih_p$ with initial condition $\phih_p(0) = p$, the
real part of the bosonic action evaluates to
$(\Morse(\phih_p(+\infty)) - \Morse(p))/\hbar$.  Since $\Morse$ is
monotonically increasing along the flow, the contribution to the path
integral from $\phih_p$ vanishes unless $\phih_p(+\infty)$ is a
critical point of $\Morse$.

For this reason, let us choose the boundary condition at $s = +\infty$
in such a way that $\phi (+\infty)$ lies in a submanifold
$\BLag_\infty$ of the critical locus $\crit(\Morse)$ of $\Morse$, and
define $\BLag$ to be the set of all points $p$ such that the gradient
flow $\phih_p$ reaches $\BLag_\infty$ at $s = +\infty$:
\begin{equation}
  \BLag = \{p \in \BX \mid \phih_p(+\infty) \in \BLag_\infty\} \,.
\end{equation}
For $\BLag$ thus constructed, the condition \eqref{eq:boundedness} is
satisfied since $\Morse|_\BLag$ is bounded above by the constant
$\Morse|_{\BLag_\infty}$.

For $\BLag$ to be a Lagrangian submanifold, $\BLag_\infty$ must obey
an additional condition.  Since $\Morse = 2\Re(e^{-\iu\alpha} \BW)$,
we have $\crit(\Morse) = \crit(\BW)$.  The critical locus is therefore
a complex submanifold of $\BX$ and itself K\"ahler.  A necessary and
sufficient condition is that $\BLag_\infty$ is a Lagrangian
submanifold of $\crit(\BW)$.

To see that this is sufficient, note that the gradient flow generated
by $\Morse$ coincides with the Hamiltonian flow generated by $\LA$ (if
$\LA$ is globally defined, to be precise) because
$\iota_{\del_\sh \phi } \omega = g^{\mu\nu} \del_\nu\Morse
\omega_{\mu\rho} \, \rmd \phi ^\rho = \del_\nu\Morse \LI^\nu{}_\rho \,
\rmd \phi ^\rho = -\rmd\LA$.  It follows that $\omega$ is preserved
along the flow:
$\CL_{\del_\sh \phi} \omega = (\rmd \iota_{\del_\sh \phi} +
\iota_{\del_\sh \phi} \rmd) \omega = 0$.%
\footnote{This is where the K\"ahler condition $\rmd\omega = 0$ is
  necessary.  In the almost symplectic case, the argument goes through
  provided $\iota_{\del_\sh \phi} \rmd\omega = 0$.}
Since $\omega$ is preserved, we can evaluate $\omega(v,w)$ for any
$v$, $w \in T_p\BLag$ by pushing the vectors forward along the flow
$\phih\colon [0,\infty) \times \BX \to \BX$.  Following the flow, we
eventually reach a fixed point, where we have
$\omega(\phih(+\infty)_* v, \phih(+\infty)_* w) = 0$ if $L_\infty$ is
Lagrangian.  Hence, $\omega(v,w) = 0$.  That $\BLag$ is
middle-dimensional follows from the fact that the Hessian of $\Morse$
has the same number of positive and negative eigenvalues because
$\Morse$ is the real part of a holomorphic function.

The necessity of the above condition is clear since $\BLag_\infty$ is
itself contained in $\BLag$, and $\BLag_\infty$ must be a
middle-dimensional submanifold of $\crit(\BW)$ in order for $\BLag$ to
be middle-dimensional in $\BX$.

On a Lagrangian brane constructed in this way, one can define $\Morse$
at $p \in \BLag$ in terms of the value of $\Morse$ at the fixed point
$\phih_p(+\infty) \in \BLag_\infty$ by integrating the gradient flow
equation.  The value of $\LA$ is constant along the flow.  Since $\BW$
is constant on $\BLag_\infty$, defining $\BW$ on $\BLag$ amounts to
choosing a single complex number as the value of $\BW$ on each
connected component of $\BLag_\infty$.  This constant may be absorbed
in $W_0$, so does not introduce further ambiguity.

\section{Rozansky--Witten theory, A-model and quantum mechanics}
\label{sec:RW-A-QM}

As a first application of the results obtained above, in this section
we establish the equivalences between the $\Omega$-deformed
Rozansky--Witten theory on $\D \times \M_1$, the A-model on
$\I \times \M_1$ and quantum mechanics on $\M_1$.  These equivalences
connect two approaches to quantization, namely brane
quantization~\cite{Gukov:2008ve} and quantization by the
$\Omega$-deformation~\cite{Yagi:2014toa}.

\subsection{Rozansky--Witten theory}

Rozansky--Witten theory~\cite{Rozansky:1996bq} is a three-dimensional
topological field theory, and may be obtained by topological twist
from $\CN = 4$ supersymmetric sigma model, for which the target space
must be hyperk\"ahler.  However, just as the B-model can be
constructed for non-K\"ahler target spaces, the target space of
Rozansky--Witten theory can be more generally a complex symplectic
manifold.

A complex symplectic manifold $Y$ is a complex manifold endowed with a
holomorphic symplectic form $\Omega$, a closed nondegenerate
$(2,0)$-form.  The real and imaginary parts of $\Omega$ are real
symplectic forms $\omega_J$ and $\omega_K$:
\begin{equation}
  \Omega = \omega_J + \iu\omega_K \,.
\end{equation}
We denote the complex structure of $Y$ by $I$.

There exists an almost complex structure $K$ such that $\omega_K$ is
compatible with $K$ and $-\omega_K K$ is positive.  Moreover, one can
choose $K$ in such a way that $IK = -KI$.  Then, $J = KI$ is also an
almost complex structure, and $\omega_J$ is compatible with $J$ since
$J^t \omega_J J = -J^t \omega_J IK = J^t \omega_K K = -J^t K^t
\omega_K = I^t \omega_K = \omega_J$.  The three almost complex
structures $I$, $J$, $K$ satisfy the quaternion relations
\begin{equation}
  I^2 = J^2 = K^2 = IJK = -1 \,.
\end{equation}

There is a metric $g$ compatible with both $(J, \omega_J)$ and
$(K,\omega_K)$:
\begin{equation}
  \label{eq:RW-g}
  g = -\omega_J J = \omega_J IK = -\omega_K K \,.
\end{equation}
This metric is also compatible with $I$ since
$I^t g I = -I^t \omega_J JI = -\omega_K K = g$, and
\begin{equation}
  \omega_I = gI
\end{equation}
is a $(1,1)$-form with respect to $I$.  In general, $\omega_I$ is not
closed and $g$ is not K\"ahler.  In fact, $Y$ is hyperk\"ahler if and
only if $\omega_I$ is closed~\cite{MR0935967}.  A hyperk\"ahler
manifold has three integrable complex structures $I$, $J$, $K$,
obeying the quaternion relations, and a metric $g$ that is K\"ahler
with respect to each of $I$, $J$, $K$.

The fields of Rozansky--Witten theory with complex symplectic target
$Y$, placed on a three-manifold $\M_3$, are
\begin{align}
  \varphi &\in \Map(\M_3, Y) \,,
  \\
  \eta &\in \Pi\Omega^0(\M_3, \varphi^*T^{0,1}Y) \,,
  \\
  \rho &\in \Pi\Omega^1(\M_3, \varphi^*T^{1,0}Y) \,.
\end{align}
The theory has supersymmetry which transform the fields as
\begin{align}
  \label{eq:RW-SUSY-varphi}
  \delta\varphi^i &= 0 \,,
  \\
  \label{eq:RW-SUSY-rho}
  \delta\rho^i &= \rmd\varphi^i \,,
  \\
  \label{eq:RW-SUSY-varphib}
  \delta\varphi^\ib &= \eta^\ib \,,
  \\
  \label{eq:RW-SUSY-eta}
  \delta\eta^\ib &= 0 \,.
\end{align}
The supercharge $Q_0$ for these transformations satisfies $Q_0^2 = 0$,
and one considers the $Q_0$-cohomology.

To write down the action for Rozansky--Witten theory, one must choose
a metric $g$ and a torsion-free connection $\nabla$ on $Y$, both
compatible with $I$, and a metric on $\M_3$.  The action is given by
\begin{multline}
  \label{eq:RW-S}
  S_{\mathrm{RW}}
  =
  \delta_{Q_0} \int_{\M_3} g_{i\jb} \rho^i \wedge \star\rmd\varphib^\jb
  -\frac{\iu}{4} \int_{\M_3}\Bigl(
  \Omega_{ij} \rho^i \wedge \rmd_\nabla\rho^j
  + \frac13 \nabla_k \Omega_{ij}
  \rmd\varphi^i \wedge \rho^j \wedge \rho^k
  \\
  - \frac13 \Omega_{ij} (R_\nabla)^j{}_{kl\mb}
  \rho^i \wedge \rho^k \wedge \rho^l \eta^\mb\Bigr).
\end{multline}
The metric $g$ appears only in the $Q_0$-exact part of the action, so
the theory does not depend on the choice of $g$.  It turns out that
the non-$Q_0$-exact part changes by a $Q_0$-exact term if $\nabla$ is
replaced by another connection.  Thus, the theory is also independent
of the choice of~$\nabla$.

\subsection{B-model formulation of Rozansky--Witten theory}

In order to apply the results from the previous section, we need to
describe Rozansky--Witten theory as a B-model.  This is possible when
the spacetime is of the form $\M_3 = \Sigma \times \M_1$, where
$\Sigma$ is a two-manifold and $\M_1$ is a one-manifold.  We denote a
coordinate on $\M_1$ by $t$.

In this case, the bosonic field
$\varphi\colon \Sigma \times \M_1 \to Y$ of Rozansky--Witten theory
may be identified with a map from $\Sigma$ to $\Map(\M_1, Y)$.  As
such, it can be the bosonic field of the B-model on $\Sigma$ with
target space
\begin{equation}
  \BX = \Map(\M_1, Y) \,.  
\end{equation}

A vector field $v$ on $X$, evaluated at $\varphi \in X$, is a section
of the bundle $\varphi^*TY$ over $\M_1$ and can be written locally as
$v(t) = v^\mu(t) \del_\mu$.  Tensor fields on $X$ can be expressed
likewise.  A complex structure $I$ and a compatible torsion-free
connection $\nabla$ on $\BX$ are naturally induced from those on $Y$:
for a vector $v$ and a tensor $w$ on $X$,
\begin{align}
  (\LI v)(t) &= Iv(t) \,, \\
  (\Bnabla_v w)(t) &= \nabla_{v(t)} w(t) \,.
\end{align}

Similarly, given a metric on $\M_1$, a metric on $\BX$ is induced from
that on $Y$: the inner product between $v$, $w \in T_\varphi\BX$ is
given by
\begin{equation}
  g(v,w) = \int_{\M_1} \rmd\th \, g\bigl(v(t), w(t)\bigr) \,,
\end{equation}
where we have chosen $\th$ in such a way that the metric on $\M_1$ is
$\rmd\th^2$.

As can be seen from the above formula for the metric, one may think of
the coordinate $\th$ as a ``continuous index.''  In diffeomorphism
invariant expressions, $\th$ should be integrated over, just as the
indices $i$, $j$, $\dotsc$ and $\ib$, $\jb$, $\dotsc$ are to be summed
over.

Keeping this in mind, we can immediately write down the standard
B-model action (the formula~\eqref{eq:OB-SC} with $V = 0$) for the
target space $X$:
\begin{equation}
  S_{\mathrm{B},\mathrm{C}}
  = \delta_{Q_0} \int_\Sigma \int_{\M_1} \rmd\th \, g_{i\jb}
    \bigl(\rho^i \wedge \star_\Sigma \rmd_\Sigma \varphib^\jb
    + \auxF^i \wedge \star_\Sigma \mu^\jb\bigr)
    \,.
\end{equation}
The Hodge star and the exterior derivative in this action are defined
with respect to $\Sigma$ since the spacetime of the B-model is
$\Sigma$ and does not include $\M_1$.

Compared to the Rozansky--Witten action~\eqref{eq:RW-S}, many terms
are missing from this action.  Crucially, all kinetic terms that
involve derivatives along $\M_1$ cannot be present here because,
again, $\M_1$ is not part of the spacetime.  These missing terms must
come from a superpotential.

A superpotential with the required property is constructed as follows.
Since the holomorphic symplectic form $\Omega$ is closed, locally we
can write it as
\begin{equation}
  \Omega = \rmd \Lambda
\end{equation}
for some holomorphic one-form $\Lambda$.  Then, for $\varphi \in \BX$,
we set
\begin{equation}
  \label{eq:RW-W}
  \BW(\varphi) = \frac{\iu}{2} \int_{\M_1} \varphi^*\Lambda \,.
\end{equation}
(The precise definition of $W$ involves the integral of $\Omega$ over
a cobordism from a reference copy of $\M_1$ inside $Y$ to
$\varphi(\M_1)$, as explained in section~\ref{sec:multivalued-W}.)  In
terms of local holomorphic Darboux coordinates $(P_m, Q^m)$ such that
$\Omega = \rmd P_m \wedge \rmd Q^m$, we may write
$\Lambda = P_m \, \rmd Q^m$.

In the present setup, the derivative $\del_\mu$ that appears in the
superpotential action~\eqref{eq:OB-SW} should be replaced by the
functional derivative $\delta/\delta\varphi^\mu(\th)$.  For the above
superpotential, the functional derivative of $\BW$ is
\begin{equation}
  \frac{\delta\BW}{\delta\varphi^i}
  = \frac{\iu}{2} \Omega_{ij} \del_\th\varphi^j \,.
\end{equation}
The equations of motion for the auxiliary fields are
\begin{equation}
  \begin{aligned}
    \star_\Sigma \auxF^i
    &= -\frac{\iu}{2} g^{i\jb} \Omegab_{\jb\kb} \del_\th\varphib^\kb \,,
    \\
    \star_\Sigma \auxFb^\ib
    &= -\frac{\iu}{2} g^{\ib j} \Omega_{jk} \del_\th\varphi^k
       - g^{\ib j} \eta^\kb \nabla_\kb g_{j\lb} \star_\Sigma \mu^\lb \,,
  \end{aligned}
\end{equation}
 and we have the on-shell supersymmetry variation
\begin{equation}
  \delta_\nabla (\star_\Sigma \mu^\ib)
  = -\frac{\iu}{2} g^{\ib j} \Omega_{jk} \del_\th\varphi^k
    - g^{\ib j} \eta^\kb \nabla_\kb g_{j\lb} \star_\Sigma \mu^\lb \,.
\end{equation}

Comparing the last equation with the supersymmetry
variations~\eqref{eq:RW-SUSY-varphi}, \eqref{eq:RW-SUSY-rho} and
\eqref{eq:RW-SUSY-varphib} in Rozansky--Witten theory, we deduce the
identification
\begin{equation}
  \star_\Sigma \mu^\ib
  = -\frac{\iu}{2} g^{\ib j} \Omega_{jk} \rho_\th^k \,.
\end{equation}
The rest of the fermions, $\eta$ and $\rho$ in the B-model, are
identified with $\eta$ and the components of $\rho$ along $\Sigma$ in
Rozansky--Witten theory.

Under this identification of the fields, we can show that the action
for the B-model matches that for Rozansky--Witten theory.  To show
this, we take the metric $g$ to be the one given by the
relations~\eqref{eq:RW-g}.  This metric satisfies
$\Omegab^t g^{-1} \Omega = 2g(1 - \iu I) = 2(1 + \iu I)^t g$, or
$\Omegab_{\kb\ib} g^{\kb l} \Omega_{lj} = 4g_{\ib j}$.  Furthermore,
we choose $\nabla$ to be a connection that preserves $\Omega$.  Such a
torsion-free connection compatible with $I$
exists~\cite{Thompson:1998vx}.

It is clear that the $Q_0$-exact terms
\begin{equation}
  \delta_{Q_0} \int_{\Sigma \times \M_1} \rmd\th \wedge
  \Bigl(g_{i\jb} \rho^i \wedge \star_\Sigma \rmd_\Sigma\varphib^\jb
  + \frac{\iu}{2} \mu^\ib \Omegab_{\ib\jb} \del_\th\varphib^\jb\Bigr) \,,
\end{equation}
contained in the B-model action, reproduce the $Q_0$-exact part of the
Rozansky--Witten action~\eqref{eq:RW-S}.  The remaining terms in the
B-model action are
\begin{multline}
  \int_{\Sigma \times \M_1} \rmd\th \wedge \biggl(
  g_{i\jb} \auxF^i \wedge_\Sigma \star\auxFb^\jb
  + \eta^\kb \nabla_\kb g_{i\jb} \auxF^i \wedge \star_\Sigma \mu^\jb
  + \frac{\iu}{2} \auxF^i \Omega_{ij} \del_\th\varphi^j
  \\
  + g_{i\jb} \Bigl(\rmd_{\nabla,\Sigma} \rho^i
    - \frac12 (R_\nabla)^i{}_{jk\lb} \eta^\lb\rho^j\wedge\rho^k
    \Bigr) \wedge \star_\Sigma \mu^\jb
    - \frac{\iu}{4} \Omega_{ij} \rho^i \wedge \nabla_\th \rho^j
  \biggr) \,,
\end{multline}
where $\rmd_{\nabla,\Sigma}$ is the operator $\rmd_\nabla$ restricted
to $\Sigma$ and $\rmd_{\nabla,\M_1} = \rmd\th \wedge \nabla_\th$.  The
terms proportional to $\auxF$ cancel by the equation of motion for
$\auxFb$.  Using the first Bianchi identity
$(R_\nabla)^\mu{}_{\nu\rho\sigma} + (R_\nabla)^\mu{}_{\rho\sigma\nu} +
(R_\nabla)^\mu{}_{\sigma\nu\rho} = 0$, which holds for any
torsion-free connection, and the identity
$\Omega_{ik} (R_\nabla)^k{}_{jl\mb} + \Omega_{kj}
(R_\nabla)^k{}_{jl\mb} = 0$, which follows from $\nabla\Omega = 0$,
one can check that the other terms reproduce the non-$Q_0$-exact part
of the Rozansky--Witten action.

\subsection{\texorpdfstring{$\Omega$-deformed Rozansky--Witten theory
    and quantum mechanics}{Ω-deformed Rozansky--Witten theory and
    quantum mechanics}}

Let us take $\Sigma$ to be a disk $\D$ with a rotation invariant
metric.  Being a B-model, Rozansky--Witten theory on $\D \times \M_1$
can be subjected to the $\Omega$-deformation described in
section~\ref{sec:B-SQM-0dSM}.  As we now show, this $\Omega$-deformed
Rozansky--Witten theory is equivalent to quantum mechanics with
vanishing Hamiltonian.

The support $\BLag$ of the brane on $\del\D$ must be a
``middle-dimensional'' submanifold of $\BX$ such that the symplectic
form $\Bomega$ on $\BX$, induced from $\omega_I$ by
\begin{equation}
  \Bomega(v,w)
  =
  \int_{\M_1} \rmd\th \, \omega_I\bigl(v(t), w(t)\bigr) \,,
\end{equation}
vanishes when pulled back to $\BLag$.  In the present
infinite-dimensional setup, this condition should be interpreted as
the requirement that the action of the complex structure exchanges the
tangent bundle and the normal bundle of $L$.  To impose this condition
in a local fashion, we take a Lagrangian submanifold $M \subset Y$ and
set
\begin{equation}
  \BLag = \Map(\M_1, M) \,.
\end{equation}
(Recall that we have defined a
Lagrangian submanifold of an almost symplectic manifold to be a
middle-dimensional submanifold on which the pullback of the almost
symplectic form vanishes.)

The boundary superpotential is given by the integral of a complex
gauge field $\Lambda_0$ on $M$:
\begin{equation}
  \BW_0(\varphi) = \frac{\iu}{2} \int_{\M_1} \varphi^* \Lambda_0 \,.
\end{equation}
For $\BW_0$ to be locally constant on $\BLag$, the boundary gauge
field must be flat:
\begin{equation}
  \rmd\Lambda_0 = 0 \,.
\end{equation}

We should also make sure that $\Morse - \Morse_0$ is bounded on
$\BLag$.  This condition actually requires $\Morse - \Morse_0 = 0$ on
$\BLag$~\cite{Witten:2010zr}.  To see this, let us consider the case
$\M_1 = \S^1$.  Then, for any point $\varphi \in \BLag$, the map
$\varphi_n$ defined by $\varphi_n(\th) = \varphi(n\th)$ with
$n \in \Z$ is also a point in $\BLag$.  Since
$\Morse(\varphi_n) - \Morse_0(\varphi_n) = n(\Morse(\varphi) -
\Morse_0(\varphi))$, if $\Morse - \Morse_0$ is nonzero somewhere in
$\BLag$, it is unbounded.

For this condition to be satisfied,
$\Im(e^{-\iu\alpha} i_M^*\Lambda/\eps)$ must be a flat connection so
that we can set
$\Im(e^{-\iu\alpha} \Lambda_0/\eps) = -\Im(e^{-\iu\alpha}
i_M^*\Lambda/\eps)$.  Equivalently, we must have
$\Im(e^{-\iu\alpha} i_M^* \Omega) = 0$, or
\begin{equation}
  \label{eq:omega_K=0}
  i_M^*\omega_{K_\alpha} = 0 \,,
\end{equation}
where $\omega_{K_\alpha} = g K_\alpha$ is the $(1,1)$-form associated
with the almost complex structure
\begin{equation}
  K_\alpha = K \cos\alpha - J \sin\alpha \,.
\end{equation}

Therefore, $M$ is a Lagrangian submanifold with respect to both
$\omega_I$ and $\omega_{K_\alpha}$.  Since $I$ and $K_\alpha$ swap
$TM$ and $NM$, the action of the almost complex structure
\begin{equation}
  J_\alpha
  = K_\alpha I
  = J \cos\alpha + K \sin\alpha
\end{equation}
leaves $TM$ invariant.  In other words, $M$ is holomorphic in
$J_\alpha$.  It follows that $\omega_{J_\alpha} = gJ_\alpha$ is
nondegenerate on $TM$, so $M$ is a symplectic manifold with symplectic
form $i_M^*\omega_{J_\alpha}$.

The quantization condition \eqref{eq:quantization} translates to the
condition
\begin{equation}
  [i_M^*\Omega] \in 2\eps H^2(M;\Z) \,,
\end{equation}
which, in view of the condition \eqref{eq:omega_K=0}, is equivalent to
\begin{equation}
  [i_M^*\omega_{J_\alpha}] \in 2\pi\hbar H^2(M;\Z) \,.
\end{equation}
This condition means that $A/\hbar$, with
\begin{equation}
  A = \Re(e^{-\iu\alpha} i_M^*\Lambda) \,,
\end{equation}
is a connection on a complex line bundle over $M$ with curvature
\begin{equation}
  \frac{1}{\hbar} \rmd A = \frac{1}{\hbar} i_M^*\omega_{J_\alpha} \,.
\end{equation}

The localization formula \eqref{eq:localization}, applied to the
present setup, gives
\begin{equation}
  \int_\BLag \vol_\BLag \exp\biggl(\frac{\iu}{\hbar}
  \int_{\M_1} \varphi^*(A - A_0)\biggr) \,,
\end{equation}
where $A_0 = \Re(e^{-\iu\alpha} \Lambda_0)$ is a flat connection.  In
terms of real Darboux coordinates $(p_\beta, q^\beta)$ on $M$ such
that
\begin{equation}
  A - A_0 = p_\beta \, \rmd q^\beta \,,
\end{equation}
we may write the above integral as
\begin{equation}
  \int_\BLag \cD p_\beta \cD q^\beta
  \exp\biggl(\frac{\iu}{\hbar}
  \int_{\M_1} p_\beta \del_\th q^\beta \rmd\th\biggr) \,.
\end{equation}
This is the path integral for quantum mechanics on $\M_1$ with
vanishing Hamiltonian, quantizing the symplectic manifold
$(M, i_M^*\omega_{J_\alpha})$.

\subsection{Reduction to the A-model}

In the above argument, we have established the equivalence between the
$\Omega$-deformed Rozansky--Witten theory and quantum mechanics by
directly writing down the localized expression for the path integral.
We could have also done this in two steps, as we did in
section~\ref{sec:B-SQM-0dSM}, via the intermediate supersymmetric
quantum mechanics which we obtain by cigar reduction.  Doing so
connects the $\Omega$-deformed Rozansky--Witten theory with the
A-model setup studied by Gukov and Witten~\cite{Gukov:2008ve}.

Let us recall the construction of the A-model~\cite{Witten:1988xj}.
The target space is a symplectic manifold $Y$, endowed with an almost
complex structure and a compatible symplectic form.  We will see that
in the present context the almost complex structure is $K_\alpha$ and
the symplectic form is $\omega_{K_\alpha}$, so we will use this
notation.  The A-model further depends on a choice of a closed real
two-form $B$ on $Y$.

The worldsheet of the A-model is a Riemann surface $C$, whose complex
structure we call $\cstr$, equipped with a metric.  On one-forms,
$\cstr$ acts from the right: if $z$ is a holomorphic coordinate on
$C$, we have $\rmd z \cstr = \iu \, \rmd z$ and
$\rmd\zb \cstr = -\iu \, \rmd\zb$.

The fields of the A-model are
\begin{align}
  \phi  &\in \Map(C, Y) \,,
  \\
  \psi &\in \Pi\Omega^0(C, \phi ^*TY) \,,
  \\
  \chi &\in \Pi\Omega^1(C, \phi ^*TY) \,,
  \\
  \auxH &\in \Omega^1(C, \phi ^*TY) \,.
\end{align}
The one-form fermion $\chi$ and auxiliary field $\auxH$ obey the
self-duality constraint
\begin{align}
  \chi &= K_\alpha \chi \cstr \,,
  \\
  \auxH &= K_\alpha \auxH \cstr \,.
\end{align}

The A-model has a supercharge $Q$ which squares to zero.  The
covariant supersymmetry variations of the fields are
\begin{equation}
  \begin{aligned}
    \delta \phi ^\mu &= \psi^\mu \,,
    \\
    \delta_{\nabla'} \psi^\mu &= 0 \,,
    \\
    \delta_{\nabla'} \chi^\mu
    &= \rmd \phi ^\mu + (K_\alpha)^\mu{}_\nu \rmd \phi ^\nu \cstr + \iu\auxH^\mu \,,
    \\
    \delta_{\nabla'} \auxH^\mu
    &= \iu\bigl(\rmd_{\nabla'} \psi^\mu
                + (K_\alpha)^\mu{}_\nu \rmd_{\nabla'} \psi^\nu \cstr\bigr)
       - \frac{\iu}{2} (R_{\nabla'})^\mu{}_{\nu\rho\sigma}
         \chi^\nu \psi^\rho \psi^\sigma
      + \psi^\rho (\nabla'_\rho K_\alpha)^\mu{}_\nu \rmd \phi ^\nu \cstr
    \,.
  \end{aligned}
\end{equation}
As in supersymmetric quantum mechanics, $\nabla'$ is a torsion-free
connection on $Y$.

The action of the A-model is given by
\begin{equation}
  S_{\mathrm{A}}
  =
  \frac{1}{\hbar} \delta \int_C \frac14 \Bigl(
  g_{\mu\nu} \chi^\mu \wedge \star
  \bigl(\rmd \phi ^\nu + (K_\alpha)^\nu{}_\rho \rmd \phi ^\rho \cstr - \iu\auxH^\nu\bigr)
  \Bigr)
  + \frac{1}{\hbar} \int_C \phi ^*(-\omega_{K_\alpha} + \iu B) \,.
\end{equation}
The second integral is a topological term and therefore $Q$-invariant.

The equation of motion for the auxiliary field $\auxH$ is
\begin{equation}
  \auxH^\mu
  =
  \frac{\iu}{2} \delta \phi ^\nu g^{\mu\sigma} \nabla'_\nu g_{\sigma\rho} \chi^\rho \,,
\end{equation}
and the on-shell supersymmetry variation for $\chi$ is
\begin{equation}
  \label{eq:on-shell-SUSY-A}
  \delta_{\nablat'} \chi
  = \rmd \phi  + K_\alpha \rmd \phi  \cstr \,,
\end{equation}
where $\nablat'$ is the metric connection \eqref{eq:Gammat'}.  If
$\auxH$ is integrated out, the bosonic part of the A-model action
becomes
\begin{equation}
  \frac{1}{\hbar} \int_C \frac12 g_{\mu\nu} \rmd \phi ^\mu \wedge \star\rmd \phi ^\nu
  + \frac{\iu}{\hbar} \int_C \phi ^*B \,.
\end{equation}

Let us return to the $\Omega$-deformed Rozansky--Witten theory.  We
place the theory on $\R \times \S^1 \times \M_1$, and view it as the
$\Omega$-deformed B-model on $\Sigma = \R \times \S^1$ with target
$X = \Map(\M_1, Y)$.  Then, we reduce the theory on $\S^1$ and compare
the resulting supersymmetric quantum mechanics with the A-model on
$C = \R \times \M_1$ with target $Y$.

The coordinates on $\R$ and $\M_1$ are $s$ and $t$, respectively.  For
notational simplicity, we take the metric on $\R$ to be $\rmd s^2$ so
that $\sh$ can be replaced everywhere by $s$.  We give $C$ a complex
structure by declaring that
\begin{equation}
  z = s + \iu t
\end{equation}
is a holomorphic coordinate.  Then, we have
\begin{equation}
  \cstr^t{}_s = -\cstr^s{}_t = 1 \,,
\end{equation}
and $C$ is locally isomorphic to the complex plane $\C$ with the
standard metric $\rmd s^2 + \rmd t^2$.

For the superpotential~\eqref{eq:RW-W}, we have
\begin{equation}
  \frac{\delta\Morse}{\delta \phi ^\mu}
  = -(\omega_{K_\alpha})_{\mu\nu} \del_\th \phi ^\nu \,,
\end{equation}
so the on-shell supersymmetry variation~\eqref{eq:SQM-SUSY-on-shell}
of $\chi$ in supersymmetric quantum mechanics is
\begin{equation}
  \delta_{\nablat'} \chi_s^\mu
  =
  \del_s \phi ^\mu + (K_\alpha)^\mu{}_\nu \del_\th \phi ^\nu \,.
\end{equation}
This coincides with the supersymmetry variation of $\chi_s$ in the
A-model.

In fact, all the supersymmetry variations of the fields match between
the circle reduction of the $\Omega$-deformed Rozansky--Witten theory
and the A-model.  (The component $\chi_\th$ of $\chi$ is determined
from $\chi_s$ by the self-duality constraint.)  The actions of the two
theories also match, with the B-field given by
\begin{equation}
  B = \omega_{J_\alpha} \,.
\end{equation}

Thus, we have found that upon reduction on a circle, the
$\Omega$-deformed Rozansky--Witten theory with complex symplectic
target $Y$ and complex structure $I$ becomes the A-model with target
$Y$, symplectic structure $\omega_{K_\alpha}$ and B-field
$\omega_{J_\alpha}$.

Next, let us consider the $\Omega$-deformed Rozansky--Witten theory on
$\D \times \M_1$, and reduce it on the circle fibers of the cigar $\D$
to obtain the A-model on $\I \times \M_1$.  We wish to identify the
boundary conditions at $s = 0$ and $\ell$.

We have already seen that the brane at $s = \ell$ must be supported on
a Lagrangian submanifold $M \subset Y$ with respect to
$\omega_{K_\alpha}$.  The boundary
conditions~\eqref{eq:SQM-BC-l-phi}--\eqref{eq:SQM-BC-l-delphi} in
supersymmetric quantum mechanics translates to the boundary conditions
\begin{align}
  \phi &\in M \,,
  \\
  \psi &\in T_\phi  M \otimes \C \,,
  \\
  \chi_s &\in N_\phi M \otimes \C \,,
  \\
  \del_s \phi &\in N_\phi M
\end{align}
in the A-model.  Here we have used the fact that $\del_\th \phi $ is
tangent to $M$, so $K_\alpha = g^{-1} \omega_{K_\alpha}$ sends it to a
vector normal to $M$.

The boundary condition for $\phi $ is the Dirichlet condition.  To
rewrite the boundary conditions for the fermions in a more familiar
form, one may switch to the notation $\psi_\pm$, commonly used in
$\CN = (1,1)$ supersymmetric sigma model.  These fermions are related
to $\psi$, $\chi_s$ by
\begin{equation}
  \begin{aligned}
  \psi
  &= \frac12 (1 + \iu K_\alpha) \psi_+ + \frac12 (1 - \iu K_\alpha) \psi_-
   = \frac12 (\psi_+ + \psi_-) + \frac{\iu}{2} K_\alpha (\psi_+ - \psi_-) \,,
  \\
  \chi_s
  &= -\frac{\iu}{4} (1 - \iu K_\alpha) \psi_+
     + \frac{\iu}{4} (1 + \iu K_\alpha) \psi_-
   = -\frac{1}{4} K_\alpha (\psi_+ + \psi_-)
     - \frac{\iu}{4} (\psi_+ - \psi_-) \,.
  \end{aligned}
\end{equation}
In terms of $\psi_\pm$, the boundary conditions for the fermions is
\begin{align}
  \psi_+ + \psi_- &\in T_\phi M \,,
  \\
  \psi_+ - \psi_- &\in N_\phi M \,.
\end{align}
These are the boundary conditions for a Lagrangian A-brane, that is, a
boundary condition in $\CN = (2,2)$ supersymmetric sigma model that
preserves half of the four supercharges, including the one used in the
construction of the A-model.

Unlike the Lagrangian brane at $s = \ell$, the boundary conditions at
$s = 0$ does not restrict the value of $\phi $.  There is a natural
candidate for the corresponding brane: a canonical coisotropic
brane~\cite{Kapustin:2001ij}, whose support is the entire target
space.

Indeed, the boundary conditions at $s = 0$ require
$(1 - \iu I) \psi = (1 - \iu I) \chi = 0$, which is equivalent to
demanding
\begin{equation}
  \psi_+ = -J_\alpha\psi_- \,.
\end{equation}
This is precisely the boundary condition on the fermions imposed by
the canonical coisotropic brane whose Chan--Paton line bundle carries
a connection with curvature $\omega_{J_\alpha}$.

On the bosonic field, the boundary condition simply requires $\phi $ to
obey the equation for a $K_\alpha$-holomorphic curve
\begin{equation}
  \del_s \phi  + K_\alpha \del_\th \phi  = 0 \,.
\end{equation}
If $K_\alpha$ is integrable, this equation says that $\phi $ is a
holomorphic map.  This is the familiar localization property of the
A-model.

Thus, we have essentially shown that the boundary condition at $s = 0$
is the canonical coisotropic brane, with the curvature of the
Chan--Paton connection being $\omega_{J_\alpha}$.  For this
identification to make sense, however, $\omega_{J_\alpha}$ must
satisfy the quantization condition on $Y$, not just on $M$:
\begin{equation}
  [\omega_{J_\alpha}] \in 2\pi \hbar H^2(Y;\Z) \,.
\end{equation}
This is simply because the canonical coisotropic brane is supported
everywhere on $Y$, and the curvature of a connection on any complex
line bundle obeys the quantization condition.

Let us assume that this quantization condition is satisfied, and
verify that the Chan--Paton connection in question is present in the
action.  Under this assumption, the action is given, up to $Q$-exact
terms, by the formula~\eqref{eq:SQM-action-D-0-f}:
\begin{equation}
  -\frac{1}{\hbar} \int_{\I \times \M_1} \omega_{K_\alpha}
  + \frac{\iu}{\hbar} \int_{\{\ell\} \times \M_1} A_0
  - \frac{\iu}{\hbar} \int_{\{0\} \times \M_1} A \,.
\end{equation}
The gauge field $A$ (extended to a one-form on all of $Y$) has
curvature $\omega_{J_\alpha}$ and is indeed the Chan--Paton connection
for the canonical coisotropic brane.  The flat gauge field $A_0$ is
the Chan--Paton connection for the Lagrangian A-brane.

To summarize, the $\Omega$-deformed Rozansky--Witten theory on
$\D \times \M_1$, reduced on the circle fibers of $\D$, is equivalent
to the A-model on $\I \times \M_1$, with the canonical coisotropic
brane with the Chan--Paton curvature $\omega_{J_\alpha}$ at $s = 0$
and a Lagrangian brane at $s = \ell$ whose support $M$ is Lagrangian
with respect to $\omega_I$ and $\omega_{K_\alpha}$ and holomorphic in
$J_\alpha$.  This A-model setup quantizes $M$, viewed as a symplectic
manifold with symplectic form $i_M^*\omega_{J_\alpha}$.

\subsection{Hamiltonians}

As we have seen, quantum mechanics arising from the $\Omega$-deformed
Rozansky--Witten theory on $\D \times \M_1$ has vanishing Hamiltonian.
The reason can be traced back to the topological invariance of
Rozansky--Witten theory, which in particular implies that translation
along the time axis $\M_1$ is trivial.

To get quantum mechanics with a nontrivial Hamiltonian, one must
modify the theory and break topological invariance on $\M_1$, while
preserving the supercharge $Q_V$.  Once Rozansky--Witten theory is
expressed as a B-model, one can readily achieve such a modification
with introduction of a nontopological term to the superpotential.

Let $w$ be an $I$-holomorphic function on $Y$, and suppose that the
Lagrangian brane $M$ is chosen in such a way that $\Im(w/\eps) = 0$ on
$M$.  (More generally, $w$ may have an explicit dependence on $t$.) We
add the following term to the superpotential~\eqref{eq:RW-W}:
\begin{equation}
  \Delta\BW(\varphi)
  = \frac{\iu}{2} \int_{\M_1} \varphi^*w \, \rmd\th \,.
\end{equation}
Then, the localized path integral becomes
\begin{equation}
  \label{eq:QM-H}
  \int_\BLag \vol_\BLag
  \exp\biggl(\frac{\iu}{\hbar}
  \int_{\M_1} \varphi^*(A - A_0 - H \rmd\th)\biggr) \,,
\end{equation}
where
\begin{equation}
  H = -\Re(e^{-\iu\alpha} w) \,.
\end{equation}
Therefore, the $\Omega$-deformed Rozansky--Witten theory, modified by
this additional superpotential, is equivalent to quantum mechanics
with Hamiltonian $H$.

Conversely, if a real function $H$ on $M$ can be extended to an
$I$-holomorphic function on $Y$, we can devise a modification of
Rozansky--Witten theory so that upon the $\Omega$-deformation it
reduces to quantum mechanics on $M$ with Hamiltonian $H$.

Although this is a perfectly good way to introduce a nonzero
Hamiltonian from the point of view of Rozansky--Witten theory, it
makes the relation to the A-model less straightforward.  As a matter
of fact, for the Hamiltonian to be realized in the A-model, $H$ must
be rather special.

In the A-model, the Hamiltonian originates, if at all, from a
superpotential.  A superpotential in the A-model is a one-form with
values in real functions on $Y$:
\begin{equation}
  Z \in \Omega^1(C, \phi ^*\CO_Y) \,.
\end{equation}
It obeys the constraint
\begin{equation}
  \del_\mu Z + \del_\nu Z (K_\alpha)^\nu{}_\mu \cstr = 0 \,.
\end{equation}
If $K_\alpha$ is integrable, this constraint means that we can write
\begin{equation}
  Z = w' \rmd z + \wb' \rmd\zb = 2\Re(w') \rmd s - 2\Im(w') \rmd t
\end{equation}
for some $K_\alpha$-holomorphic function $w'$ on $Y$.

The superpotential enters the A-model through the action
\begin{equation}
  S_{\mathrm{A},Z}
  =
  -\frac{1}{\hbar} \delta \int_C \frac12 \chi^\mu \wedge \del_\mu Z
  + \frac{1}{\hbar} \int_C \rmd Z \,.
\end{equation}
The addition of this action shifts the equation of motion for $\auxH$
by $\Delta\auxH^\mu = - \iu g^{\mu\nu} \star\del_\nu Z$, and produces
the potential
\begin{equation}
  \frac{1}{\hbar} \int_C
  \frac14 g^{\mu\nu} \del_\mu Z \wedge \star\del_\nu Z \,.
\end{equation}

For $C = \I \times \M_1$, the supersymmetry transformations of the
A-model with superpotential $Z$ coincide with those of supersymmetric
quantum mechanics on $\I$, obtained by cigar reduction of the
$\Omega$-deformed Rozansky--Witten theory with the modified
superpotential, if we identify
\begin{equation}
  Z_t = -\Im(e^{-\iu\alpha} w) \,.
\end{equation}
The actions also match, provided that  the boundary couplings
\begin{equation}
  -\frac{\iu}{\hbar} \int_{\del C} \rmd H
  - \frac{\iu}{\hbar} \int_{\{\ell\} \times \M_1} e^{-\iu\alpha} w \, \rmd t
  =
  -\frac{\iu}{\hbar} \int_{\{0\} \times \M_1} H \rmd\th
  - \frac{1}{\hbar} \int_{\{\ell\} \times \M_1} Z_t \, \rmd t
\end{equation}
are included in the A-model action.  The first term on the left-hand
side accounts for the change in the B-field in the A-model, while the
second term accounts for the change in the boundary coupling in the
$\Omega$-deformed Rozansky--Witten theory.  These boundary terms
combine with the second term in $S_{A,Z}$ to form the $I$-holomorphic
boundary coupling
\begin{equation}
  -\frac{\iu}{\hbar} \int_{\{0\} \times \M_1} e^{-\iu\alpha} w \, \rmd\th \,,
\end{equation}
which is $Q$-invariant by the boundary condition
$(1 - \iu I) \psi = 0$ at $s = 0$.

We have found that for this A-model construction to work, the
Hamiltonian must be the real part of the $I$-holomorphic function
$-e^{-\iu\alpha} w$ whose imaginary part is the imaginary part of a
$K_\alpha$-holomorphic function $-2w'$.

This is a strong constraint, but as pointed out
in~\cite{Witten:2010zr}, there is a natural way to construct a
Hamiltonian obeying this condition if $Y$ is a hyperk\"ahler manifold
that admits a Hamiltonian action of a Lie group preserving the
hyperk\"ahler structure.  Using the moment maps $\mm_I$,
$\mm_{J_\alpha}$, $\mm_{K_\alpha}$ for this Hamiltonian hyperk\"ahler
action with respect to the symplectic structures $\omega_I$,
$\omega_{J_\alpha}$, $\omega_{K_\alpha}$, one defines
$w = \iu(\mm_J + \iu\mm_K)$ and $w' = -(\mm_I + \iu\mm_{J_\alpha})/2$.
Then, $H = \mm_{K_\alpha}$ and this Hamiltonian satisfies the required
condition.

\section{Gauge symmetry}
\label{sec:gauge}

So far we have studied the relation between A-type and B-type
constructions in the case in which the theories involved do not
possess gauge symmetry.  In this section, we extend the analysis in
section~\ref{sec:B-SQM-0dSM} to incorporate gauge symmetry.  We will
formulate the $\Omega$-deformation of the gauged B-model, and
establish the equivalences between the $\Omega$-deformed gauged
B-model on a disk, supersymmetric gauged quantum mechanics on an
interval, and a zero-dimensional gauged sigma model with
complexified gauge group.

\subsection{\texorpdfstring{$\Omega$-deformed gauged
    B-model}{Ω-deformed gauged B-model}}

The $\Omega$-deformation of B-twisted gauge theories with linear
targets was formulated in~\cite{Luo:2014sva}.  Here we generalize this
formulation so that the target space is allowed to be curved.  We will
focus on the K\"ahler case to avoid being overly technical, though the
discussion that follows can be modified to accommodate non-K\"ahler
target spaces.  In any event, we will not encounter such target spaces
in the gauge theory applications we will study.

Let $G$ be a compact Lie group (or the space of maps from $\M_d$ to a
compact Lie group $H$ in higher-dimensional applications), and $G_\C$
its complexification.  We choose a basis $\{T_a\}$ of the Lie algebra
$\gf$ of $G$.  Our convention is such that $T_a$ are represented by
antihermitian matrices in a unitary representation of $G$.

The target space of the gauged B-model~\cite{Baptista:2007ap} is a
K\"ahler manifold $X$ that admits a holomorphic $G_\C$-action with $G$
acting in a Hamiltonian fashion.  We let $v_a$ denote the vector field
generated by $T_a$.  Since the $G_\C$-action preserves the complex
structure $I$, the components $v_a^i$ of $v_a$ are holomorphic
functions, while $v_a^\ib = \overline{v_a^i}$ are antiholomorphic
functions.  If $\{T_a\}$ satisfy the commutation relations
$[T_a, T_b] = f_{ab}{}^c T_c$, then $[v_a, v_b] = -f_{ab}{}^c v_c$.

The $G$-action being Hamiltonian means that there are real functions
$\{\mm_a\}$ such that $\rmd\mm_a = \iota_{v_a} \omega$, where $\omega$
is the K\"ahler form.  As a consequence, the $G$-action preserves the
K\"ahler form:
$\CL_{v_a} \omega = (\rmd\iota_{v_a} + \iota_{v_a}\rmd) \omega =
\rmd^2\mm_a = 0$.  Since $\CL_{v_a} I = 0$ and the K\"ahler metric $g$
is given by $g = -\omega I$, the $G$-action is an isometry.  The
functions $\{\mm_a\}$ define the moment map $\mm \colon X \to \gf^*$
by $\mm_a = \langle\mm, T_a\rangle$.  We assume that $\mm$ is
$G$-equivariant, that is,
$\langle\mm(g \cdot x), \xi\rangle = \langle\mm(x),
\Ad_{g^{-1}}\xi\rangle$ for any $g \in G$, $\xi \in \gf$.

In addition to the chiral multiplet
fields~\eqref{eq:B-varphi}--\eqref{eq:B-Gb}, the gauged B-model has a
gauge field which is locally a one-form valued in $\gf$,
\begin{equation}
  A \in \Omega^1\bigl(\Sigma, \gf\bigr) \,,
\end{equation}
as well as bosonic fields
\begin{align}
  \sigma &\in \Omega^1\bigl(\Sigma, \gf\bigr) \,,
  \\
  \auxD &\in \Omega^0\bigl(\Sigma, \gf\bigr)
\end{align}
and fermionic fields
\begin{align}
  \alpha &\in \Pi\Omega^0\bigl(\Sigma, \gf\bigr) \,,
  \\
  \lambda &\in \Pi\Omega^1\bigl(\Sigma, \gf\bigr) \,,
  \\
  \zeta &\in \Pi\Omega^2\bigl(\Sigma, \gf\bigr) \,,
\end{align}
all in the adjoint representation.  These fields form a vector
multiplet.

It is convenient to define complex gauge fields
\begin{align}
  \CA &= A + \iu\sigma \,,
  \\
  \CAb &= A - \iu\sigma \,.
\end{align}
The field strengths of $A$ is $F = \rmd A + A \wedge A$.  In a like
manner, we define $\CF = \rmd\CA + \CA \wedge \CA$ and
$\CFb = \rmd\CAb + \CAb \wedge \CAb$.

The $\Omega$-deformed supercharge $Q_V$ acts on the vector multiplet
by
\begin{align}
  \label{eq:OGB-SUSY-A}
  \delta\CA &= \iota_V\zeta \,,
  \\
  \delta\CAb &= 2\lambda - \iota_V\zeta \,,
  \\
  \delta\lambda &= \iota_V F - \iu \rmd_A\iota_V\sigma \,,
  \\
  \delta\zeta &= \CF \,,
  \\
  \delta\alpha &= \auxD \,,
  \\
  \delta \auxD &= \iota_V\rmd_\CA\alpha \,.
\end{align}
The exterior derivatives coupled to the gauge fields act on the vector
multiplet in the usual manner.  For instance,
$\rmd_\CA\alpha = \rmd\alpha + [\CA, \alpha]$.

On the chiral multiplet, $Q_V$ acts by
\begin{align}
  \delta\varphi^i &= \iota_V\rho^i \,,
  \\
  \delta\rho^i &= \rmd_\CA\varphi^i + \iota_V\auxG^i \,,
  \\
  \delta\auxG^i &= \rmd_\CA\rho^i - \zeta^a v_a^i \,,
  \\
  \delta\varphib^\ib &= \eta^\ib \,,
  \\
  \delta\eta^\ib &= \iota_V \rmd_\CA\varphib^\ib \,,
  \\
  \delta\mu^\ib &= \auxGb^\ib \,,
  \\
  \label{eq:OGB-SUSY-Gb}
  \delta\auxGb^\ib &= \rmd_\CA\iota_V \mu^\ib \,.
\end{align}
The action of $\rmd_\CA$ on the chiral multiplet fields is determined
by $v_a$.  We have
\begin{align}
  \rmd_\CA\varphi^i &= \rmd\varphi^i + \CA^a v_a^i \,,
  \\
  \rmd_\CA\rho^i &= \rmd\rho^i + \CA^a \del_j v_a^i \wedge \rho^j \,,
\end{align}
and similarly,
\begin{align}
  \rmd_\CA\varphib^\ib &= \rmd\varphib^\ib + \CA^a v_a^\ib \,,
  \\
  \rmd_\CA\mu^\ib &= \rmd\mu^\ib + \CA^a \del_\jb v_a^\ib  \wedge \mu^\jb \,.
\end{align}

In the $\Omega$-deformed gauged B-model, the $G$-action is gauged.  If
$\veps$ is a $\gf$-valued function on $\Sigma$, the infinitesimal
gauge transformation $\delta_\veps$ with parameter $\veps$ acts by
\begin{align}
  \delta_\veps\varphi^i &= \veps^a v_a^i \,,
  \\
  \delta_\veps\rho^i &= \veps^a \del_j v_a^i \rho^j
\end{align}
and so on.  The operator $\rmd_\CA$ does not commute with gauge
transformations: although $\rmd_\CA\varphi^i$ transforms covariantly,
\begin{align}
  \delta_\veps(\rmd_\CA\varphi^i)
  &= \veps^a \del_j v_a^i \rmd_\CA\varphi^j \,,
\end{align}
the gauge transformation of $\rmd_\CA\rho^i$ contains a nonhomogeneous
term:
\begin{equation}
  \delta_\veps(\rmd_\CA\rho^i)
  = \veps^a \del_j v_a^i \rmd_\CA\rho^i
    + \veps^a \del_k\del_j v_a^i \rmd_\CA\varphi^k \wedge \rho^j \,.
\end{equation}

To construct a gauge covariant derivative, we need to pick a
connection $\nabla$.  Recall that $\nabla$ should be torsion-free and
preserve the complex structure.  Moreover, we require it to be
$G$-invariant:
\begin{equation}
  (\CL_{v_a} \Gamma)^\mu_{\nu\rho}
  = \nabla_\nu \nabla_\rho v_a^\mu -  (R_\nabla)^\mu{}_{\rho\nu\sigma} v_a^\sigma
  = 0 \,.
\end{equation}
This equation can be rewritten as
\begin{equation}
  v_a(\Gamma^\mu_{\nu\rho})
  = \del_\sigma v_a^\mu \Gamma^\sigma_{\nu\rho}
    - \del_\nu v_a^\sigma \Gamma^\mu_{\sigma\rho}
    - \del_\rho v_a^\sigma \Gamma^\mu_{\nu\sigma}
    - \del_\nu\del_\rho v_a^\mu \,.
\end{equation}
The left-hand side is the gauge transformation of $\Gamma$ with
parameter $\veps = T_a$, whereas the right-hand side is the action of
the diffeomorphism on $\Gamma$ generated by $T_a$.  In other words,
this condition ensures that a gauge transformation acts on $\Gamma$ as
the corresponding diffeomorphism.

Since the $G$-action is an isometry and $X$ is K\"ahler, one choice of
$\nabla$ is the Levi-Civita connection.  The above condition can be
readily checked: from the Killing equation
$ \nabla_\mu (v_a)_\nu + \nabla_\nu (v_a)_\mu = 0$, one gets
$\nabla_k v_a^i = -g^{i\lb} g_{\mb k} \nabla_\lb v_a^k$ and
\begin{equation}
  \nabla_j \nabla_k v_a^i
  = -g^{i\lb} g_{k\mb} \nabla_j \nabla_\lb v_a^\mb
  = g^{i\lb} g_{\mb k} [\nabla_\lb, \nabla_j] v_a^\mb
  = g^{i\lb} g_{\mb k} R^\mb{}_{\nb\lb j} v_a^\nb
  = R^i{}_{kj\nb} v_a^\nb \,,
\end{equation}
where the equations $\nabla_j v_a^\mb = \del_j v_a^\mb = 0$ is used in
the second equality.

Using $\nabla$, we define the gauge covariant derivative
$\rmd_{\CA,\nabla}$ by coupling $\rmd_\CA$ to the pullback of
$\nabla$.  For example,
\begin{equation}
  \rmd_{\CA,\nabla} \rho^i
  = \rmd_\CA\rho^i + \rmd_\CA\varphi^j \Gamma^i_{jk}  \wedge \rho^k \,.
\end{equation}
In the gauge transformation of $\rmd_{\CA,\nabla}$, the nonhomogeneous
terms coming from $\rmd_\CA$ and $\rmd_\CA\varphi^j \Gamma^i_{jk}$
cancel.  Thus, $\rmd_{\CA,\nabla}$ commutes with the gauge symmetry.

The supersymmetry transformations satisfy
\begin{equation}
  \delta_{Q_V}^2 = \rmd_\CA \iota_V + \iota_V \rmd_\CA \,,
\end{equation}
except when $\delta_{Q_V}^2$ acts on $\auxG$, $\auxGb$, which yields
extra terms involving second derivatives of $v_a$.  These terms are
present because $\auxG$, $\auxGb$ are not tensors.  Rather, the
auxiliary fields $\auxF$, $\auxFb$ are tensors, and $\auxG$, $\auxGb$
differ from them by terms depending on $\Gamma$; see the
definitions~\eqref{eq:auxF} and~\eqref{eq:auxFb}.  The above relation
holds on $\auxF$, $\auxFb$.

The diffeomorphism covariant version of the supersymmetry
transformations is
\begin{align}
  \label{eq:OGB-SUSY-covariant-varphi}
  \delta\varphi^i &= \iota_V\rho^i \,,
  \\
  \delta_\nabla\rho^i
  &= \rmd_\CA\varphi^i + \iota_V \auxF^i \,,
  \\
  \delta_\nabla \auxF^i
  &= \rmd_{\CA,\nabla}\rho^i - \zeta^a v_a^i
     - \biggl(\frac13 (R_\nabla)^i{}_{jkl}  \iota_V\rho^l
       + \frac12 (R_\nabla)^i{}_{jk\lb}   \eta^\lb\biggr)
       \rho^j \wedge \rho^k \,,
  \\
  \delta\varphib^\ib &= \eta^\ib \,,
  \\
  \delta_\nabla\eta^\ib &= \iota_V \rmd_\CA \varphib^\ib \,,
  \\
  \delta_\nabla\mu^\ib &= \auxFb^\ib,
  \\
  \label{eq:OGB-SUSY-covariant-Fb}
  \delta_\nabla\auxFb^\ib
  &= \rmd_{\CA,\nabla} \iota_V\mu^\ib
  - \biggl((R_\nabla)^\ib{}_{\jb\kb l} \iota_V\rho^l
  + \frac12 (R_\nabla)^\ib{}_{\kb\jb\lb} \eta^\lb\biggr)
  \eta^\jb \mu^\kb \,.
\end{align}
The worldsheet derivatives that appear on the right-hand sides are all
gauge covariant, though $\delta_\nabla$ itself does not commute with
gauge transformation.

The action of the $\Omega$-deformed gauged B-model consists of three
pieces:
\begin{equation}
  S_{\mathrm{\Omega GB}}
  = S_{\mathrm{\Omega GB}, \mathrm{V}}
    + S_{\mathrm{\Omega GB}, \mathrm{C}} + S_{\mathrm{\Omega GB}, W} \,.
\end{equation}
The first piece is the action for the vector multiplet and given by
\begin{equation}
  \label{eq:OGB-SV}
  S_{\mathrm{\Omega GB},\mathrm{V}}
  =
  \delta_{Q_V} \int_\Sigma \Tr\biggl(
  \biggl(-\star\auxD + \frac{2\iu}{1+\|V\|^2} (\rmd_A\star\sigma
   + \star\iota_\Vb\rmd_A\iota_V\sigma)\biggr) \alpha
  - \frac{1}{1+\|V\|^2} \zeta \star\CFb\biggr) \,,
\end{equation}
where $\Tr$ is a negative semidefinite bilinear form on $\gf$.
We have chosen this particular integrand so that after the auxiliary
fields are integrated out, the bosonic part of the action takes a nice
form, free of cross terms.  The second one is the chiral multiplet
action:
\begin{equation}
  \label{eq:OGB-SC}
  S_{\mathrm{\Omega GB},\mathrm{C}}
  =
  \delta_{Q_V} \int_\Sigma \Bigl(
  g_{\ib j}
  \bigl(\rmd_\CAb \varphib^\ib + \iota_\Vb\auxFb^\ib\bigr)
  \wedge \star\rho^j
  + g_{\ib j} \mu^\ib \star\auxF^j
  + \star\iu\mm_a \alpha^a
  \Bigr) \,.
\end{equation}
The third is the superpotential action and takes the same form as in
the nongauged case:
\begin{equation}
  \label{eq:OGB-SW}
  S_{\mathrm{\Omega GB},W}
  = \int_\Sigma \Bigl(
    \auxF^i \del_i \BW
    + \frac{1}{2} \rho^i \wedge \rho^j \nabla_i\del_j \BW
    - \delta_{Q_V}\bigl(\mu^\ib \del_\ib\BWb\bigr)
    \Bigr)
    - \int_{\del\Sigma} \BW \frac{\rmd\theta}{V^\theta}
    \,.
\end{equation}
The superpotential $W$ must be gauge invariant:
\begin{equation}
  v_a(W) = 0 \,.
\end{equation}
This condition implies that $\del_i W$ and $\nabla_i \del_j W$
transform under the gauge symmetry as tensors, and the superpotential
action is gauge invariant.

\subsection{Reduction to supersymmetric gauged quantum mechanics}

Taking $\Sigma = \R \times \S^1$ and keeping only the zero modes along
$\S^1$ in the $\Omega$-deformed gauged B-model, we get a topologically
twisted supersymmetric gauged quantum mechanics on $\R$.  We recall
that the metric on $\R \times \S^1$ is
$\rmd\sh^2 + |\eps|^{-2} \rmd\theta^2$ and
$V = \eps \del_\theta = \del_\thetah$.

Supersymmetric gauged quantum mechanics that we obtain is constructed
from a vector multiplet consisting of fields
\begin{align}
  A &\in \Omega^1\bigl(\R, \gf\bigr) \,,
  \\
  \sigma &\in \Omega^1\bigl(\R, \gf\bigr) \,,
  \\
  \tau &\in \Omega^0\bigl(\R, \gf \otimes \C\bigr) \,,
  \\
  \auxD &\in \Omega^0\bigl(\R, \gf\bigr)
  \\
  \alpha &\in \Pi\Omega^0\bigl(\R, \gf\bigr) \,,
  \\
  \lambda &\in \Pi\Omega^1\bigl(\R, \gf\bigr) \,,
  \\
  \kappa &\in \Pi\Omega^0\bigl(\R, \gf\bigr) \,,
  \\
  \xi &\in \Pi\Omega^1\bigl(\R, \gf\bigr) \,,
\end{align}
and a chiral multiplet with
fields~\eqref{eq:SQM-chiral-phi}--\eqref{eq:SQM-chiral-psi}.

Supersymmetry transforms the vector multiplet fields as
\begin{align}
  \delta A &= \lambda  \,,
  \\
  \delta\sigma &= \xi \,,
  \\
  \delta\tau &= 0 \,,
  \\
  \delta\taub &= \kappa \,,
  \\
  \delta\lambda &= -\rmd_A\tau \,,
  \\
  \delta\xi &= [\tau,\sigma] \,,
  \\
  \delta\kappa &= [\tau, \bar\tau] \,,
  \\
  \delta\alpha &= \auxD \,,
  \\
  \delta\auxD &= [\tau, \alpha] \,,
\end{align}
and the chiral multiplet fields as
\begin{align}
  \label{eq:SGQM-SUSY-phi}
  \delta \phi ^\mu &= \psi^\mu \,,
  \\
  \delta_{\nabla'} \psi^\mu &= \tau^a v_a^\mu \,,
  \\
  \delta_{\nabla'} \chi^\mu
  &= \rmd_A \phi ^\mu +  \sigma^a I^\mu{}_\nu v_a^\nu + \iu\auxH^\mu \,,
  \\
  \label{eq:SGQM-SUSY-H}
  \begin{split}
  \delta_{\nabla'} \auxH^\mu
  &= \iu \, \rmd_{A,\nabla'}\psi^\mu
    + \iu\lambda^a v_a^\mu
    - \xi^a I^\mu{}_\nu v_a^\nu
    \\ &\qquad
    - \sigma^a I^\mu{}_\nu \nabla'_\rho v_a^\nu \psi^\rho
    - \iu\tau^a \nabla'_\nu v_a^\mu \chi^\nu
     - \frac{\iu}{2} (R_{\nabla'})^\mu{}_{\nu\rho\sigma} \chi^\nu \psi^\rho \psi^\sigma \,.
  \end{split}
\end{align}
We have the relation $\delta_Q^2 = \delta_\tau$, where the
right-hand side is the infinitesimal gauge transformation with
parameter $\tau$.

The action for supersymmetric gauged quantum mechanics is
\begin{multline}
  \label{eq:SGQM-S}
  S_{\mathrm{SGQM}}
  =
  \frac{2}{\hbar} \delta_Q \int_\R \Tr\Bigl(
  \star\bigl(
  -\auxD + \iu \star\rmd_A\star\sigma - [\tau, \bar\tau]\bigr)\alpha
  + \frac12(\lambda + \iu\xi) \star\rmd_\CAb\taub\Bigr)
  \\
  +\frac{1}{\hbar} \delta_Q \int_\R \biggl(
  \frac12 g_{\mu\nu} \chi^\mu \star(\rmd_A \phi ^\nu - \iu\auxH^\nu)
  - \star\sigma^a \chi^\mu \del_\mu\mm_a
  \\ \qquad\qquad\qquad\qquad
  + \star(g_{\mu\nu} + \iu\omega_{\mu\nu}) \taub^a v_a^\mu \psi^\nu
  + \star 2\iu\alpha^a \mm_a
  \biggr)
  \\
  + \frac{1}{\hbar} \int_\R \bigl(
  - \delta_Q(\chi^\mu \del_\mu\Morse) +\rmd\Morse + \iu\phi ^*\fg\bigr)
  \,.
\end{multline}
The potential $h$ is assumed to be invariant under the action of
$G_\C$, not only $G$.  This condition is satisfied if $h$ is the real
part of a gauge invariant holomorphic function, which is the case when
the theory is the reduction of the $\Omega$-deformed gauged B-model.
Note that due to this assumption, supersymmetric gauged quantum
mechanics knows about the complex structure of the K\"ahler target
space $X$.

Integrating out the auxiliary fields, one finds that the bosonic part
of the action is
\begin{multline}
  \label{eq:SGQM-S-bosonic}
  \frac{1}{\hbar} \int_\R \biggl(
  \frac12 \|\rmd_A\star\sigma\|^2 + \|\rmd_A\tau\|^2
  + \frac12 \|[\tau,\taub]\|^2 + \|[\sigma,\tau]\|^2
  \\
  + \frac12 \|\rmd_A\phi\|^2 + \frac12 \|\rmd h\|^2
  + \frac12 \|\mm\|^2
  + \|\sigma^a v_a\|^2
  + \|\tau^a v_a\|^2
  \\
  - \star \iu [\tau,\taub]^a \mm_a
  - \star \iu \tau^a \taub^b v_a^\mu \del_\mu\mm_b
  - \rmd\bigl(\iu\Tr(\star\sigma [\tau,\taub])
  + \star\sigma^a \mm_a\bigr)
  + 2\sigma^a Iv_a(h)
  \biggr) \,,
\end{multline}
where the norms of various tensors are defined with the metrics on
$\R$ and $X$ as well as the positive bilinear form $-\Tr$ on $\gf$.
The first two terms in the last line cancel by the equivariance of the
moment map, which implies
\begin{equation}
  \label{eq:mm-equivariance}
  v_a(\mm_b) = -f_{ab}{}^c \mm_c \,,
\end{equation}
whereas the very last term vanishes since $h$ is $G_\C$-invariant.
The integral of the total derivative terms vanishes by appropriate
boundary conditions.

The on-shell supersymmetry transformations are given by
\begin{align}
  \delta\alpha
  &= \frac{\iu}{2} \star \rmd_A \star\sigma
    - \frac12 [\tau,\taub] - \frac{\iu}{2} \mm^\vee \,,
  \\
  \label{eq:SGQM-Q-psi-onshell}
  \delta_{\nablat'}\chi
  &= \rmd_A\phi + \sigma^a Iv_a - g^{-1} \rmd h \,,
\end{align}
where $\mm^\vee$ is the dual of $\mm$ with respect to the metric $-\Tr$
on $\gf$.

As asserted already, upon reduction on $\S^1$, the $\Omega$-deformed
gauged B-model becomes supersymmetric gauged quantum mechanics.  The
vector multiplet fields of the latter are expressed in terms of the
zero modes of the vector multiplet fields of the former as follows:
\begin{align}
  \label{eq:SGQM-OGB-A}
  \star A
  &= e^{-\iu\alpha} \iota_V \star A_0 \,,
  \\          
  \star\sigma
  &= e^{-\iu\alpha} \iota_V \star \sigma_0 \,,
  \\
  \tau
  &= \iota_V \CA_0 \,,
  \\
  \star\lambda
  &= e^{-\iu\alpha} \iota_V\star\lambda_0 \,,
  \\
  \kappa
  &= 2\iota_\Vb\lambda_0 \,,
  \\
  \star\xi
  &= \iu e^{-\iu\alpha} \iota_V \star\lambda_0
     + \iu e^{\iu\alpha} \star\zeta_0 \,,
  \\
  \alpha &= \alpha_0 \,,
  \\
  \label{eq:SGQM-OGB-D}
  \auxD &= \auxD_0 \,.
\end{align}
The left-hand sides refer to fields in supersymmetric quantum
mechanics, the right-hand sides the $\Omega$-deformed gauged B-model.
The chiral multiplet fields are identified as in the
relations~\eqref{eq:phi-varphi}--\eqref{eq:psi-mu} and
\begin{align}
  \auxH^i
  &= -\iu(\rmd_{A_0}\varphi_0^i
     + \iu\sigma_0^a v_a^i + 2 \iota_V \auxF_0^i) \,,
  \\
  \auxH^\ib
  &= +\iu(\rmd_{A_0}\varphib_0^\ib
     - \iu\sigma_0^a v_a^i + 2 \iota_\Vb \auxFb_0^\ib) \,.
\end{align}

Under the above identification of fields, the supersymmetry
transformations of the $\Omega$-deformed gauged B-model reproduce
those of supersymmetric gauged quantum mechanics.  Likewise, the
action for the zero modes of the $\Omega$-deformed B-model coincides
with the action for supersymmetric gauged quantum mechanics.

To be precise, if supersymmetric gauged quantum mechanics is obtained
by circle reduction, the real part of $\tau$ is periodic due to the
gauge symmetry.  A better characterization of $\tau$ is in terms of
the holonomy of $\CA$ around $\S^1$, which is valued in $G$ rather
than $\gf$.  That said, the topology of the target space does not
really matter in the applications that we will consider, for in the
end the path integral localizes to a locus where $\tau = 0$.

\subsection{Boundary conditions}

Let us take $\Sigma = \D$ and discuss the boundary conditions for
supersymmetric gauged quantum mechanics on $\I$, obtained by cigar
reduction of the $\Omega$-deformed gauged B-model.

In the $\Omega$-deformed gauged B-model, we impose boundary conditions
on $\del\D$ such that the gauge symmetry is unbroken.  Then, the gauge
field should satisfy the Neumann condition $F = 0$.  We choose a gauge
such that
\begin{equation}
  A_s = 0
\end{equation}
on the boundary.  Then, this condition reads
\begin{equation}
  \del_s A_\theta = 0 \,.
\end{equation}

We also require the boundary conditions to be $Q_V$-invariant.  Taking
the supersymmetry variations of the above conditions, we get
\begin{equation}
  \lambda_s = \del_s\lambda_\theta = 0 \,,
\end{equation}
and taking further supersymmetry variations gives
\begin{equation}
  \del_s\sigma_\theta = 0 \,.
\end{equation}

Since the boundary values of $A_\theta$ and $\sigma_\theta$ are not
fixed, to be compatible with the equation of motion for $\auxD$, the
boundary value of $\auxD$ should not be fixed either.  Since $\alpha$
and $\auxD$ are paired by supersymmetry, both $\alpha$ and $\auxD$
obey the Neumann condition:
\begin{equation}
  \del_s\alpha = \del_s\auxD = 0 \,.
\end{equation}

Since half of the fermions obey the Neumann condition, the other half
should obey the Dirichlet condition, and we get
\begin{equation}
  \zeta = 0 \,.
\end{equation}
Supersymmetry then tells that $\sigma_s$ should also obey the
Dirichlet condition:
\begin{equation}
  \sigma_s = 0 \,.
\end{equation}

The boundary conditions for the chiral multiplet are not affected by
the coupling to the vector multiplet, and specified by a submanifold
$L \subset X$.  The only additional condition compared to the
nongauged case is that $L$ must be $G_\C$-invariant, which implies
that $v_a$ is tangent to $L$ on the boundary.

Upon reduction to supersymmetric gauged quantum mechanics, the above
boundary conditions for the vector multiplet become
\begin{equation}
  A = \sigma = \lambda = \xi
  = \rmd\tau = \rmd\auxD = \rmd\kappa = \rmd\alpha = 0 \,.
\end{equation}
In other words, all one-form fields and the derivatives of all
zero-form fields vanish at $s = \ell$.

The boundary at $s = 0$ comes from the tip of the cigar.  Looking at
the identifications~\eqref{eq:SGQM-OGB-A}--\eqref{eq:SGQM-OGB-D}, we
deduce
\begin{equation}
  A = \sigma = \tau = \lambda =  \kappa = 0
\end{equation}
at $s = 0$.  The remaining fermions should obey the Neumann condition:
\begin{equation}
  \rmd\alpha = \rmd\star\xi = 0 \,.
\end{equation}
The boundary conditions for the chiral multiplet is the same as in the
nongauged case.

Note that the condition $A = 0$ is really a gauge fixing condition,
rather than a boundary condition, because a gauge transformation can
shift the boundary value of $A$.  However, the corresponding gauge
transformation in the $\Omega$-deformed gauged B-model is singular at
the tip of the cigar.  Thus, the cigar geometry picks the gauge
$A = 0$ at $s = 0$.

\subsection{Reduction to zero-dimensional gauged sigma model}

Shrinking the interval $\I$ to a point, we can reduce supersymmetric
gauged quantum mechanics to a zero-dimensional gauge theory.  Let us
determine this theory.

To simplify the analysis we choose the gauge $A = 0$.  We integrate
the auxiliary fields out, rescale the metric of $\I$ by $u^{-2}$, and
take the limit $u \to \infty$.  Then, all bosonic fields get frozen to
constants.  By the boundary conditions at $s = 0$, the bosonic fields
in the vector multiplet are set to zero.  By the boundary conditions
at $s = \ell$, the bosonic field $\phi$ of the chiral multiplet must
be valued in $L$.

When the gauge symmetry is absent, we have chosen the support $L$ of
the brane at $s = \ell$ to be a Lagrangian submanifold of $X$.  To
understand what sort of submanifold $L$ should be in the presence of
gauge symmetry, let us rescale the target metric $g$ by $u$ at the
same time as the rescaling of the metric on $\I$.  This entails
rescaling of the moment map $\mm$ by $u$, and further constrains
$\phi$ to be a zero of $\mm$:
\begin{equation}
  \label{eq:mm=0}
  \mm = 0 \,.
\end{equation}

If we gauge the global gauge symmetry (which preserves the gauge
$A = 0$ and, due to the equivariance of $\mm$, the above equation),
then the theory is effectively described by one without gauge
symmetry, whose target space is
\begin{equation}
  \SX = \mm^{-1}(0)/G \,.
\end{equation}
The space $\SX$ is the symplectic quotient $X \sslash G$ of $X$, and
is K\"ahler.  Hence, we should take $L$ to be a submanifold whose
image in $\SX$ is a Lagrangian submanifold $\SL$.

The submanifold $L$ must also be $G_\C$-invariant, and this condition
can be naturally satisfied.  A key fact is that by a complex analytic
version of the Kempf--Ness theorem, $X \sslash G$ is isomorphic to the
quotient $X^{\mathrm{ss}}/G_\C$ of the set $X^{\mathrm{ss}}$ of
semistable points of $X$ by $G_\C$.  (A point in $X$ is said to be
semistable if the closure of its $G_\C$-orbit intersects
$\mm^{-1}(0)$.)  This fact shows that we may take $L$ to be the
preimage of $\SL$ under the projection
$\pi\colon X^{\mathrm{ss}} \to X^{\mathrm{ss}}/G_\C$:
\begin{equation}
  L = \pi^{-1}(\SL) \,.
\end{equation}
This is the support of a Lagrangian brane in the presence of gauged
symmetry.

With this choice of $L$, the same argument as before immediately tells
us that the fermion zero modes from the chiral multiplet that survive
the boundary conditions are the constant modes of $\psi^\ib \del_\ib$
and $\chib^\ib \del_\ib$ that descend to zero vectors in $\SX$.  In
other words, they must be in the kernel of $\pi_*$.  The kernel is
spanned by vector fields generated by $G_\C$ and, on $L$, sits inside
$TL \otimes \C$.  Since $\star\chi$ must be normal to $L$ at
$s = \ell$, the zero mode of $\chi$ must vanish.  However, the zero
mode of $\psi$ may take the form
\begin{equation}
  \psi_0 = \beta_0^a v_a^\ib \del_\ib \,,
\end{equation}
with $\beta_0$ being an arbitrary constant valued in $\Pi\gf$.

There is another fermion zero mode that pairs up with $\beta_0$.  By
looking at the boundary conditions, we see that the zero mode of
$\alpha$ can be an arbitrary constant $\alpha_0 \in \Pi\gf$.
The only nonvanishing term in the action that contains these fermion
zero modes is
\begin{equation}
  \frac{1}{\hbar} \int_\R
  2\iu \star\beta^b v_b^\ib \del_\ib \mm_a \alpha_0^a \,.
\end{equation}
We can rewrite this term as
\begin{equation}
  \frac{1}{\hbar}
  \int_\R \bigl(
  -\star\beta_0^b (Iv_b)(\mm_a) \alpha_0^a
  + \iu\beta_0^b v_b(\mm_a) \alpha_0^a\bigr) \,.
\end{equation}
By the relation~\eqref{eq:mm-equivariance} the second term in the
integrand vanishes on $\mm^{-1}(0)$, to which the path integral
localizes.  The first term has a nice interpretation.

The bosonic terms in the action for the zero-dimensional theory are
those that appear in the formula~\eqref{eq:localization}, and are
$G_\C$-invariant.  Let us think of $G_\C$ as a complex gauge group.
The isomorphism $\mm^{-1}(0)/G \iso X^{\mathrm{ss}}/G_\C$ essentially
says that the localization equation~\eqref{eq:mm=0} is a partial gauge
fixing condition: it breaks the gauge group $G_\C$ down to $G$.  The
first term in the above integral produces the Faddeev--Popov
determinant associated with this partial gauge fixing, with
$\alpha_0$, $\beta_0$ playing the roles of ghost fields.

Therefore, the $\Omega$-deformed gauged B-model on $\del\D$, and the
corresponding supersymmetric gauged quantum mechanics on $\I$, are
equivalent to a zero-dimensional gauged sigma model with target $L$
and gauge group $G_\C$.  The path integral for this zero-dimensional
theory takes the same form as the formula~\eqref{eq:localization}, but
with $L$ replaced by $\SL$ and $W - W_0$ understood as a function on
$\SL$.

\section{Gauge theory applications}
\label{sec:gauge-applications}

In the final section, we discuss examples of A-type and B-type
constructions which realize gauge theories of dimension $d = 1$ to
$6$.  The theories constructed here are gauged quantum mechanics,
gauged symplectic bosons, and Chern--Simons theory and its
higher-dimensional variants.

\subsection{Gauged quantum mechanics}
\label{sec:GQM}

Let us begin with $d = 1$.  This example is a generalization of the
example studied in section~\ref{sec:RW-A-QM}, and relates the
$\Omega$-deformed gauged Rozansky--Witten theory, the gauged A-model
and gauged quantum mechanics.  We will restrict ourselves to the case
when the target spaces are hyperk\"ahler.

The target space of gauged Rozansky--Witten
theory~\cite{Kapustin:2010pk} with gauge group $H$ is a hyperk\"ahler
manifold $Y$ with a triholomorphic $H_\C$-action such that the
$H$-action is tri-Hamiltonian.  Corresponding to the three complex
structures $I$, $J$, $K$ of $Y$, there are three symplectic structures
$\omega_I$, $\omega_J$, $\omega_K$ and three moment maps $\mm_I$,
$\mm_J$, $\mm_K$.

Placed on $\D \times \M_1$, gauged Rozansky--Witten theory can be
formulated as a gauged B-model on~$\D$.  The B-model description
singles out one of the complex structures of $Y$, which we take to be
$I$.

In the B-model description, we have a chiral multiplet whose scalar
field takes values in $Y$, as well as a chiral multiplet whose scalar
field is the component $\CA_t$ of a complex linear combination of a
three-dimensional gauge field and a $\hf$-valued one-form field.
Hence, the target space of the gauged B-model is
\begin{equation}
  X = \Map(\M_1, Y \times \hf_\C) \,.
\end{equation}
The gauge group is
\begin{equation}
  G = \Map(\M_1, H) \,,
\end{equation}
acting on $X$ by the pointwise $H$-action on $Y$ and $\hf_\C$.  This
is the group of gauge transformations for a gauge theory on $\M_1$
with gauge group $H$.  (More precisely, these are local descriptions
of $X$ and $G$ if the gauge bundle for gauged Rozansky--Witten theory
is nontrivial.)

The superpotential is the gauge invariant version of the
functional~\eqref{eq:RW-W}.  For simplicity, let us assume that the
holomorphic symplectic form $\Omega = \omega_J + \iu\omega_K$ for $I$
can be written as $\Omega = \rmd\Lambda$ for some $H$-invariant
holomorphic one-form $\Lambda$.  Using $\Lambda$, we can define
$\mm_J$ and $\mm_K$ by $(\mm_J + \iu\mm_K)_a = -\iota_{v_a} \Lambda$
since
$\rmd(\mm_J + \iu\mm_K)_a = -(\CL_{v_a} - \iota_{v_a} \rmd)\Lambda =
\iota_{v_a} \Omega$.  Then,
\begin{equation}
  \label{eq:GRW-W}
  W
  = \frac{\iu}{2} \int_{\M_1}
    \bigl(\varphi^* \Lambda - \CA^a (\mm_J + \iu\mm_K)_a\bigr)
\end{equation}
is the gauge invariant superpotential.

We apply the $\Omega$-deformation to this gauged B-model and perform
cigar reduction.  Then, we obtain supersymmetric gauged quantum
mechanics on $\I$ with target $X$, and this is equivalent to the
gauged A-model~\cite{Baptista:2007ap} on $\I \times \M_1$ with target
$Y$ and symplectic form
$\omega_{K_\alpha} = \omega_K \cos\alpha - \omega_J \sin\alpha$.  A
generalization of the canonical coisotropic brane appears at $s = 0$.
At $s = \ell$, we have a brane supported on an $H_\C$-invariant
submanifold $M \subset Y$ such that its image in $Y \sslash H$ is
Lagrangian.  This brane comes from a brane in the $\Omega$-deformed
gauged B-model whose support is $L = \Map(\M_1, M)$.

Finally, shrinking $\I$ to a point, we get a zero-dimensional gauged
sigma model with target $L$ and gauge group $G_\C$.  The action of the
model is given by $W$.  This is gauged quantum mechanics on $\M_1$
with target $M$ and gauge group $H_\C$.

Integrating over $\CA$ sets
\begin{equation}
  \mm_J + \iu\mm_K = 0 \,,
\end{equation}
so this gauged quantum mechanics may also be regarded as nongauged
quantum mechanics with target space
\begin{equation}
  \bigl(M \cap (\mm_J + \iu\mm_K)^{-1}(0)\bigr)/H_\C \,,
\end{equation}
which is a submanifold of the hyperk\"ahler quotient
\begin{equation}
  Y \ssslash H
  = \bigl(\mm_I^{-1}(0) \cap \mm_J^{-1}(0) \cap \mm_K^{-1}(0)\bigr)/H
  \iso (\mm_J + \iu\mm_K)^{-1}(0)/H_\C \,.
\end{equation}
As we have seen in section~\ref{sec:RW-A-QM}, for the whole
construction to work we must choose $M$ in such a way that this
submanifold of $Y \ssslash H$ is Lagrangian with respect to $\omega_I$
and $\omega_{K_\alpha}$, and symplectic with respect to
$\omega_{J_\alpha}$.

\subsection{Gauged symplectic bosons}
\label{sec:VA}

Next, we consider an example with $d = 2$.  In this example, a chiral
CFT on a Riemann surface $\M_2$ is realized by an $\CN = 2$
supersymmetric gauge theory on $\D \times \M_2$~\cite{Oh:2019bgz,
  Jeong:2019pzg} and by an $\CN = 4$ supersymmetric gauge theory on
$\I \times \M_2$~\cite{Gaiotto:2017euk, Costello:2018fnz}.

Let us take a four-dimensional $\CN = 2$ supersymmetric gauge theory,
constructed from a vector multiplet for gauge group $H$ and a
hypermultiplet in a complex symplectic representation $Y$ of $H$.  The
theory has eight supercharges $Q^A_\alpha$, $\Qb_{A\dot\alpha}$, where
$\alpha = \pm$, $\dot\alpha = \dot\pm$ are spinor indices, and
$A = \pm$ is an index for an R-symmetry group $\SU(2)_R$ under which
$Q^A_\alpha$ and $\Qb_{A\alpha}$ transform in the fundamental and the
antifundamental representations, respectively.  An R-symmetry group
$\U(1)_r$ rotates $Q^A_\alpha$ with charge $r = +\frac12$ and
$\Qb^A_{\dot\alpha}$ with $r = -\frac12$.

We place the theory on $\D \times \M_2$ and apply Kapustin's
twist~\cite{Kapustin:2006hi}.  Under the rotation group $\U(1)_\D$ of
$\D$, the supercharges $Q^A_+$, $\Qb^A_{\dot-}$ have spin
$M_\D = +\frac12$ and $Q^A_-$, $\Qb^A_{\dot+}$ have $M_\D = -\frac12$.
Under the rotation group $\U(1)_{\M_2}$ of $\M_2$, $Q^A_+$,
$\Qb^A_{\dot+}$ have $M_{\M_2} = +\frac12$ and $Q^A_-$,
$\Qb^A_{\dot-}$ have $M_{\M_2} = -\frac12$.  The Kapustin twist
replaces the rotation generators $M_\D$, $M_{\M_2}$ with the twisted
rotation generators%
\footnote{The $\U(1)_r$ R-symmetry is anomalous if the one-loop beta
  function is nonzero.  Nevertheless, one can make sense of the
  following construction by introducing to the action a term that
  breaks the rotation invariance on $\D$~\cite{Oh:2019bgz}. This term
  is chosen in such a way that the anomaly in $\U(1)_r$ cancels the
  explicit violation of $\U(1)_\D$.  This is done at the cost of
  losing gauge invariance, but gauge invariance is restored if an
  appropriate surface defect is inserted at the center of $\D$.  After
  the reduction to a theory on $\I \times \M_2$, the surface defect
  becomes a boundary theory localized at $s = 0$.  Such boundary
  theories are important ingredients in the construction
  of~\cite{Costello:2018fnz}.}
\begin{align}
  M_\D' &= M_\D + r \,,
  \\
  \label{eq:M_M_2'}
  M_{\M_2}' &= M_{\M_2} + R \,,
\end{align}
where $R$ is the generator of a subgroup $\U(1)_R \subset \SU(2)_R$
such that $Q^\pm_\alpha$, $\Qb^\pm_{\dot\alpha}$ have
$R = \pm\frac12$.

Since $Q^+_-$, $Q^-_+$, $\Qb^+_{\dot-}$ and $\Qb^-_{\dot+}$ have
$M_{\M_2}' = 0$, these supercharges become scalars on $\M_2$ after the
twist, and as such are preserved even when $\M_2$ is curved.  Two of
them have $M_\D = +\frac12$ and the other two have $M_\D = -\frac12$,
so they generate $\CN = (2,2)$ supersymmetry on $\D$.  In the language
of $\CN = (2,2)$ supersymmetry, $\U(1)_R$ is called the vector
R-symmetry, while $\U(1)_r$ is the axial R-symmetry, as can be seen
from the fact that $\U(1)_r$ rotates the scalars in the vector
multiplet but $\U(1)_R$ does not.

The replacement of $M_\D$ with $M_\D'$ is the B-twist of $\CN = (2,2)$
supersymmetry, which by definition twists the rotation group with the
axial R-symmetry.  Among the four supercharges, $Q^+_-$ and
$\Qb^+_{\dot-}$ are scalars with respect to $M_\D'$.  The linear
combination
\begin{equation}
  \label{eq:KT-Q}
  Q_0 = Q^+_- + \Qb^+_{\dot-}
\end{equation}
squares to zero, and by taking the $Q_0$-cohomology we get a
topological theory on $\D$ since the components of the stress tensor
along $\D$ are $Q_0$-exact.  It turns out that the generator of
antiholomorphic translations on $\M_2$ is also $Q_0$-exact, so the
$Q_0$-cohomology defines a holomorphic--topological theory on
$\D \times \M_2$, which we will refer to as Kapustin theory.

As a gauged B-model on $\D$, Kapustin theory has a chiral multiplet
whose scalar field is valued in $Y$ and another chiral multiplet whose
scalar is the component $A_\zb$ of the gauge field along the
antiholomorphic direction of $\M_2$.  The target space of the gauged
B-model is therefore
\begin{equation}
  X = \Map(\M_2, Y \times \hf_\C) \,,
\end{equation}
and the gauge group is
\begin{equation}
  G = \Map(\M_2, H) \,.
\end{equation}

The hypermultiplet consists of a pair of chiral multiplets in
conjugate representations.  In terms of the scalar fields $q$ and
$\qt$ of these chiral multiplets, the superpotential is given
by~\cite{Oh:2019bgz}
\begin{equation}
  W = \int_{\M_2} \qt \delb_A q \,,
\end{equation}
up to an overall factor which can be absorbed by a rescaling of
fields.  Note that taking $\M_2 = \M_1 \times \S^1$ and performing
dimensional reduction on $\S^1$ reduces $W$ to the
superpotential~\eqref{eq:GRW-W} for gauged Rozansky--Witten theory.

From this B-model description, we see that if we turn on the
$\Omega$-deformation using the rotation symmetry of $\D$, Kapustin
theory on $\D \times \M_2$ becomes equivalent to a chiral CFT on
$\M_2$, described by the action $W$.  This CFT is known as the system
of gauged symplectic bosons with values in $Y$.  The integration cycle
for the path integral is specified by a Lagrangian brane placed on
$\del\D$.

Reducing the $\Omega$-deformed Kapustin theory on the circle fibers of
the cigar $\D$, we arrive at the corresponding A-type theory on
$\I \times \M_2$.  By construction, this is a holomorphic--topological
theory on $\I \times \M_2$.  As we will argue shortly, it is a certain
twist of the circle reduction of the parent four-dimensional $\CN = 2$
supersymmetric gauge theory, which is a twist of a three-dimensional
$\CN = 4$ supersymmetric gauge theory.  This is a nontrivial statement
as one might have expected that the $\Omega$-deformation would affect
the reduced theory.  For the moment let us accept this statement as a
fact, and identify the relevant twist.

A three-dimensional $\CN = 4$ supersymmetric gauge theory has an
R-symmetry group $\SU(2)_H \times \SU(2)_C$ and eight supercharges
$Q_\alpha^{A\dot A}$, where $\alpha = \pm$ is a spinor index,
$A = \pm$ is an $\SU(2)_H$ index and $\dot A = \dot\pm$ is an
$\SU(2)_C$ index.  If the theory is constructed by reduction from four
dimensions, $\SU(2)_H$ comes from $\SU(2)_R$, whereas $\U(1)_r$
becomes a subgroup $\U(1)_C \subset \SU(2)_C$.  In the case at hand,
the supercharges are identified as $Q_\alpha^{A\dot+} = Q^A_\alpha$
and $Q_\alpha^{A\dot-} = \Qb^A_{\dot\alpha}$.

Formulated on $\I \times \M_2$, the theory admits two distinct
twists~\cite{Gaiotto:2016wcv}.  One may twist it by replacing the
rotation group $\U(1)_{\M_2}$ with the diagonal subgroup of
$\U(1)_{\M_2} \times \U(1)_{\mathrm{C}}$.  This is called the C-twist.
According to the definition~\eqref{eq:M_M_2'} of the twisted rotation
generator $M_{\M_2}'$, relevant for the A-type construction is the
H-twist, which twists $\U(1)_{\M_2}$ with a subgroup
$\U(1)_H \subset \SU(2)_H$.

The H-twist makes four supercharges scalars on $\M_2$.  To see what
linear combination of these should be used for the construction of the
holomorphic--topological theory, let us take $\M_2 = \C$ and consider
placing Kapustin theory on $\I \times \S^1 \times \C$.  We think of
$\I \times \S^1$ as modeling the flat cylinder part of $\D$.

On this spacetime all supercharges are unbroken, and the Lie
derivative by $V = e^{\iu\alpha}\del_\thetah$ acts on differential
forms simply by acting on their coefficient functions with $V$.  Then,
the $\Omega$-deformed supercharge $Q_V$ can be written, up to an
overall factor, as%
\footnote{The supersymmetry transformation laws for the
  $\Omega$-deformed gauged B-model may be obtained from those
  in~\cite{Witten:1993yc} by substitution $\del_0 = \iu\del_\thetah$,
  $\del_1 = \del_\sh$ and $\epsb_\pm = \pm\iu/2$,
  $\eps_\pm = -(V^\sh \pm \iu V^\thetah)/2$.  This gives
  $Q_V \propto \Qb_+ + \Qb_- +(V^\thetah - \iu V^\sh) Q_- + (V^\thetah
  + \iu V^\sh) Q_+$.  According to~\cite{Oh:2019bgz}, this supercharge
  is identified with
  $Q^+_- + \Qb^+_{\dot-} + (V^\thetah - \iu V^\sh) Q^-_+ - (V^\thetah
  + \iu V^\sh) \Qb^-_{\dot+}$ in the Kapustin twist.}
\begin{equation}
  \label{eq:H-twist-QV}
  Q_V
  = Q^+_- + \Qb^+_{\dot-} + e^{\iu\alpha}(Q^-_+ - \Qb^-_{\dot+}) \,,
\end{equation}
because this linear combination of supercharges satisfies
$Q_V^2 \propto \CL_V$.  Upon dimensional reduction in the
$\theta$-direction, $Q_V$ descends to the supercharge
\begin{equation}
  Q = Q_-^{+\dot+} + Q_-^{+\dot-}
      + e^{\iu\alpha}(Q_+^{-\dot+} - Q_+^{-\dot-})
\end{equation}
in an $\CN = 4$ supersymmetric gauge theory on $\I \times \C$.  This
is the supercharge whose cohomology defines the
holomorphic--topological theory.

The above argument explains why the $\Omega$-deformation
``disappears'' when a B-twisted supersymmetric field theory is reduced
on a circle, that is, the circle reduction of the $\Omega$-deformed
B-twisted theory is a twist of the reduction of the undeformed theory,
not a deformation thereof.  The reason is that the $\Omega$-deformed
supercharge $Q_V$ already exists in the undeformed theory, and becomes
the supercharge $Q$ for the twisted theory obtained by the reduction.
The only thing the $\Omega$-deformation can do is to change the action
by $Q_V$-invariant terms, which merely corresponds to a $Q$-invariant
deformation of the reduced theory.  We have fixed the action for the
reduced theory by asking it to have the standard bosonic
terms~\eqref{eq:SGQM-S-bosonic}.

In fact, we can characterize the $Q$-cohomology using only a half of
$Q$, and it is illuminating to do so.  This characterization is based
on the fact that the $Q_V$-cohomology of operators is isomorphic to
the $Q_V$-cohomology of states by conformal invariance (which the
$Q_V$-cohomology possesses), and by unitarity the latter is isomorphic
to the space of $Q_V$-harmonic states.  If we define
\begin{align}
  \QQ_V &=  Q^+_- + e^{\iu\alpha} Q^-_+ \,,
  \\
  \QQb_V &=  \Qb^+_{\dot-} - e^{\iu\alpha}\Qb^-_{\dot+} \,,
\end{align}
then $Q_V = \QQ_V + \QQb_V$ and
\begin{equation}
  \{Q_V, Q_V^\dagger\}
  = 2\{\QQ_V, \QQ_V^\dagger\}
  = 2\{\QQb_V, \QQb_V^\dagger\} \,,
\end{equation}
where the hermiticity condition is given by
$(Q^+_\alpha)^\dagger = -\Qb^-_{\dot\alpha}$ and
$(Q^-_\alpha)^\dagger = \Qb^+_{\dot\alpha}$.  Therefore, the
$Q_V$-cohomology is isomorphic to both $\QQ_V$-cohomology and
$\QQb_V$-cohomology.

Let us choose $\QQ_V$.  Under the dimensional reduction this
supercharge becomes
\begin{equation}
  \QQ = Q_-^{+\dot+} + e^{\iu\alpha} Q_+^{-\dot+} \,,
\end{equation}
and the $Q$-cohomology is isomorphic to the $\QQ$-cohomology.  The
supercharge $\QQ$ is used in the discussions of holomorphic boundary
conditions in~\cite{Gaiotto:2016wcv, Gaiotto:2017euk,
  Costello:2018fnz} (up to a flip of the spinor index).

There is something remarkable about $\QQ$: it is proportional to a
$\U(1)_H$ rotation of the supercharge $Q_-^{+\dot+} + Q_+^{-\dot+}$
for the A-twist~\cite{Blau:1996bx}, in which the rotation group
$\SU(2)$ is replaced by the (anti)diagonal subgroup of
$\SU(2) \times \SU(2)_H$.  Consequently, the $\QQ$-cohomology of the
twisted theory on $\I \times \M_2$ is actually fully topological in
the bulk, not topological on $\I$ and holomorphic on $\M_2$ as one
might have expected.

This is consistent with the relation between gauged Rozansky--Witten
theory and the gauged A-model, discussed in section~\ref{sec:GQM}.  If
we take $\M_2 = \M_1 \times \S^1$ and perform reduction on $\S^1$, the
supercharge $Q_-^+ + \Qb_{\dot-}^+$ for Kapustin theory on
$\D \times \M_2$ reduces to the supercharge
$Q_-^{+\dot+} + Q_+^{+\dot-}$ for an $\CN = 4$ supersymmetric gauge
theory on $\D \times \M_1$.  This is the supercharge for the B-twist,
so Kapustin theory reduces to gauged Rozansky--Witten theory.  The
cigar reduction of the $\Omega$-deformed gauged Rozansky--Witten
theory on $\D \times \M_1$ is the gauged A-model on $\I \times \M_1$,
which has two-dimensional topological invariance, not just topological
on $\I$ and on $\M_1$ separately.

At first sight, it might be puzzling that one obtains something
holomorphic like a chiral CFT out of a fully topological theory.  The
resolution to this puzzle is that the three-dimensional topological
invariance is broken at the center of $\D$ where $\QQ_V$ reduces to
$Q_-^+$.  On the boundary at $s = 0$, boundary conditions allow for
additional holomorphic observables.

Thus, we have found that the A-type theory on $\I \times \M_2$ that
realizes the gauged symplectic bosons with values in $Y$ is the
A-twist of an $\CN = 4$ supersymmetric gauge theory.  This theory has
a vector multiplet for gauge group $H$ and a hypermultiplet valued in
$Y$, originating from their counterparts in four dimensions.  The
vector multiplet has three scalar fields transforming as a triplet
under $\SU(2)_C$.  They come from the complex scalar in the
four-dimensional $\CN = 2$ vector multiplet and the component
$A_\theta$ of the four-dimensional gauge field.  The hypermultiplet
contains a pair of scalar fields $q^A$, transforming as a double under
$\SU(2)_H$.  In terms of the four-dimensional fields, $q^+ = q$ and
$q^- = \qt^\dagger$.

Let us determine the boundary conditions on these fields at $s = 0$.

If we describe the theory as supersymmetric gauged quantum mechanics
on $\I$ with target $X$ and gauge group $G$, from the
three-dimensional vector multiplet we get a one-dimensional vector
multiplet and a one-dimensional $\gf$-valued chiral multiplet whose
scalar field is $A_\zb$.  At $s = 0$, all bosonic fields of the
one-dimensional vector multiplet are set to zero, while $A_\zb$
satisfies the gradient flow equation~\eqref{eq:SQM-BC-0-phi}.  In the
three-dimensional terms, these conditions amount to the Dirichlet
condition on the vector multiplet scalars and a deformed Neumann
condition
\begin{equation}
  F_{\sh\zb} \propto e^{\iu\alpha} \bigl((q^+)^\dagger T^a  q^-\bigr) T_a
\end{equation}
on the gauge field.

The three-dimensional hypermultiplet gives rise to a pair of chiral
multiplets in supersymmetric gauged quantum mechanics.  At $s = 0$,
the boundary condition for the scalar fields of these multiplets
demands
\begin{equation}
  \del_\sh q^- \propto e^{-\iu\alpha} \delb_A q^+ \,.
\end{equation}

The boundary conditions just described are consistent with those
discussed in~\cite{Gaiotto:2017euk, Costello:2018fnz}, where it was
found that symplectic bosons arise from boundaries in this
three-dimensional $\CN = 4$ supersymmetric gauge theory.

\subsection{Chern--Simons theory}
\label{sec:CS}

Now we discuss an example with $d = 3$.  In this example,
Chern--Simons theory on a three-manifold $\M_3$ is realized by
maximally supersymmetric Yang--Mills theories on
$\D \times \M_3$~\cite{Luo:2014sva} and
$\I \times \M_3$~\cite{Witten:2010cx, Witten:2010zr}.

Five-dimensional $\CN = 2$ super Yang--Mills theory, like any other
maximally supersymmetric Yang--Mills theories, can be constructed from
ten-dimensional super Yang--Mills theory by dimensional reduction.
Let $(x^1, \dotsc, x^{10})$ be coordinates on $\R^{10}$, and reduce
super Yang--Mills theory on $\R^{10}$ in the directions of $x^6$,
$\dotsc$, $x^{10}$.  Then, the components $A_6$, $\dotsc$, $A_{10}$ of
the ten-dimensional gauge field are turned into five scalar fields
$\phi_6$, $\dotsc$, $\phi_{10}$, valued in the Lie algebra $\hf$ of
the gauge group $H$.  They transform in the vector representation of
the R-symmetry group $\Spin(5)_R$, which originates from the rotation
symmetry of $\R^5$ on which the reduction is performed.  The sixteen
supercharges of the theory transform in the spinor representation of
$\Spin(5)_R$.

We place this theory on $\D \times \M_3$, and twist the rotation group
$\SU(2)_{\M_3}$ on $\M_3$ by the subgroup $\SU(2)_R$ of $\Spin(5)_R$
rotating $\phi_8$, $\phi_9$ and $\phi_{10}$.  The supercharges
transform as doublets under $\SU(2)_{\M_3}$ and $\SU(2)_R$, so a
quarter of the sixteen supercharges become singlets under the diagonal
subgroup of $\SU(2)_{\M_3} \times \SU(2)_R$.  These four supercharges
are scalars under the twisted rotation group on $\M_3$, and can be
preserved even when $\M_3$ is curved.  They have spins $\pm\frac12$
under $\U(1)_\D$, so generate $\CN = (2,2)$ supersymmetry on $\D$.

Let $\U(1)_r$ be the subgroup of $\Spin(5)_R$ rotating $\phi_6$ and
$\phi_7$.  From the point of view of $\CN = (2,2)$ supersymmetry,
$\U(1)_r$ is the axial R-symmetry rotating the two scalars in the
vector multiplet.  Further twisting the theory on $\D$ with $\U(1)_r$,
we get a gauged B-model.

In the twisted theory, $\phi_8$, $\phi_9$, $\phi_{10}$ are the
components of a one-form on $\M_3$.  They combine with $A_3$, $A_4$,
$A_5$ to form a $Q$-invariant complex gauge field $\CA$ on $\M_3$:
\begin{equation}
  \CA
  = (A_3 + \iu \phi_8) \rmd x^3
    + (A_4 + \iu \phi_9) \rmd x^4
    + (A_5 + \iu \phi_{10}) \rmd x^5 \,.
\end{equation}
The components of $\CA$ are the scalar fields of three chiral
multiplets.  Hence, the target space $X$ of the gauged B-model is the
space of complex gauge fields on $\M_3$.  The gauge group $G$ is the
group of gauge transformations on $\M_3$, and $G_\C$ acts naturally on
$X$ by complex gauge transformations.  If the gauge bundle is trivial,
we have
\begin{equation}
  X = \Omega^1(\M_3, \hf_\C)
\end{equation}
and
\begin{equation}
  G = \Map(\M_3, H) \,.
\end{equation}
The $G$-invariant K\"ahler metric on $X$ is given by
\begin{equation}
  g(v,w)
  = -\frac{1}{2g_{\mathrm{5d}}^2}
    \int_{\M_3} \Tr(v \wedge \star\wb + \vb \wedge \star w) \,,
\end{equation}
where $v$, $w$ are $\hf_\C$-valued one-forms on $\M_3$ and
$g_{\mathrm{5d}}$ is the gauge coupling of five-dimensional $\CN = 2$
super Yang--Mills theory.

The superpotential $W$ is a gauge invariant functional, given by the
integral of a holomorphic function of $\CA$ over $\M_3$.  For the
total action to be second order in derivatives after the auxiliary
fields are integrated out, $W$ must be first order.  In view of these
requirements, the only candidate for $W$ is the integral of the
Chern--Simons form
\begin{equation}
  \ChS(\CA)
  =
  \Tr\biggl(\CA \wedge \rmd \CA + \frac23 \CA \wedge \CA \wedge \CA\biggr)
\end{equation}
over $\M_3$, up to an overall normalization factor.  The overall
factor is meaningful in this example since $W$ is not homogeneous and
cannot be absorbed by a rescaling of $\CA$.  The inhomogeneity of $W$
also explains why the vector R-symmetry is absent in the B-model
description.

The absolute value of the normalization of $W$ is fixed by the
requirement that integrating out the auxiliary fields reproduces the
kinetic term
\begin{equation}
  -\frac{1}{4g_{\mathrm{5d}}^2} \int_{\D \times \M_3} \Tr(F \wedge \star F)
\end{equation}
for the five-dimensional gauge field.  This is satisfied if we take
\begin{equation}
  \label{eq:CS-W}
  W = \frac{1}{4g_{\mathrm{5d}}^2} \int_{\M_3} \ChS(\CA) \,.
\end{equation}
The phase of $W$ is not so important because shifting the phase has
the same effect as the action of the broken vector R-rotation that
leaves $\CA$ intact.  Different choices of the phase correspond to
different ways to identify the supercharge of the twisted
five-dimensional theory and that of the gauged B-model.

From this superpotential, we learn that the $\Omega$-deformation of
the twisted $\CN = 2$ super Yang--Mills theory on $\D \times \M_3$
with gauge group $H$ is Chern--Simons theory with gauge group $H_\C$,
described by the action
\begin{equation}
  -\frac{\pi\iu}{2\eps g_{\mathrm{5d}}^2} \int_{\M_3} \ChS(\CA) \,.
\end{equation}

Let us turn to the corresponding A-type construction.  The A-type
theory is the reduction of the above theory in the direction of
$x^2 = \theta$.  This is a topological twist of $\CN = 4$ super
Yang--Mills theory on $\I \times \M_3$, with gauge coupling
\begin{equation}
  g_{\mathrm{4d}} = \sqrt{\frac{|\eps|}{2\pi}} g_{\mathrm{5d}} \,.
\end{equation}
As we now argue, the relevant twist is the
GL-twist~\cite{Yamron:1988qc, Marcus:1995mq, Kapustin:2006pk}, with
its $\C\P^1$-valued parameter equal to
\begin{equation}
  \label{eq:GL-t}
  t = \frac{1 - \cos\alpha}{\sin\alpha} \,.
\end{equation}

The GL-twist of $\CN = 4$ super Yang--Mills theory on a four-manifold
replaces the rotation group $\Spin(4)$ with the diagonal subgroup of
$\Spin(4) \times \Spin(4)_R$, where the second factor is a subgroup of
the R-symmetry group $\Spin(6)_R$.  In the present notation,
$\Spin(6)_R$ rotates $\phi_6$, $\dotsc$, $\phi_{10}$ and $\phi_2$
which comes from $A_2$, and $\Spin(4)_R$ rotates $\phi_6$, $\phi_8$,
$\phi_9$, $\phi_{10}$.  Thea twist turns the latter four scalars into
the components of the one-form
\begin{equation}
  \phi
  = \phi_6 \rmd x^1 + \phi_8 \rmd x^3
    + \phi_9 \rmd x^4 + \phi_{10} \rmd x^5 \,.
\end{equation}
(The coordinate on $\I$ is $x^1 = s$ and those on $\M_3$ are
$(x^3, x^4, x^5)$.)

The GL-twisted theory has two scalar supercharges $Q_\ell$ and $Q_r$,
and one picks a linear combination
\begin{equation}
  Q = Q_\ell + t Q_r
\end{equation}
to define the topological theory, where $t$ takes values in $\C\P^1$.
The supersymmetry transformations generated by $Q$ may be found
in~\cite{Kapustin:2006pk}.  What is important to us is that there is a
two-form fermion $\chi$ whose supersymmetry variation is given by
\begin{align}
  \label{eq:delta-chi+}
  \delta\chi^+
  &= (F - \phi \wedge \phi)^+ + t(\rmd_A\phi)^+ \,,
  \\
  \label{eq:delta-chi-}
  \delta\chi^-
  &= t(F - \phi \wedge \phi)^- - (\rmd_A\phi)^- \,,
\end{align}
where the superscripts $+$ and $-$ denote the self-dual and
anti-self-dual parts, respectively.

We claim that when $t$ is real, the GL-twisted $\CN = 4$ super
Yang--Mills theory on $\I \times \M_3$ can be reformulated as
supersymmetric gauged quantum mechanics on $\I$ with target $X$ and
gauge group $G$, with the identification
\begin{equation}
  \sigma_s = \phi_6  \,.
\end{equation}
Furthermore, the potential $\Morse$ is determined by the
superpotential~\eqref{eq:CS-W} via the relation~\eqref{eq:Morse},
which shows that the theory is the cigar reduction of the
$\Omega$-deformed B-model described above.

One way to see this is to compare the supersymmetry variations of
fermions that give rise to flow equations.  If we make the complex
gauge field
$\CA = (A_s + \iu\sigma_s) \rmd s + \CA_3 \rmd x^3 + \CA_4 \rmd x^4 +
\CA_5 \rmd x^5$ and let $\CF$ be the curvature of $\CA$, then we can
write the supersymmetry variation~\eqref{eq:SGQM-Q-psi-onshell} as
\begin{equation}
  \delta\chi_\sh
  = \iota_{\del_\sh} \CF + e^{\iu\alpha} \star_{\M_3} \CFb \,,
\end{equation}
where we consider $\chi_\sh$ as a one-form
$\chi_\sh^3 \rmd x^3 + \chi_\sh^4 \rmd x^4 + \chi_\sh^5 \rmd x^5$.
This is equivalent to the supersymmetry
variations~\eqref{eq:delta-chi+} and~\eqref{eq:delta-chi-}, with $t$
given by the formula~\eqref{eq:GL-t}.

Another way to argue for this claim is to consider compactification to
two dimensions.

Five-dimensional $\CN = 2$ super Yang--Mills theory has a unique full
twist, which replaces the rotation group $\Spin(5)$ with the diagonal
subgroup of $\Spin(5) \times \Spin(5)_R$.  The topological twist we
have performed is compatible with this twist.  If we take
$\M_3 = \R \times \M_2$ for some compact Riemann surface $\M_2$, then
in the infrared, we get a fully twisted $\CN = 4$ supersymmetric sigma
model on $\D \times \R$ with a hyperk\"ahler target space.  (Twisting
along $\M_2$ breaks half of supersymmetry.)  This is Rozansky--Witten
theory, and its target space is the moduli space $\SM_H$ of the
Hitchin equations on $\M_2$~\cite{Gaiotto:2009hg}.  The
$\Omega$-deformation and cigar reduction then produce the A-model on
$\I \times \R$ with target $\SM_H$.

The GL-twisted $\CN = 4$ super Yang--Mills theory for real $t$, upon
compactification on $\M_2$, indeed reduces to the A-model with target
$\SM_H$~\cite{Kapustin:2006pk}.  In $\Omega$-deforming
Rozansky--Witten theory with target $\CM_H$, we have picked a complex
structure in which the complex gauge field $A + \iu\phi$ on $\M_2$
defines holomorphic coordinates.  If we call it $I$, then according
to~\cite{Kapustin:2006pk}, the complex structure of the A-model thus
obtained can be written in terms of the other two complex structures
$J$, $K$ as
\begin{equation}
  \frac{1 - t^2}{1 + t^2} K - \frac{2t}{1 + t^2} J
  = K \cos\alpha - J \sin\alpha \,.
\end{equation}
This agrees with the relation between the $\Omega$-deformed
Rozansky--Witten theory and the A-model which we found in
section~\ref{sec:RW-A-QM}.

Thus, we conclude that the GL-twisted $\CN = 4$ super Yang--Mills
theory on $\I \times \M_3$ with gauge group $H$ is equivalent to
Chern--Simons theory on $\M_3$ with gauge group $H_\C$, assuming that
appropriate boundary conditions are imposed.  The boundary conditions
at $s = 0$ are such that when $\M_3 = \R \times \M_2$, they correspond
to the canonical coisotropic brane in the A-model with target $\SM_H$.
This conclusion is in accordance with the results obtained
in~\cite{Gukov:2008ve, Witten:2010cx, Witten:2010zr}.

\subsection{Higher-dimensional Chern--Simons theories}
\label{sec:HDCS}

There are higher-dimensional analogs of the above A-type and B-type
constructions, in which higher-dimensional variants of Chern--Simons
theories are realized by maximally supersymmetric Yang--Mills theories
in dimensions five and up.

These Chern--Simons theories are all related by dimensional reduction.
The one of highest dimension is six-dimensional Chern--Simons theory,
also known as holomorphic Chern--Simons theory~\cite{Witten:1992fb}.
This is a holomorphic theory that can be formulated on a Calabi--Yau
threefold $\M_6$.

Suppose that we have $\M_6 = \M_4 \times \R \times \S^1$ for some
Calabi--Yau twofold $\M_4$, and reduce the theory on $\S^1$.  The
holomorphy on the cylinder $\R \times \S^1$ implies that the reduced
theory is topological on $\R$.  What we get is a
holomorphic--topological theory, called five-dimensional Chern--Simons
theory~\cite{Costello:2016nkh}, placed on $\M_5 = \M_4 \times \M_1$
with $\M_1 = \R$.

If we further take $\M_4 = \M_2 \times \R \times \S^1$ and perform
reduction on $\S^1$, we get four-dimensional Chern--Simons
theory~\cite{Costello:2013zra, Costello:2013sla, Costello:2017dso} on
$\M_4 = \M_2 \times \M_2'$ with $\M_2' = \R \times \M_1$, which is
holomorphic on $\M_2$ and topological on $\M_2'$.

Finally, taking $\M_2 = \R \times \S^1$ and reducing on $\S^1$, we
reach the ordinary, fully topological, Chern--Simons theory on
$\M_3 = \R \times \M_2'$.

Since six-dimensional Chern--Simons theory sits at the top of the
hierarchy, let us begin by discussing the constructions of this
theory.  We will be concise here; some of the assertions made below
are explained in more detail in appendix~\ref{sec:8dSYM}.

The B-type construction uses eight-dimensional super Yang--Mills
theory on $\D \times \M_6$.  This is made topological on $\D$ by
twisting of the rotation group $\U(1)_\D$ with the R-symmetry
$\U(1)_R$.  The need for a $\U(1)$ R-symmetry with which the theory is
twisted along $\D$ explains why one cannot go higher than eight
dimensions in the B-type constructions.

Covariantly constant spinors on $\D \times \M_6$ are linear
combinations of the products of covariantly constant spinors on $\D$
and those on $\M_6$.  The Calabi--Yau threefold $\M_6$ has two
covariantly constant spinors, one for each chirality.  If $\D$ is
flat, there are also two on $\D$.  Accordingly, the theory has four
supercharges when $\D$ is flat.  Under $\U(1)_\D \times \U(1)_R$, two
of these supercharges have positive chirality on $\M_6$ and weights
$(\pm\frac12, \pm\frac12)$.  The other two have negative chirality on
$\M_6$ and weights $(\pm\frac12,\mp\frac12)$.  (The supercharges of
ten-dimensional super Yang--Mills theory are spinors of positive
chirality.)  These supercharges generate $\CN = (2,2)$ supersymmetry
on $\D$.  As $\U(1)_R$ rotates the scalars of the $\CN = (2,2)$ vector
multiplet, $\U(1)_R$ is the axial R-symmetry and the twist on $\D$ is
the B-twist.

As a gauged B-model, the theory has three chiral multiplets whose
scalar fields are the antiholomorphic components $A_{\bar3}$,
$A_{\bar4}$, $A_{\bar5}$ of the gauge field with respect to local
coordinates $(z^3, z^4, z^5)$ on $\M_6$.  These scalars make up a
$(0,1)$-form
\begin{equation}
  \CA
  = A_{\bar3} \rmd\zb^{\bar3}
    + A_{\bar4} \rmd\zb^{\bar4} + A_{\bar5} \rmd\zb^{\bar5}
\end{equation}
on $\M_6$ with values in $\hf_\C$, so the target space of the gauged
B-model is
\begin{equation}
  X = \Omega^{0,1}(\M_6, \hf_\C) \,.
\end{equation}
The gauge group is
\begin{equation}
  G = \Map(\M_6, H) \,,
\end{equation}
and $G_\C$ acts on $X$ by complex gauge transformations.  The
$G$-invariant K\"ahler metric is given by
\begin{equation}
  g(v,w)
  = -\frac{1}{2g_{\mathrm{8d}}^2}
    \int_{\M_6} \Tr(v \wedge \star \wb + \vb \wedge \star w)
\end{equation}
for $\hf_\C$-valued $(0,1)$-forms $v$, $w$ on $\M_6$, where
$g_{\mathrm{8d}}$ is the gauge coupling.

This example may be thought of as the complexification of the B-type
construction of Chern--Simons theory discussed in
section~\ref{sec:CS}, in the sense that the real coordinates on $\M_3$
in that example are replaced by complex coordinates on $\M_6$ here.
Correspondingly, the superpotential is essentially given by the
Chern--Simons functional constructed from $\CA$.  The Chern--Simons
form is a three-form, so to be integrated over $\M_6$, it must be
wedged with the holomorphic volume form $\Omega_3$ of $\M_6$:
\begin{equation}
  \label{eq:HDCS-W}
  W
  = \frac{2\iu}{g_{\mathrm{8d}}^2} \int_{\M_6} \Omega_3 \wedge \ChS(\CA) \,.
\end{equation}
Here we have normalized $\Omega_3$ in such a way that
$\Omega_3 \wedge \Omegab_3$ equals $-\iu/8$ times the volume form of
$\M_6$.  The overall normalization is explained in
appendix~\ref{sec:8dSYM}.

The above superpotential is the action for holomorphic Chern--Simons
theory.  Therefore, the $\Omega$-deformation of eight-dimensional
super Yang--Mills theory on $\D \times \M_6$ is holomorphic
Chern--Simons theory on $\M_6$.

The corresponding A-type construction is based on seven-dimensional
super Yang--Mills theory on $\I \times \M_6$.  The theory may be
reformulated as supersymmetric gauged quantum mechanics on $\I$ with
target $X$ and gauge group $G$, and in this description we know what
the boundary conditions should be.

Now we consider dimensional reductions to five- and four-dimensional
Chern--Simons theories.

First, let us take $\M_6 = \M_4 \times \R \times \S^1$ and reduce the
theory on $\S^1$ to get super Yang--Mills theory on
$\D \times \M_4 \times \R$.  If we write $z^3 = (x^3 - \iu x^8)/2$,
$z^4 = (x^4 - \iu x^9)/2$ and $z^5 = (x^5 - \iu x^{10})/2$, the
superpotential becomes
\begin{equation}
  W
  = \frac{1}{g_{\mathrm{7d}}^2} \int_{\M_4 \times \R}
    \Omega_2 \wedge \ChS(\CA) \,,
\end{equation}
where $\Omega_2$ is the holomorphic volume form of the Calabi--Yau
twofold $\M_4$, defined by $\Omega_3 = \Omega_2 \wedge \rmd z_3$, and
\begin{equation}
  \CA
  = A_{\bar3} \rmd z^3 + A_{\bar4} \rmd x^4
    + (A_5 + \iu\phi_{10}) \rmd x^5 \,,
\end{equation}
with the scalar $\phi_{10}$ coming from $A_{10}$.

This superpotential is the action for five-dimensional Chern--Simons
theory, so the $\Omega$-deformation of seven-dimensional super
Yang--Mills theory on $\D \times \M_4 \times \R$ is five-dimensional
Chern--Simons theory on $\M_4 \times \R$.  The A-type construction
involves six-dimensional $\CN = (1,1)$ super Yang--Mills theory on
$\I \times \M_4 \times \R$.

Taking $\M_4 = \M_2 \times \R \times \S^1$ and reducing the
seven-dimensional theory on $\S^1$, we get six-dimensional
$\CN = (1,1)$ super Yang--Mills theory on
$\D \times \M_2 \times \R^2$.  The superpotential for the gauged
B-model is
\begin{equation}
  W
  = \frac{\iu}{2g_{\mathrm{6d}}^2} \int_{\M_2 \times \R^2}
    \Omega_1 \wedge \ChS(\CA) \,,
\end{equation}
with $\Omega_2 = \Omega_1 \wedge \rmd z^4$ and
\begin{equation}
  \CA
  = A_{\bar3} \rmd z^3 + (A_4 + \iu\phi_9) \rmd x^4
    + (A_5 + \iu\phi_{10}) \rmd x^5 \,.
\end{equation}

With the $\Omega$-deformation turned on, this setup yields
four-dimensional Chern--Simons theory on
$\M_2 \times \R^2$~\cite{Costello:2018txb}.  The A-type construction
makes use of five-dimensional $\CN = 2$ super Yang--Mills theory on
$\I \times \M_2 \times \R^2$.  This setup is essentially the same as
the one proposed in~\cite{Ashwinkumar:2018tmm, Ashwinkumar:2019mtj}.

Taking $\M_2 = \R \times \S^1$ and reducing these theories on $\S^1$,
we get back to the A-type and B-type constructions for Chern--Simons
theory on $\R^3$, described in section~\ref{sec:CS}.

One can analyze the above sequence of dimensional reduction and deduce
the relevant twist for each of the theories that appear in these
constructions.  A comprehensive list of twists and the relations
between them under dimensional reduction may be found
in~\cite{Elliott:2020ecf}.

The B-type constructions use $(5+n)$-dimensional maximally
supersymmetric Yang--Mills theories on
$\D \times \M_{2n} \times \R^{3-n}$.  Starting from $n = 3$ and going
down to $n = 2$, $1$ and $0$, we find that they are twisted along
$\D \times \R^{3-n}$ with the $\Spin(5-n)$ R-symmetry.  In particular,
for $n = 0$, we have the fully twisted $\CN = 2$ super Yang--Mills
theory on $\D \times \R^3$.

On the A-type side, we have $(4+n)$-dimensional maximally
supersymmetric Yang--Mills theories on
$\I \times \M_{2n} \times \R^{3-n}$.  As we argued in
section~\ref{sec:CS}, for $n = 0$, we get the GL-twist on
$\I \times \R^3$ with the parameter $t$ determined by the phase of the
$\Omega$-deformation parameter $\eps$.  Tracing back the sequence of
dimensional reduction, we deduce that the theories are twisted along
$\I \times \R^{3-n}$ with an $\Spin(4-n)$ subgroup of the R-symmetry
group.

\section*{Acknowledgments}

We would like to thank Meer Ashwinkumar for useful correspondences and
Dylan Butson for explaining his forthcoming work.  NI gratefully
acknowledges support from NSF Grant PHY-1911298.  The research of JY
is supported by the Perimeter Institute for Theoretical Physics.
Research at Perimeter Institute is supported in part by the Government
of Canada through the Department of Innovation, Science and Economic
Development Canada and by the Province of Ontario through the Ministry
of Colleges and Universities.

\appendix

\section{\texorpdfstring{Eight-dimensional super Yang--Mills theory as
    $\CN = (2,2)$ supersymmetric gauge theory}{Eight-dimensional super
    Yang--Mills theory as N = (2,2) supersymmetric gauge theory}}
\label{sec:8dSYM}

In this appendix we formulate eight-dimensional super Yang--Mills
theory as an $\CN = (2,2)$ supersymmetric gauge theory when part of
spacetime is a Calabi--Yau threefold.  Since the theory is the
dimensional reduction of ten-dimensional super Yang--Mills theory, we
will first rewrite the supersymmetry of the latter theory in the form
of $\CN = (2,2)$ supersymmetry.

To begin with, let us recall spinors in ten dimensions.  We work in
Euclidean signature.

Spinors in $\R^{10}$ are representations of the Clifford algebra,
generated by the gamma matrices $\Gamma_I$, $I = 1$, $\dotsc$, $10$,
satisfying the relations
\begin{equation}
  \{\Gamma_I, \Gamma_J\} = 2\delta_{IJ} \,.
\end{equation}
The chirality operator $-\iu\Gamma_1 \dotsm \Gamma_{10}$ has
eigenvalues $\pm 1$.

Introduce the complex coordinates $z^i = x^{2i-1} + \iu x^{2i}$,
$i = 1$, $\dotsc$, $5$.  Then, the corresponding gamma matrices
\begin{align}
  \gamma_i &= \frac12 (\Gamma_{2i-1} - \iu\Gamma_{2i}) \,,
  \\
  \gammab_i &= \frac12 (\Gamma_{2i-1} + \iu\Gamma_{2i}) \,
\end{align}
obey the relations
\begin{equation}
  \{\gamma_i, \gammab_j\} = \delta_{ij} \,.
\end{equation}
Thus, one may think of $\gammab_i$ as fermion creation operators and
$\gamma_i$ as annihilation operators.

In this language, spinors are states in the fermionic Fock space built
on the vacuum by the action of the creation operators.  The chirality
operator can be expressed in terms of the fermion number operators
$\gammab_i \gamma_i$ as
\begin{equation}
  -\iu\Gamma_1 \dotsm \Gamma_{10}
  = \prod_{i=1}^5 (2\gammab_i \gamma_i - 1) \,.
\end{equation}
Spinors of positive chirality have an odd number of excitations, and
those of negative chirality have an even number of excitations.  They
have sixteen components.

A concise way of presenting spinors on $\C^5$ is to use differential
forms.  In this presentation, spinors are linear combinations of
$(p,0)$-forms, with $p = 0$, $\dotsc$, $5$.  The action of the gamma
matrices are operations changing the form degree by one:
\begin{align}
  \gamma_i &= \iota_{\del_i} \,,
  \\
  \gammab_i &= \rmd z^i \wedge \,.
\end{align}
Positive chirality spinors and negative chirality ones are odd forms
and even forms, respectively.  The product $\alphab \beta$ of two
spinors $\alpha$, $\beta$ is defined by
\begin{equation}
  \alphab \beta = (\alpha^t \wedge \beta)^{\mathrm{top}} \,,
\end{equation}
where the transpose acts as
\begin{equation}
  (\rmd z^{i_1} \wedge \dotsb \wedge \rmd z^{i_p})^t
  = \rmd z^{i_p} \wedge \dotsb \wedge \rmd z^{i_1}
\end{equation}
and $\alpha^{\mathrm{top}}$ is the top component of $\alpha$.

Now we consider super Yang--Mills theory on $\C^5$.  The theory has a
gauge field $A$ and a spinor field $\Psi$, and supersymmetry
transforms them by
\begin{align}
  \label{eq:10dSYM-SUSY-A}
  \delta A_I &= \iu\epsb \Gamma_I \Psi \,,
  \\
  \label{eq:10dSYM-SUSY-Psi}
  \delta\Psi &= \frac14 [\Gamma_I, \Gamma_J] F^{IJ} \eps \,.
\end{align}
Both the fermion $\Psi$ and the supersymmetry parameter $\eps$ are
spinors of positive chirality.

If the spacetime is not flat, $\eps$ must be a covariantly constant
spinor.  We are interested in the case when the spacetime is
$\C \times \M_6 \times \C$, where $\M_6$ is a Calabi--Yau threefold.
In this case, covariantly constant spinors are either zero-forms or
top forms on $\M_6$ since they are the only forms that are invariant
under the holonomy group $\SU(3)$.  Hence, using the holomorphic
volume form $\Omega_3$ of $\M_6$, we can write
\begin{equation}
  \eps
  =
  \eps_- \rmd z^1 + \eps_+ \rmd z^5
  - \epsb_+ \Omega_3 + \epsb_- \rmd z^1 \wedge \Omega_3 \wedge \rmd z^5 \,.
\end{equation}
We also write $\Psi$ as
\begin{multline}
  \Psi
  =
  -\lambda_- \rmd z^1 - \lambda_+ \rmd z^5
  + \lambdab_+ \Omega_3 - \lambdab_- \rmd z^1 \wedge \Omega_3 \wedge \rmd z^5
  \\
  + \sum_{k=2}^4 \bigl(
      -\iu\psib^k_+ \rmd z^k
      + \iu\psib^k_- \rmd z^1 \wedge \rmd z^k \wedge \rmd z^5
      + \iu\psi^k_+ \iota_{\del_k} \Omega_3 \wedge \rmd z^5
      + \iu\psi^k_- \rmd z^1 \wedge  \iota_{\del_k} \Omega_3\bigr) \,,
\end{multline}
where we have chosen complex coordinates $(z^2, z^3, z^4)$ on $\M_6$
such that
\begin{equation}
  \Omega_3 = \rmd z^2 \wedge \rmd z^3 \wedge \rmd z^4 \,.
\end{equation}

Plugging the above expressions for $\eps$ and $\Psi$ into the
supersymmetry variation~\eqref{eq:10dSYM-SUSY-A} of $A$, we get
\begin{align}
  \label{eq:10dSYM-SUSY-A1}
  \delta A_1
  &= -\iu\eps_-\lambdab_- - \iu\epsb_-\lambda_- \,,
  \\
  \delta A_{\bar1}
  &= \iu\eps_+\lambdab_+ + \iu\epsb_+\lambda_+ \,,
  \\
  \delta A_5
  &= -\iu\eps_+\lambdab_- - \iu\epsb_-\lambda_+ \,,
  \\
  \delta A_{\bar5}
  &= -\iu\epsb_+\lambda_- - \iu\eps_-\lambdab_+ \,,
  \\
  \delta A_k
  &= -\epsb_+\psib^k_- + \epsb_-\psib^k_+ \,,
  \\
  \delta A_\kb
  &= \eps_+\psi^k_- - \eps_-\psi^k_+ \,.
\end{align}
From the supersymmetry variation~\eqref{eq:10dSYM-SUSY-A} of $\Psi$,
we find
\begin{align}
  \delta\lambda_+
  &= \eps_+ (2F_{1\bar1} - 2F_{5\bar5} - F^k{}_k) + 4\eps_- F_{\bar15} \,,
  \\
  \delta\lambda_-
  &= -\eps_- (2F_{1\bar1} - 2F_{5\bar5} + F^k{}_k) - 4\eps_+ F_{1\bar5} \,,
  \\
  \delta\lambdab_+
  &= \epsb_+ (2F_{1\bar1} + 2F_{5\bar5} + F^k{}_k) + 4\epsb_- F_{\bar1\bar5} \,,
  \\
  \delta\lambdab_-
  &= -\epsb_- (2F_{1\bar1} + 2F_{5\bar5} - F^k{}_k) - 4\epsb_+ F_{15} \,,
  \\
  \delta\psi^k_+
  &= 4\iu\epsb_- F_{\bar1\kb}
     + 4\iu\epsb_+ F_{5\kb}
     - 2\iu\eps_+ \veps^{klm} F_{lm} \,,
  \\
  \delta\psi^k_-
  &= 4\iu\epsb_+ F_{1\kb}
     - 4\iu\epsb_- F_{\bar5\kb}
     - 2\iu\eps_- \veps^{klm} F_{lm} \,,
  \\
  \delta\psib^k_+
  &= -4\iu\eps_- F_{\bar1 k}
     - 4\iu\eps_+ F_{\bar5 k}
     + 2\iu\epsb_+ \veps^{klm} F_{\lb\mb} \,,
  \\
  \label{eq:10dSYM-SUSY-psibk-}
  \delta\psib^k_-
  &= -4\iu\eps_+ F_{1k}
     + 4\iu \eps_- F_{5k}
     + 2\iu \epsb_- \veps^{klm} F_{\lb\mb} \,.
\end{align}
Here $\veps^{klm}$, $k$, $l$, $m = 2$, $3$, $4$, are the components of
a completely antisymmetric tensor on $\M_6$, with $\veps^{234} = 1$.

We compare these equations with $\CN = (2,2)$ supersymmetry
transformation laws.%
\footnote{The following formulas are obtained from $\CN = 1$
  supersymmetry transformations in $3+1$ dimensions by dimensional
  reduction in the $x^1$- and $x^2$-directions.  Compared to the
  formulas in~\cite{Witten:1993yc}, we have made the rescaling
  $\sigma \to \sqrt2 \sigma$, $\phi \to \sqrt2 \phi$ and
  $\auxF \to \sqrt2\auxF$.}
An $\CN = (2,2)$ vector multiplet transforms under supersymmetry as
\begin{align}
  \delta A_\pm
  &= 2\iu\eps_\pm \lambdab_\pm + 2\iu\epsb_\pm \lambda_\pm \,,
  \\
  \delta\sigma
  &= -\iu\epsb_+ \lambda_- - \iu\eps_- \lambdab_+ \,,
  \\
  \delta\sigmab
  &= -\iu\eps_+ \lambdab_- - \iu\epsb_- \lambda_+ \,,
  \\
  \delta\lambda_+
  &= -\eps_+(F_{01} - 2[\sigma,\sigmab] - \iu\auxD)
     + 2\eps_-  D_+ \sigmab \,,
  \\
  \delta\lambda_-
  &= \eps_-(F_{01} - 2[\sigma,\sigmab] + \iu\auxD)
     + 2\eps_+ D_- \sigma\,,
  \\
  \delta\lambdab_+
  &= -\epsb_+(F_{01} + 2[\sigma,\sigmab] + \iu\auxD)
     + 2\epsb_- D_+ \sigma \,,
  \\
  \delta\lambdab_-
  &= \epsb_-(F_{01} + 2[\sigma,\sigmab] - \iu\auxD)
     + 2\epsb_+ D_- \sigmab\,,
  \\
  \begin{split}
  \delta\auxD
  &=
    \eps_+(D_-\lambdab_+ + 2 [\sigma, \lambdab_-])
    + \eps_-(D_+\lambdab_- + 2 [\sigmab, \lambdab_+])
    \\&\qquad
    - \epsb_+(D_-\lambda_+ + 2 [\sigmab, \lambda_-])
    - \epsb_-(D_+\lambda_- + 2 [\sigma, \lambda_+]) \,.
  \end{split}
\end{align}
An $\CN = (2,2)$ chiral multiplet transforms as
\begin{align}
  \delta\varphi
  &= \eps_+\psi_- - \eps_-\psi_+ \,.
  \\
  \delta\psi_+
  &= 2\iu\epsb_- D_+ \varphi
      + 4\iu \epsb_+ \sigmab \varphi + 2\eps_+\auxF \,,
  \\
  \delta\psi_-
  &= -2\iu\epsb_+ D_- \varphi
     - 4\iu\epsb_-\sigma \varphi + 2\eps_- \auxF \,,
  \\
  \delta\auxF
  &= -\iu\epsb_+(D_-\psi_+ + 2\sigmab\psi_- - 2\iu\lambdab_-\varphi)
     - \iu\epsb_-(D_+\psi_- + 2\sigma\psi_+ + 2\iu\lambdab_+\varphi) \,.
  \\
  \delta\varphib
  &= -\epsb_+\psib_- + \epsb_-\psib_+ \,.
  \\
  \delta\psib_+
  &= -2\iu\eps_- D_+ \varphib
      - 4\iu \eps_+ \sigma \varphib + 2\epsb_+\auxFb \,,
  \\
  \delta\psib_-
  &= 2\iu\eps_+ D_- \varphib
     + 4\iu\eps_-\sigmab \varphib + 2\epsb_- \auxFb \,,
  \\
  \delta\auxFb
  &= -\iu\eps_+(D_-\psib_+ + 2\sigma\psib_- + 2\iu\lambda_-\varphib)
     - \iu\eps_-(D_+\psib_- + 2\sigmab\psib_+ - 2\iu\lambda_+\varphib) \,.
\end{align}
These are the supersymmetry transformations in Minkowski spacetime
$\R^{1,1}$ with coordinates $(x^0, x^1)$; we have defined
$A_\pm = A_0 \pm A_1$, $\del_\pm = \del_0 \pm \del_1$ and
$D_\pm = \del_\pm + [A_\pm, \ ]$.

Taking Wick rotation $x^0 \mapsto -\iu x^2$ into account, we see that
the supersymmetry
variations~\eqref{eq:10dSYM-SUSY-A1}--\eqref{eq:10dSYM-SUSY-psibk-}
are precisely of the form of $\CN = (2,2)$ supersymmetry
transformations if we identify
\begin{align}
  \sigma &= A_{\bar5} \,,
  \\
  \auxD &= \iu F^k{}_k \,,
  \\
  \varphi^k &= A_\kb \,,
  \\
  \auxF^k &= -\iu\veps^{klm} F_{lm} \,.
\end{align}
Therefore, the scalars $\varphi^k$ of three chiral multiplets, labeled
by $k = 2$, $3$, $4$, are the components of the one-form
\begin{equation}
  \CA = A_\kb \rmd \zb^\kb \,.
\end{equation}

Let us reduce the theory to eight-dimensional super Yang--Mills theory
on $\C \times \M_6$.  This turns the components $A_5$, $A_{\bar5}$ of
the gauge field to scalars $\sigmab$, $\sigma$.  As an $\CN = (2,2)$
supersymmetric gauge theory, the target space of the theory is the
space of $\hf_\C$-valued $(0,1)$-forms on $\M_6$, where $\hf$ is the
Lie algebra of the gauge group.  We need to determine the
superpotential.

To conform to the notations used in the main body of the paper, we
should relate the $\CN = (2,2)$ supersymmetry transformations just
described to the supersymmetry
transformations~\eqref{eq:OGB-SUSY-A}--\eqref{eq:OGB-SUSY-Gb} for the
$\Omega$-deformed gauged B-model.  The $\Omega$-deformed B-twisted
supersymmetry is realized by the following supersymmetry parameters:
\begin{align}
  \epsb_\pm &= \pm\frac{\iu}{2} \,,
  \\
  \eps_\pm &= -\frac12 (V^1 \pm \iu V^2) \,.
\end{align}
It is easy to identify various fields across the two notations; in
particular, we have the identification
\begin{equation}
  \auxF = -\iu \auxF_{12} \,.
\end{equation}
The right-hand side is the component of the two-form auxiliary field
$\auxF$ in the $\Omega$-deformed B-model.

If we normalize the kinetic term of super Yang--Mills theory as
\begin{equation}
  -\frac{1}{2g_{\mathrm{8d}}^2}
  \int_{\C \times \M_6} \Tr(F \wedge \star F) \,,
\end{equation}
then the part of the B-model action that contains the auxiliary fields
$\auxF$, $\auxFb$ is
\begin{equation}
  \int_{\C \times \M_6}
  \biggl(
  -\frac{2}{g_{\mathrm{8d}}^2}
  \Tr(\auxF \wedge \star\auxF)
  +
  \auxF^k \frac{\delta\BW}{\delta A_\kb}
  - \auxFb^\kb \frac{\delta\BWb}{\delta A_k}\biggr)
  \,.
\end{equation}
Choosing a basis $\{T_a\}$ of $\hf$ such that
$\Tr(T_a T_b) = -\delta_{ab}$, we can write the equation of motion for
$\auxF$ as
\begin{equation}
  \frac{\delta\BWb}{\delta A_k^a}
  = \frac{2}{g_{8d}^2} \auxF_{12}^{ka}
  = \frac{2}{g_{8d}^2} \veps^{klm} F_{lm}^a \,,
\end{equation}
Since
\begin{equation}
  \begin{split}
    \delta \int_{\M_6} \Omegab_3 \wedge \ChS(\CAb)
    &= 2 \int_{\M_6} \Omegab_3 \wedge \Omega_3 \Tr(\veps^{klm} \delta A_k F_{lm})
    \\
    &= -16\iu \int_{\M_6} \star \Tr(\veps^{klm} \delta A_k F_{lm}) \,,
  \end{split}
\end{equation}
we deduce
\begin{equation}
  W = \frac{\iu}{8g_{\mathrm{8d}}^2} \int_{\M_6} \Omega_3 \wedge \ChS(\CA) \,.
\end{equation}

In section~\ref{sec:HDCS}, we make a different choice for the chiral
multiplet scalars.  There, the real coordinates on $\M_6$ are
$(x^3, x^4, x^5,x^8, x^9, x^{10})$, and we treat $A_{\bar3}$,
$A_{\bar4}$, $A_{\bar4}$ in the complex coordinates
$z^3 = (x^3 - \iu x^8)/2$, $z^4 = (x^4 - \iu x^9)/2$ and
$z^5 = (x^5 - \iu x^{10})/2$ as chiral multiplet scalars.  This means
that the volume form of $\M_6$ is
$(-\rmd x^3 \wedge \rmd x^8) \wedge (-\rmd x^4 \wedge \rmd x^9) \wedge
(-\rmd x^5 \wedge \rmd x^{10})$, and the holomorphic volume form
$\Omega_3 = \rmd z^2 \wedge \rmd z^3 \wedge \rmd z^4$ is normalized in
such a way that $\Omega_3 \wedge \Omegab_3$ is $-\iu/8$ of the volume
form.  The equation for motion for the auxiliary field is modified to
$\auxF^k = -2\iu\veps^{klm} F_{lm}$, but in the action there is an
extra factor of $1/4$ multiplying the quadratic term in the auxiliary
fields, so we obtain the superpotential~\eqref{eq:HDCS-W}.

\providecommand{\href}[2]{#2}\begingroup\raggedright\endgroup
\end{document}